\documentclass[11pt,a4paper]{article}
\usepackage{graphicx,amsmath,natbib,bm,url}
\usepackage[utf8]{inputenc} 
\usepackage[hyperfootnotes=false]{hyperref}
\usepackage{adjustbox}
\usepackage{subcaption}
\usepackage{authblk}
\usepackage{graphicx}
\usepackage{tikz}

\usepackage{color}  
\definecolor{darkblue}{rgb}{0.15,0,0.37}
\definecolor{darkred}{rgb}{0.35,0,0.08}
\definecolor{mygrey}{rgb}{0.85,0.85,0.85} 
\definecolor{babyblueeyes}{rgb}{0.63, 0.79, 0.95}
\definecolor{bleudefrance}{rgb}{0.19, 0.55, 0.91}

\hypersetup{
	colorlinks=true,
	hypertexnames=true,
	colorlinks,		
	linkcolor={bleudefrance},
	citecolor={bleudefrance},
	urlcolor={bleudefrance},
	pdfstartview={FitV},
	unicode,
	breaklinks=true
} 

\usepackage{microtype,todonotes}
\usepackage[a4paper,text={15.5cm, 26.5cm},centering]{geometry} 
\usepackage{paralist}
\usepackage{url}

\usepackage{mathpazo} 

\usepackage{longtable, booktabs, tabularx} 
\usepackage{caption,fixltx2e}  
\usepackage[flushleft]{threeparttable}  

\usepackage[onehalfspacing]{setspace} 

\usepackage[marginal]{footmisc} 
\usepackage{pdflscape} 

\usepackage[capposition=top]{floatrow} 

\usepackage{color, colortbl}
\usepackage{multirow}

\newcommand\blfootnote[1]{%
	\begingroup
	\renewcommand\thefootnote{}\footnote{#1}%
	\addtocounter{footnote}{-1}%
	\endgroup
}

\usepackage{fancyhdr}
\usepackage{lastpage}
\usepackage[labelfont={bf,it},textfont=it]{caption}
\date{}

\begin{document}

\vspace*{0.5cm}

\thispagestyle{empty} 

\begin{center}
{\LARGE Save the Farms: Nonlinear Impact of Climate Change on Banks' Agricultural Lending}\\[0.4cm]
{\large Teng Liu}\\[0.4cm]
{\normalsize \textit{September 1, 2024}}\\ [0.1cm]
{\normalsize (\textit{For latest version, click \href{https://www.dropbox.com/s/simioylaaxbxozi/climate_farm_bank.pdf?dl=0}{here}}})\\[0.2cm]
\end{center}

\vspace*{1cm}
\begin{center}
	\begin{minipage}{0.9\textwidth}
		\textbf{Abstract.} 
The agricultural sector is particularly susceptible to the impact of climate change. In this paper, I investigate how vulnerability to climate change affects U.S. farms' credit access, and demonstrates that such impact is unequally distributed across farms. I first construct a theoretical framework of bank lending to farms faced with climate risks, and the model helps discipline ensuing empirical analyses that use novel panel datasets at county and at bank levels. I find that higher exposure to climate change, measured by temperature anomaly, reduces bank lending to farms. Such impact is persistent, nonlinear, and heterogeneous. Small and medium farms almost always experience loss of loan access. In comparison, large farms see less severe credit contraction, and in some cases may even see improvement in funding. While small banks carry the burden of continuing to lend to small farms, their limited market share cannot compensate for the reduction of lending from medium and large banks. These results suggest that factors such as farm size and bank type can amplify the financial impact of climate change.
				\\[0.3cm]

		\begin{footnotesize}
		\end{footnotesize}
	\end{minipage}
\end{center}

\blfootnote{\textit{\textbullet \phantom{a} University of California, Santa Cruz.  \href{mailto:tedliu@ucsc.edu}{tedliu@ucsc.edu}. I am grateful for the financial support of the Food System Research Fund (FSRF) and the Hammett Fellowship of UC Santa Cruz. I would also like to thank Galina Hale, Grace Gu, Jeremy West, Elliott Campbell, and Stacy Philpott for their guidance. I also appreciate the feedback from the participants of the 2022 OCC Symposium on Climate Risk in Banking and Finance.}}

\clearpage
\section{Introduction}

The acceleration of climate change in the United States has materialized as increasing frequency and severity of extreme weather conditions, especially since year 2010.\footnote{U.S. Climate Extremes Index (CEI)} Such weather extremes and disasters contribute to economic losses: the National Oceanic and Atmospheric Administration (NOAA) estimate that since 1980, the U.S. has experienced large-scale climate disasters with loss of over \$1.875 trillion, and over 48\% of which occurred during 2010-2020.\footnote{Billion-Dollar Weather and Climate Disasters} The agricultural sector is especially vulnerable to the risks of climate change and disasters, and the issue has received considerable academic and policy attention. For instance, \cite{USDA2012} assess that increasing temperature and extreme precipitation can reduce U.S. farms' crop productivity. Despite increasing understandings of the economic implications of climate change, not enough attention has focused on the role of financial institutions in this context, particularly with respect to the U.S. agricultural sector. 

At the same time, commercial banks play a nontrivial role in financing farms' production and operations, accounting for around 40\% of total U.S. farm debt as of 2020.\footnote{See this article by \href{https://www.americanprogress.org/issues/economy/reports/2021/01/14/494574/promoting-climate-resilient-agricultural-rural-credit/}{Willingham (2021)}} Yet U.S. farms seem to be facing increasing financial challenges since 2015, as demonstrated by the growing shares of non-performing loans.\footnote{See Figure \ref{f1} in the Appendix} A number of studies examine the effect of natural disasters on general bank lending, such as  \cite{Ivanov2020}, \cite{Koetter2020}, \cite{GARBARINO2021100874}, \cite{Schuwer2018}, \cite{Petkov2019}, \cite{BREI2019232}, and \cite{CORTES2017182}.\footnote{In particular, \cite{Ivanov2020} is similar in that they examine how banking networks amplify the effect of natural disasters. Two key differences with my paper: 1).  \cite{Ivanov2020} use the restricted access Shared National Credit data, focusing on syndicated loans. 2). Their paper does not explicitly measure the vulnerability to climate change} Yet few existing studies specifically focus on bank loans to the U.S. agricultural sector. 


My paper contributes to the literature by examining how farms' credit access is affected by their exposure to climate change and disasters. Compared with companies in other sectors, farms absorb the financial impact of climate change more directly: farms that have crop or livestock failures due to climate change will experience revenue loss, thus are likely to default on their bank loans. It is also generally not feasible for farms to relocate. At the same time, understanding whether the agricultural sector has adequate financing is important to build a resilient food system in response to climate change.\footnote{ Answering the credit access question is important, as it lays the foundation for further understanding U.S. farms' climate resilience: after a climate-related disaster, it is possible that farms in U.S. counties with more bank lending recover faster than counties with less bank financing, and they may also adapt better to climate change. Thus there may be systematic differences in counties' adaptability to climate risks, which can partly be explained by bank lending.} To the best of my knowledge, this paper is one of the first studies that detect the impact of climate change vulnerability on agricultural lending, and the results pay careful attention to the distribution of such effects. More specifically, the analyses reveal that the impact varies by farm size, income area, agricultural region, and bank size. The most similar paper to my work is a concurrent working paper by \cite{Islam2022}. The main difference of my paper is that I focus on contrasting the differential lending outcomes on large versus small farms, which offers richer explanation of bank lending behavior, and more importantly pointing to the inequality of financial access that could be amplified by climate change.

In order to help answer the research question, I first construct a two-period model of bank lending to farms. The framework provides qualitative predictions on the directions of impact of climate risks, and guides the interpretation of results in the empirical section. In this simple, stylized economy, the farm seeks a loan from the bank, in anticipation of larger financing needs due to climate change. When evaluating whether to provide a loan, the bank assesses the farm's expected payoff, taking into account factors such as the farm's productivity. If assuming that a larger farm has higher productivity, the model predicts that the bank is more likely to lend to the larger farm than small farm, even if their exposure to climate change is identical. 

To uncover differences in farms' credit access empirically, I focus on exploiting county-level and bank-level variations. The key variable in this paper is banks' lending to farms, which is measured using data from the Community Reinvestment Act (CRA) housed at the Federal Financial Institutions Examination Council (FFIEC). While a few papers like \cite{CRAbord} have used such data to examine questions related to local banking, CRA agricultural lending is a novel dataset to use for questions related to climate change. It is publicly available, and offers geographic and temporal variations that are necessary to identify the effects of climate change.  Then the bank lending data are merged with measures of climate change vulnerability. The main measurement in this paper is the county-specific temperature anomaly, typically used in the scientific literature to proxy for climate change. 

Following existing studies such as \cite{Burke2015}, I use nonlinear econometric specifications to estimate the effects of climate change on bank lending to farms. The analysis differentiates between lending to large farms and that to smaller farms, as these groups may be fundamentally different in their production structure and financial capabilities. The results from Section \ref{countyresults} at the county-level confirm the predictions from the conceptual framework: given the same county-specific temperature anomaly, it is more likely for smaller farms to be denied loans than large farms.\footnote{While such overall patterns are statistically significant for all U.S. counties, there is also a range of regional heterogeneity. For example, the divergent effects on small and large are most clearly seen in northern states such as the Dakotas and southern states such as Mississippi and Louisiana.} Additionally, farms located in Census-designated middle income areas experience the impact most significantly, compared with those in lower and high income areas. 

Further, the bank-level estimation in Section \ref{banklevel} is consistent with county-level results, and also reveals that the size of banks also plays a role in lending patterns. Large banks in general are less willing to lend, small-mid banks lend less to small farms and more to large farms, and very small banks lend more to small farms. Due to the dominance of large banks in smaller loans, on which small farms most likely rely, the reduction of this type of lending may explain the credit access loss that small farms experience at the county aggregate level. 

Based on the econometric estimation, I also conduct scenario analysis of the impact of future climate scenarios at county and bank levels. More specifically, I calculate marginal effects of a given temperature anomaly (of a given climate scenario) on lending patterns. As a climate scenario becomes more pessimistic (higher temperature), the magnitudes of effects become larger, as do the divergent impact across farm types, suggesting the  widening gap of financial access due to climate change. 

The paper proceeds as follows. Section \ref{model} delineates the two-period model and its qualitative predictions. Section \ref{datasection} describes the datasets, how temperature measurements are used, and summary statistics. Section \ref{econometrics} shows the econometric framework, and presents the results at the county-aggregate level and bank-level. Section \ref{conclusion} concludes and discusses the implications of the results. 

\section{Conceptual Framework}\label{model}

\subsection*{Timeline and Environment}
\begin{figure}[!]
	\centering\includegraphics[width=10cm]{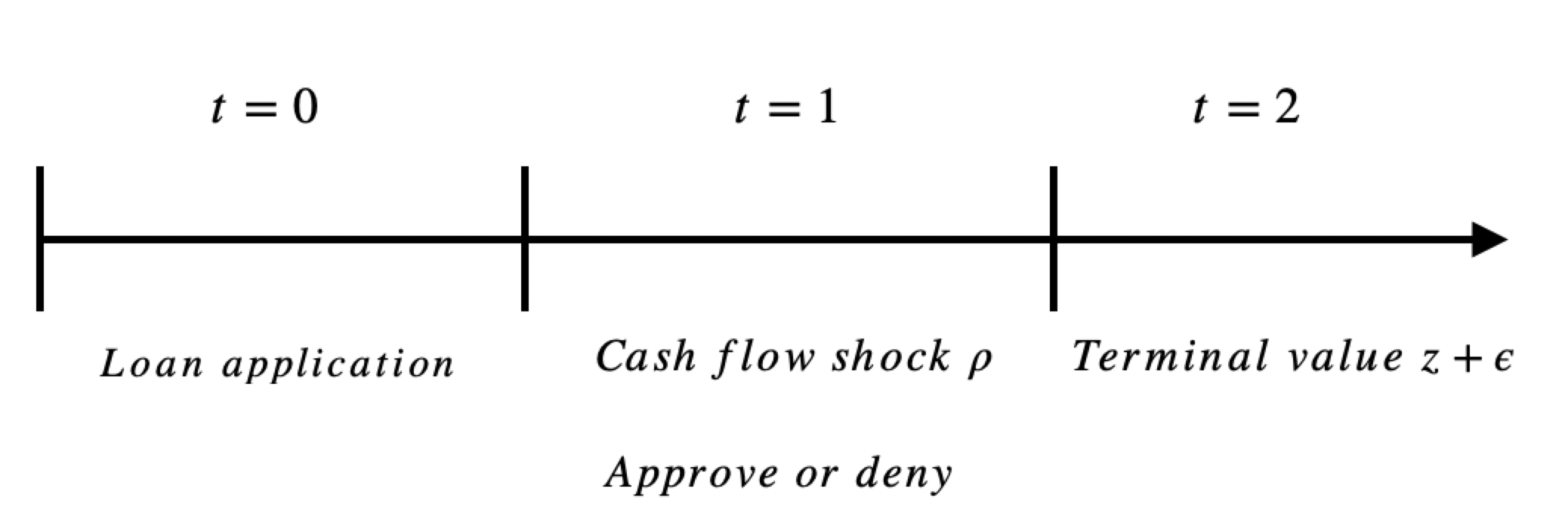}
	\caption{Timeline of Bank Lending to Farms}\label{ff2}
\end{figure}

To illustrate the link between climate change and bank lending to farms, this section provides a conceptual framework\footnote{The current framework assumes one is agnostic about the bank type.  As illustrated in the empirical section, bank size/type plays a role in lending decisions. In future studies, it is worth further developing this model to consider the role of bank type} of loan contracting based on \cite{CHODOROWREICH2021} and \cite{Holmstrom1998}, with applications to the distributional assumptions specific to climate change shocks and the importance of farm sizes. In a stylized economy, there is one bank and one farm. The farmer faces uncertain output and lives for only 2 periods. The timeline of the framework is shown in Figure \ref{ff2}. At $t=0$, the farm applies for a loan (or credit line) with the bank, and this loan is to be used next period. At $t=1$, due to climate change, the cash-flow shock $\rho$ materializes: to continue operations, the farmer needs to inject $\rho$ per unit of asset, and  $\rho 	\sim F=\mathcal{N}\left(\mu, \sigma^{2}\right)$. When  $\rho > 0$, it means the farm has cash needs.  The bank approves or denies the loan, based on the cash-flow shock and uncertain terminal value. If approved, the credit line is $\hat{\rho}$. If the farm has been denied credit, it has to shut down.\footnote{In other words, we assume that the farm has no other way to raise alternative capital} At $t=2$, the farm yields payoff $z + \epsilon$ per unit of asset. The shock $\epsilon$ is uncorrelated with $\rho$ and has mean zero, and is observable at $t=1$ if the bank pays a monitoring cost of $\xi$. 

In other words, there are two states in the system: $\rho$ and $\epsilon$. It is worth discussing further the key parameters in the framework, and why it is a simple yet realistic representation of the shocks from climate risks. The two parameters that describe the distribution of $\rho$ are mean $\mu$ and standard deviation $\sigma$. Science literature generally uses temperature anomaly to measure climate change, and some studies suggest that in recent decades, the bell curve of temperature has changed. More specifically, on a global scale, the mean of temperature anomaly has shifted to the right, and the tails of the curve have become `fatter'.\footnote{For a graphical illustration, see Figure 4 of \cite{Hansen2012}} In other words, the temperature is becoming higher than normal, and it is more likely for extreme temperature to occur. Relating this to the framework, it means that the size of the cash-flow shock that a farmer faces may become bigger and more extreme than historical averages, which has implications for whether and how much the bank provides credit. 

Moreover, $z$ is the parameter that contributes to the farm's terminal value per unit of asset at $t=2$. One can interpret this parameter as the fundamentals or productivity of a farm. Besides, a financial friction exists in the economy such that when applying for credit, the farm can only pledge a fraction, indicated by $\theta$ of their terminal value as collateral. 

For simplicity, this framework only focuses on discretion as the contractual form. Under discretion, the bank can choose to monitor the farm at $t=1$. In other words, the bank can terminate the loan/credit line upon observing cash-flow shock $\rho$ or $\epsilon$ shock to terminal value. Put another way, even if the initial loan contract is established at $t=0$, the fund is not disbursed until the next period. Then at  $t=1$, the bank observes the cash-flow shock $\rho$. The bank can also choose to monitor the shock $\epsilon$ to farm by paying a cost of $\xi$. After considering the potential payoff, the bank can choose to abandon the loan contract. In the literature, commitment is another possible contractual form, but empirical studies such as \cite{Sufi2009} show that bad shocks can modify or end credit line. More importantly, the commitment contract is less insightful in illustrating how credit provision varies with both the states of $\rho$ and $\epsilon$, as the bank grants credit only if $\rho < \theta z$ (and monitoring does not exist under the commitment contract).\footnote{More specifically, under commitment, the solution to $\hat{\rho}=\mu+\sigma h^{-1}\left(\frac{\mu-\theta z}{\sigma}\right)$, where $h^{-1}$ is the inverse of $h(x)=\phi(x) / \Phi(x)$, the ratio of standard normal probability density function to cumulative density function. For derivation details, see \cite{CHODOROWREICH2021}.}

\subsection*{Equilibrium}
To solve for the equilibrium
under the discretion contract, the bank simultaneously decides on whether to monitor and whether to approve or deny the loan. The expected value of monitoring is $V^M$, and of not monitoring $V^N$. 

The bank's lending decision can be characterized as the following:
\begin{equation}
	V=\max \{V^M, V^N\}
\end{equation}
where 
\begin{equation}
	V^{N}=\max \{\theta z-\rho, 0\}
\end{equation}
and 
\begin{equation}
	V^{M}=\mathbb{E}[\max \{\theta(z+\epsilon)-\rho, 0\}]-\xi
\end{equation}

Thus, with monitoring, the bank's decision to lend or not is based on minimizing their losses, namely lending only occurs if
$
\rho<\theta(z+\epsilon)
$.

More detail on the monitoring decision is found in \cite{CHODOROWREICH2021}, and the main insight is that monitoring only makes sense when the cash-flow shock is in a range, or $\rho \in[\underline{\rho}, \bar{\rho}]$. In other words, when the cash-flow shock $\rho$ is extremely small, to the left of $\underline{\rho}$, the bank will approve the loan and there is no value in monitoring. On the other hand, when  $\rho$ is extremely large, to the right of $\bar{\rho}$, the bank always denies the loan and there is no value in monitoring. 

To derive close-form solutions for the these critical values $\underline{\rho}$ and $\bar{\rho}$, assume for simplicity that $\epsilon$ takes on 3 values $\{-e, 0, e\}$ with probability $\{q, 1-2q, q\}$. 

When the cash-flow shock $\rho$ is sufficiently small, $\theta(z-e)<\rho<\theta z$, the expected payoff of monitoring is:
\begin{equation}
	V^{M}=\underbrace{\theta z-\rho}_{V^{N}}+\underbrace{q[\rho-(\theta(z-e)]}_{\text {Monitoring value }}-\underbrace{\xi}_{\text {Monitoring cost }}
\end{equation}
such that the bank would prefer monitoring if and only if
\begin{equation}\label{ef1}
	\rho>\underline{\rho}:=\theta(z-e)+\xi / q
\end{equation}

When the cash-flow shock $\rho$ is sufficiently large, $\theta z<\rho<\theta(z+e)$, the expected payoff of monitoring is:
\begin{equation}
	V^{M}=\underbrace{0}_{V^{N}}+\underbrace{q[\theta(z+e)-\rho]}_{\text {Monitoring value }}-\underbrace{\xi}_{\text {Monitoring cost }}
\end{equation}
where $V^{N}=0$ because without monitoring, the expected payoff $\theta z - \rho$ is negative, and the bank would automatically reject the loan, such that the bank would prefer monitoring if and only if
\begin{equation}\label{ef2}
	\rho<\bar{\rho}:=\theta(z+e)-\xi / q
\end{equation}

\begin{figure}[ht!]
	\centering
	\caption{Lending Regions under Bank Discretion: Small-Medium versus Large Farms}\label{ff1} 
	\centering\includegraphics[width=9cm]{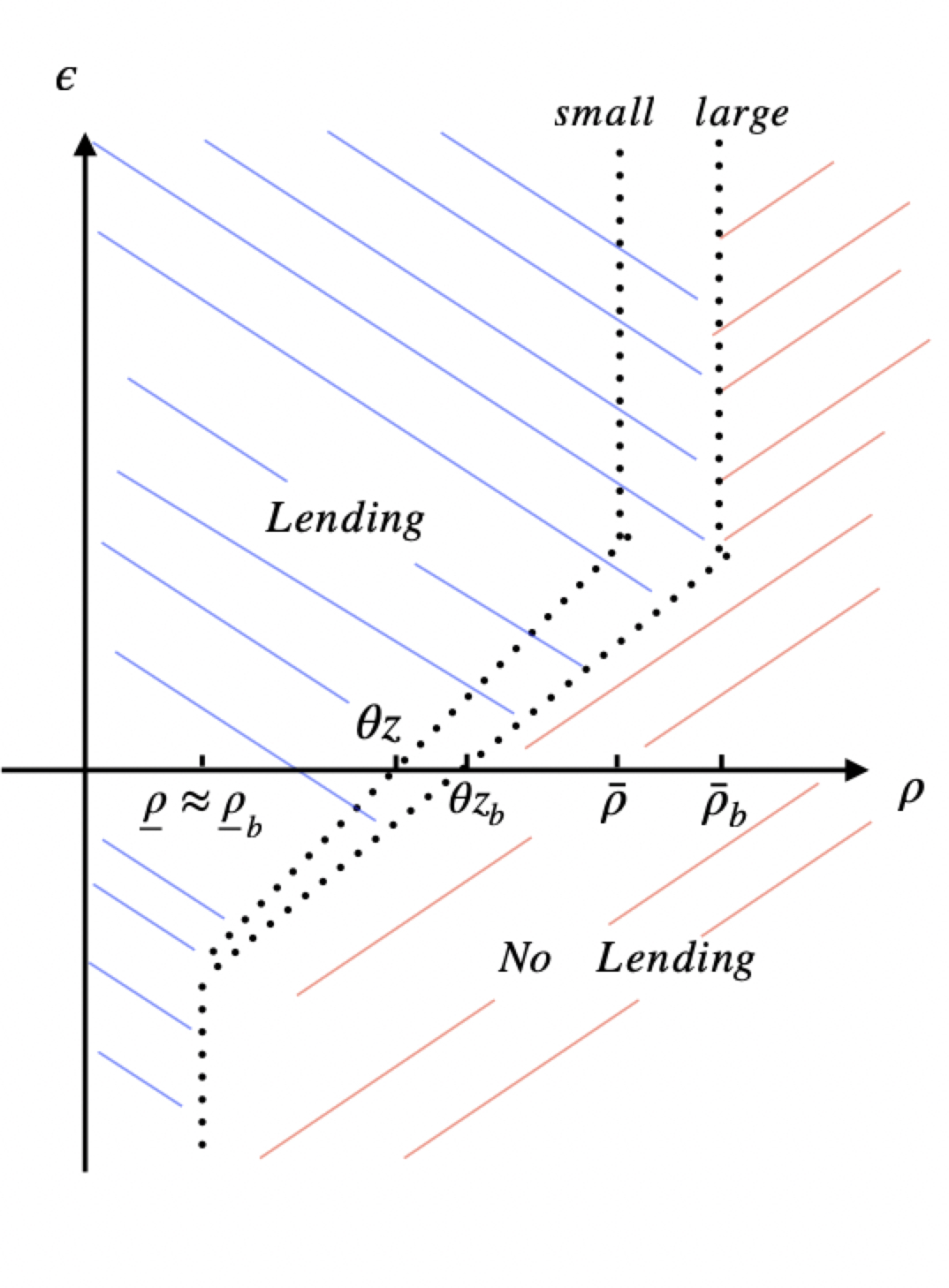}
	\vspace{10pt}
	{\scriptsize 	\\
		\textit{Note}: Compared with smaller farms, the lending region to large farms is wider. $\rho$ stands for the cash-flow shock caused by climate risks as measured by temperature anomaly; $\epsilon$ is an independent shock that contributes to a farm's terminal value. The subscript $b$ refers to the scenario in which the farm is large. \par}
\end{figure}

By combining and generalizing the above results, the bank's lending decisions can be graphically illustrated in Figure \ref{ff1}. To summarize, the lending decisions depend on both the states $\rho$ and $\epsilon$. The horizontal axis illustrates the degree of cash-flow shock: the more to the right, the higher a farm's cash needs are. The vertical axis shows the degree of fundamentals that determine the terminal value of the farm (put simply, the positive $\epsilon$ illustrates good state, while negative $\epsilon$ refers to bad state). $\theta z$ is the farm's pledgeable  terminal value as collateral. 

The graph shows that the bank will lend to the farm in two broad cases:  
\begin{inparaenum}[1)]
	\item $\epsilon$ is in good state, and $\rho$ is not too large;
	\item $\epsilon$ is in bad state, but $\rho$ is very small.
\end{inparaenum}
On the other hand, the bank will deny lending to the farm in two broad cases: 
\begin{inparaenum}[1)]
	\item $\epsilon$ is in good state, but $\rho$ is too large;
	\item $\epsilon$ is in bad state, and $\rho$ is sufficiently large.
\end{inparaenum}

So far the analysis assumes that there is only one type of farm in the economy. However, studies such as \cite{CHODOROWREICH2021} illustrate that firm size matters for the terms of contract for bank loans. More specifically, within the U.S. farm system, it is possible that there are fundamental differences between large and smaller farmers that affect the key parameters in this stylized framework. For example, regardless of the shock $\epsilon$, large farms have higher $z$ than smaller farms. In fact, there is evidence of differing productivity between large and small farms, as documented by the Economic Research Service (\href{https://www.ers.usda.gov/amber-waves/2018/december/productivity-increases-with-farm-size-in-the-heartland-region/}{ERS}), United States Department of Agriculture (USDA). Moreover, there is emerging evidence from microeconomic studies such as \cite{Etwire2022} suggesting that farm size has a positive relationship with the decision to adopt climate adaptation measures. Put another way, when faced with the identical climate induced-shock $\rho$, large farms' adaptation capability ensures they have larger terminal value. Finally, due to their market power or name recognition, large farms may have more publicly available information about themselves, thus the bank spends less $\xi$ monitoring large farms. 

Assuming that large farms have larger $z$ and smaller $\xi$, define the critical values for large farms as $\underline{\rho}_b$ and $\bar{\rho}_b$.
We can derive from Equations \ref{ef1} and \ref{ef2} that  $\underline{\rho}_b \approx  \underline{\rho}$ and $\bar{\rho}_b > \bar{\rho}$. As seen in Figure \ref{ff1}, the critical range is wider for large farms. Moreover, when faced with large shocks, it is more likely for large farms to be approved for a loan, and the lending region is bigger. Put simply, large farms have more `room to maneuver.' Motivated by the predictions of this conceptual framework, the rest of the paper examines the empirical links between climate shocks and bank lending to farms. 


\newpage
\section{Data and Measurements}\label{datasection}
The paper employs three main types of data: climate risks and disasters; bank
lending to farms; agriculture production and sales. At the baseline, the data are observed at U.S. county level and on an annual frequency. I also expand the granularity of the analysis by incorporating additional dimensions of Census tract areas and bank entities. 

\noindent \textbf{\textit{Bank Financing}}

Accurately capturing lending to U.S. farms is a challenge, as such data at loan level tend to be confidential. While bank call reports include information about farm loans, such data are only available at bank institution level, and not at bank branch level. Therefore, these data do not have geographic variations that are necessary to identify how climate change vulnerability affects farm lending in different regions. To the best of my knowledge, the Community Reinvestment Act (CRA) dataset\footnote{See \url{https://www.ffiec.gov/cra/craproducts.htm}} is the only publicly available source that provides a relatively comprehensive view of farm lending in the United States. 

The CRA is a federal law enacted to encourage banks to lend to rural as well as low- and middle-income communities.\footnote{\cite{Keenan2019}} The dataset used in this paper covers years 1996 to 2019, and loans to small farms by county as well as by bank institutions. For the purpose of the paper, I use information of small farm loan originations, both in terms of the total number and amount of loans. Additionally, the original CRA database is disaggregated by several overlapping levels of measurements: loan size, banking institution, and income characteristics of recipients. The analysis focuses on total of loans in each county. In other words, the unit of analysis is the total amount of CRA financing in each county. It results in a panel dataset at county level across years. 

More specifically, two types of CRA variables are used: number of CRA loans; total amount of such loans. 
Additionally, I categorize the loans based on farm groups: \begin{inparaenum}[1)]
	\item loans to small-medium size farms (less than \$1 million gross revenue);
	\item loans to large farms (more than \$1 million gross revenue).
\end{inparaenum}

The CRA dataset identifies loans to farms with less than \$1 million of gross annual revenue. Thus the funding that goes to large farms can be derived from this variable. This categorization is broadly consistent with the classifications of the Economic Research Service (ERS) of \cite{USDA2021}. The ERS classification is based on farm gross revenue as well: \begin{inparaenum}[1)]
	\item small farms have less than \$350 thousand gross revenue;
	\item midsize farms have \$350-999 thousand gross revenue;
	\item large farms have more than \$1 million gross revenue.
\end{inparaenum}

While the CRA dataset is insightful, it is limited in its scope and coverage. First, it does not encompass all the farm lending that banks provide, as not all U.S. financial institutions are subject to CRA reporting requirements. For example, credit unions, which are not backed by the Federal Deposit Insurance Corporation (FDIC), are not included in the dataset. Moreover, the dataset is only for farm lending that is small in terms of amount. According to the CRA reporting guidelines shown in \cite{FFIEC2}, a small farm loan is defined as one with amount less than \$500,000, and can be either for farm land or production purpose. While CRA lending goes to farms of all sizes, it is likely that the lending to large farms are not representative of all the financial flows to such farms.

\noindent \textbf{\textit{Climate Change Measures}}

To measure vulnerability to climate change, I use two types of variables:  \begin{inparaenum}[1)]
	\item county-level climate-related disasters;
	\item county-level records of extreme climate conditions.
\end{inparaenum}
The data on climate-related disasters come from the Disaster Declarations Summary by the Federal Emergency Management Agency (FEMA).\footnote{See \url{https://www.fema.gov/openfema-dataset-disaster-declarations-summaries-v2}} Publicly available, this dataset records the dates and types of natural disasters (e.g., hurricane; flood) at the county level. The year coverage is 1953 to present. However, extreme weather events are not equivalent to climate change. 

Thus I also use the temperature and precipitation data that NOAA collects from year 1895 to present. More specifically, I use NOAA Monthly U.S. Climate Divisional Database (NClimDiv). Using such data, I measure climate anomaly that each county experiences in each year compared to its long-run average (e.g., 30 years).\footnote{In the climate literature, there does not seem to be consensus on what the ``correct" baseline is, studies use anywhere between the past 30 years and the preindustrial years as reference periods. See \cite{Moore4905}} Essentially this anomaly variable is the difference between yearly observed value and the county's mean value in record.\footnote{The original NOAA data are monthly observations, and I took average of the monthly observations to obtain the annual observations} Finally, I also categorize counties into climate regions using the 2012 International Energy Conservation Code created by the International Code Council. 

It is worth discussing the primary choice of temperature measurement in this paper---maximum temperature (observed daily but extrapolated to yearly for analysis in the paper). While mean temperature and its increase are often discussed in relation to climate change, it is problematic as an empirical measure as it has a clear trend. Additionally, temperature extremes are likely more indicative of the unexpected ``shocks" posed by climate change, therefore likely have more exogenous variation. For example, it is likely easier for agents to predict mean temperature than temperature extremes based on historical observations---the mean temperature is the first moment, whereas temperature extremes are related to higher moments. In other words, temperature extremes are likely to be unanticipated by economic agents, thus truly exogenous to the economic behavior one tries to explain. 

Thus the choice comes down to maximum versus minimum temperature, and the maximum measurement is the primary metric used in the paper. The main reason for this choice is that at least within the United States, record high temperatures are becoming more common than record low temperatures. For instance, one report by the \href{https://www.epa.gov/climate-indicators/climate-change-indicators-high-and-low-temperatures}{Environmental Protection Agency} (EPA) shows that the frequency distribution of extreme highs and lows are uneven in the past few decades; in particular, in the 2000s extreme highs had twice as many occurrences as extreme lows. One way of interpreting this unevenness is that it is more likely for climate change to materialize as maximum than minimum temperature. Besides the factor of frequency, studies by \cite{Vose2017} for the National Climate Assessment project that throughout the century, the intensity of extreme highs are going to increase, while that of extreme lows will decrease---more severe heat waves, and less severe cold waves. Finally, recent studies such as \cite{Diffenbaugh2021} that examine how climate change affects the economics of agriculture use daily maximum temperature as the primary measurement. However, in this paper, minimum temperature and precipitation data are included as controls or used as alternative measurements in robustness tests. 

\newpage
\noindent \textbf{\textit{Agricultural and Other Banking Variables}}

Estimations in the paper are supplemented by variables that illustrate the characteristics of agricultural production and banking in each county. The farm-related data mainly come from USDA Agricultural Census, which is generally conducted every 5 years. Additionally, USDA ERS also provides classifications such as farm areas that are helpful to understand regional heterogeneity of the impact of climate on bank lending. Finally, I use the Summary of Deposit (SOD) dataset from FDIC to control for each county's banking characteristics such as number of bank branches.

\subsection*{Summary Statistics}
The analysis in the paper makes an effort to distinguish between the lending to large and to smaller farms. At the same time, CRA lending goes primarily to non-large farms. Thus it is first useful to understand the distribution of farm size and why one should care about the insights from analyzing the CRA lending data. 

Using the USDA  Agricultural Census 1997 through 2017, I have tabulated the distribution of farm size in Table (\ref{t1a}). In the United States, large farms dominate the value of total market production (\cite{USDA}). But as seen in Table (\ref{t1a}), in terms of number of operations, the vast majority of farms are not large. In fact, small farms on average account for over 80\% of the country's total farms in the past two decades.\footnote{The sales categories in the Census go from \$100k to \$249k, and then \$250k to \$499k. Thus the Census data do not provide a clear cutoff between small and midsize farms---the threshold is \$350k according to ERS.} Across the years observed, it seems that farms with over \$500 thousand gross sales have increased in shares---in fact, their shares have exactly doubled between 1997 and 2017. At the same, farms in the intermediate range (\$100k to 499k) have dwindled, with the smallest farms (less than \$100k) seeing a small increase across the years observed. While it requires more rigorous testing, it seems the distribution of farm size, in terms of number of operations, has become more bimodal. In summary, based on the Census data, small farms are critical parts of the U.S. agricultural system.

\begin{table}[htbp]
	\caption{Size and Sales Distribution of U.S. Farms}\label{t1a}
	\centering
	\begin{center}
		\begin{tabular}{llrrrrr}
			\toprule
			\toprule
			&       & \multicolumn{5}{c}{\textbf{Share of farms (Percent)}} \\
			\midrule
			\textit{\textbf{Farm size}} & \textit{\textbf{Sales category}} & \textbf{1997} & \textbf{2002} & \textbf{2007} & \textbf{2012} & \textbf{2017} \\
			\midrule
			\textit{Small} & \textit{$<$ \$100k} & 81.9  & 85.3  & 83.8  & 81.5  & 82.1 \\
			\midrule
			\textit{Small to Midsize} & \textit{\$100k-499k} & 14.5  & 11.3  & 10.9  & 11.1  & 10.7 \\
			\midrule
			\textit{Midsize} & \textit{\$500k-999k} & 2.2   & 2.1   & 2.8   & 3.6   & 3.4 \\
			\midrule
			\textit{Large} & \textit{$>$ \$1 million} & 1.4   & 1.3   & 2.5   & 3.8   & 3.8 \\
			\bottomrule
			\bottomrule
		\end{tabular}
	\end{center}
	\begin{tablenotes}
		\tiny
		\item \textit{Source}: USDA Census 1997-2017, Table 2 ``Market Value of Agricultural Products Sold Including Landlord's Share and Direct Sales"
	\end{tablenotes}
\end{table}%

Tables (\ref{t1}) provides summary statistics of CRA loans at the county level in terms of number of loans, and the amount of loans. The table is grouped by lending to large and to small-medium farms.
Overall there are 3,106 counties in the CRA dataset across 24 years (1996-2019). As shown by Table (\ref{t1}), due to the nature of CRA, the overall magnitudes of lending to small-medium farms are bigger than to large farms. Section \ref{tsum} provides additional summary statistics grouped by the loan size thresholds.

\begin{figure}[!]
	\centering\includegraphics[width=15cm]{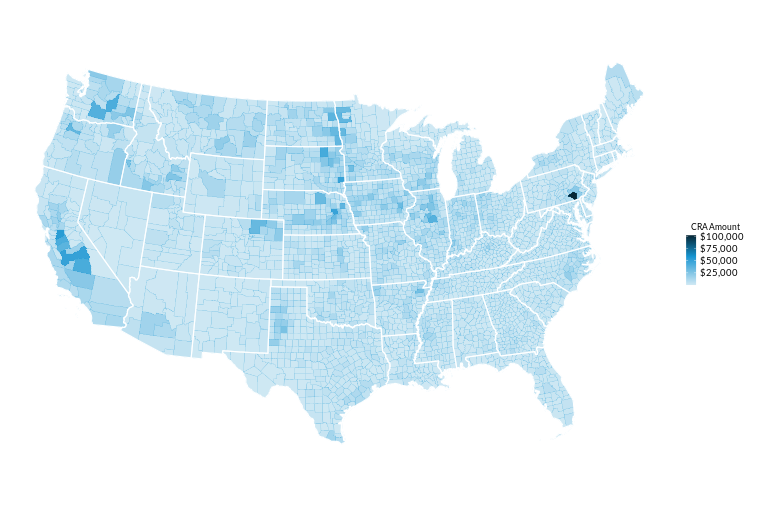}
	\caption{Yearly Average Total Amount of CRA Farm Loans, 1996-2019, in thousand 2015 USD}\label{f4}
	{\scriptsize Source: \cite{ffiec}}
\end{figure}

\begin{figure}[!]
	\centering\includegraphics[width=15cm]{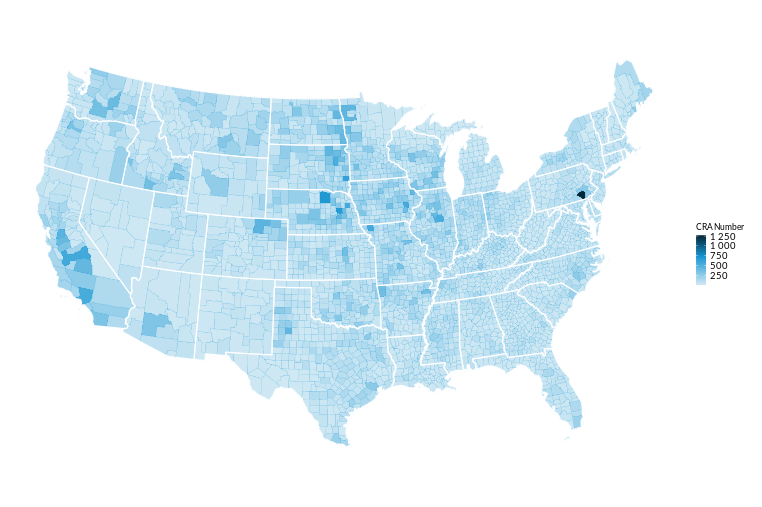}
	\caption{Yearly Average Number of CRA Farm Loans, 1996-2019}\label{f5}
	{\scriptsize Source: \cite{ffiec}}
\end{figure}

\begin{figure}[!]
	\centering\includegraphics[width=15cm]{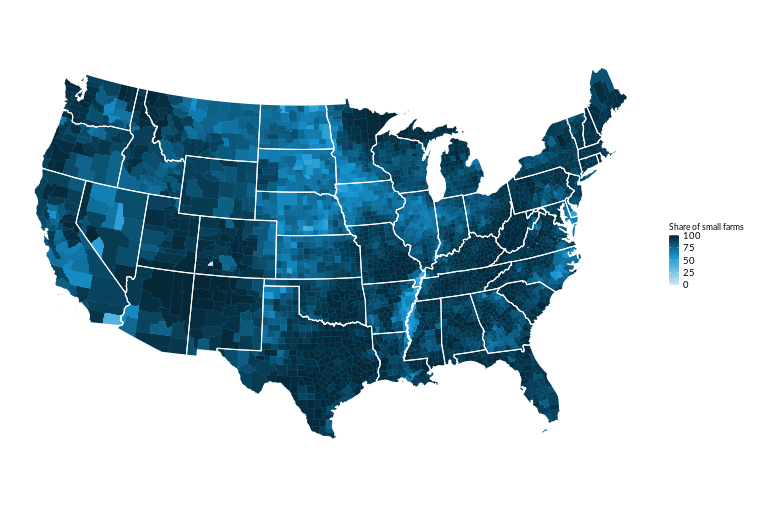}
	\caption{Share of Small Farms (gross sales $<$ \$250k), 2012-2017}\label{f6}
	{\scriptsize Source: USDA Census}
\end{figure}

To illustrate the geospatial distribution of CRA lending, I have mapped data through Figures (\ref{f4}) and (\ref{f5}) that illustrate the average county values during 1996-2019. At first glance, regions and states that are traditionally major agricultural producers (e.g., parts of California and the Midwest) also tend to receive more financing, both in terms of number and amount of loans. For the rest of the country, CRA lending seems relatively evenly distributed. 
While Table (\ref{t1a}) summarizes the size distribution of farms across years, Figure (\ref{f6}) illustrates the distribution across space. Comparing it with the CRA maps, it seems there are parallels between the high share of larger farms and the bank lending, though the patterns do not hold for all counties. 

\begin{table}[ht!]
	\centering
	\caption{Summary Statistics}\label{t1}
	\begin{adjustbox}{width=\columnwidth,center}
		\begin{tabular}{rlccccc}
			\toprule
			\multicolumn{1}{l}{\textbf{Variable}} &       & \multicolumn{1}{l}{\textbf{Mean}} & \multicolumn{1}{l}{\textbf{Std. Dev.}} & \multicolumn{1}{l}{\textbf{Min}} & \multicolumn{1}{l}{\textbf{Max}} & \multicolumn{1}{l}{\textbf{Observations}} \\
			\midrule
			\midrule
			\multicolumn{1}{l}{\textit{\textbf{Number of loans}}} & overall & 17.07 & 26.69 & 1.00  & 1030.00 & N =   64253 \\
			\multicolumn{1}{l}{\textit{\textbf{to large farms}}} & between &       & 18.24 & 1.00  & 245.67 & n =    3102 \\
			& within &       & 18.75 & -166.97 & 979.99 & T-bar = 20.7 \\
			\midrule
			\multicolumn{1}{l}{\textit{\textbf{Amount of loans }}} & overall & 1080.59 & 2226.96 & 1.01  & 40557.89 & N =   63983 \\
			\multicolumn{1}{l}{\textit{\textbf{to large farms}}}  & between &       & 1748.34 & 3.64  & 27361.08 & n =    3101 \\
			\multicolumn{1}{l}{\textit{\textbf{(thousand 2015 USD)}}}   & within &       & 1236.82 & -13643.49 & 26748.78 & T-bar =  20.63 \\
			\midrule
			\multicolumn{1}{l}{\textit{\textbf{Number of loans}}} & overall & 52.87 & 81.44 & 1.00  & 1623.00 & N =   70914 \\
			\multicolumn{1}{l}{\textit{\textbf{to small-medium farms}}}  & between &       & 64.85 & 1.00  & 1101.08 & n =    3105 \\
			& within &       & 48.25 & -416.13 & 1020.78 & T-bar = 22.8 \\
			\midrule
			\multicolumn{1}{l}{\textit{\textbf{Amount of loans}}} & overall & 2890.80 & 4867.74 & 1.01  & 140596.10 & N =   70749 \\
			\multicolumn{1}{l}{\textit{\textbf{to small-medium farms}}}  & between &       & 4212.58 & 1.06  & 85225.58 & n =    3104 \\
			\multicolumn{1}{l}{\textit{\textbf{(thousand 2015 USD)}}}   & within &       & 2322.11 & -48932.92 & 58261.29 & T-bar = 22.8 \\
			\bottomrule
			\bottomrule
		\end{tabular}%
	\end{adjustbox}
\end{table}%

\section{Econometric Analysis}\label{econometrics}
In this section I present econometric analysis testing the hypothesis that farms located in U.S. counties more exposed to climate change tend to receive less bank financing. The analysis uses two measures of climate change: disasters; climate condition anomaly. In the main section, I focus on presenting the results of temperature anomaly (the estimation using natural disaster variables are in the appendix). The econometric specifications in the paper focus on the nonlinearity of temperature effect, following the standard approach in the literature such as \cite{Diffenbaugh2021} and \cite{Schlenker2009} on agriculture and climate change.

\subsection{Climate Change Measured by Temperature Anomaly: County Level}{\label{countyresults}}
In this section, I use temperature and precipitation data from NOAA to measure climate change. Following the framework as in \cite{Burke2015}, the baseline regression is the following:
\begin{equation}\label{e2} 
	y_{it}=\beta_{1} T_{i t} + \beta_{2} T^2_{i t}+ u_{i}+ \eta_{t} +\lambda_{rt} +\gamma_{i} trend + \gamma_{i2} trend^2+e_{i t}
\end{equation}

Here the variable $T_{i t}$ stands for anomaly of maximum temperature observed in a county in each year. To calculate this variable, I take the difference between each county's observation in a year and the same county's average maximum temperature in the past 30 years. There are two reasons for this choice of calculation. First, compared with mean value, temperature anomaly is a preferred metric for climate scientists to measure climate change.\footnote{For example, see NOAA's description of its dataset \href{https://www.ncdc.noaa.gov/monitoring-references/faq/anomalies.php}{Global Surface Temperature Anomalies}} Additionally, the process of calculating the anomaly is equivalent to ``centering" the variable $T_{i t}$, so that the issue of multicollinearity is minimized when including the quadratic term $T^2_{i t}$. The inclusion of the quadratic term is similar to that in \cite{Burke2015}, so that the nonlinear effect of temperature is accounted for. 

Table (\ref{t5real}) shows the results of the baseline regression. Columns (1)-(2) show the results of CRA lending to large farms, while Columns (3)-(4) showing lending to small-medium farms. 
The signs of coefficients are the opposite for these two groups in Table (\ref{t5real}). More specifically, the linear effect of high temperature anomaly is negative for large farms, and it is positive for small-medium farms. In contrast, the quadratic effects for the two groups are the opposite. The results here suggest that even given the same climate change risks, large and small farms face divergent paths of financing outcomes. 

The results are robust to excluding 2008 and 2009 observations, as shown in Section \ref{d2}. Additionally, the results are robust to alternative year intervals of temperature anomaly, as seen in Section \ref{d3}. More specifically, the current measure of anomaly is based on the deviation from the 30-year mean, but the results are consistent when the temperature anomalies are calculated with respect to means of the following year intervals as well: 50 years, 70 years, 100 years, and years 1895 to 2019. 

%

\begin{table}[ht]
	\caption {CRA Farm Loans and Climate Vulnerability, County Total} \label{t5real} 
	\begin{adjustbox}{width=\columnwidth,center}
		\begin{tabular}{lcccc} \hline
 & (1) & (2) & (3) & (4) \\
VARIABLES & Num. to large farms & Amount to large farms & Num. to small-mid farms & Amount to small-mid farms \\ \hline
 &  &  &  &  \\
High temperature anomaly & -0.14* & -9.36 & 1.48*** & 56.58*** \\
 & (0.08) & (6.13) & (0.19) & (11.86) \\
High temperature anomaly (square) & 0.01 & 1.98 & -0.85*** & -27.49*** \\
 & (0.03) & (2.04) & (0.07) & (4.31) \\
 &  &  &  &  \\
Observations & 72,834 & 72,834 & 72,834 & 72,834 \\
R-squared & 0.194 & 0.094 & 0.134 & 0.057 \\
County FE & Yes & Yes & Yes & Yes \\
Year FE & Yes & Yes & Yes & Yes \\
Region x Year FE & Yes & Yes & Yes & Yes \\
 Robust SE & Cluster & Cluster & Cluster & Cluster \\ \hline
\multicolumn{5}{c}{ Robust standard errors in parentheses} \\
\multicolumn{5}{c}{ *** p$<$0.01, ** p$<$0.05, * p$<$0.1} \\
\end{tabular}

	\end{adjustbox}
	\begin{tablenotes}
		\scriptsize
		\vspace{5pt}
		\item \textit{Note}: Regression specification for this table is Equation (\ref{e2}), 
		$
		y_{it}=\beta_{1} T_{i t} + \beta_{2} T^2_{i t}+ u_{i}+ \eta_{t} +\lambda_{rt} +\gamma_{i} trend + \gamma_{i2} trend^2+e_{i t}
		$
	\end{tablenotes}
\end{table}

\begin{figure}[ht!]
	\centering
	\caption{Nonlinear Effect of Temperature on CRA Lending}\label{f3} 
	\begin{subfigure}{7cm}
		\centering\includegraphics[width=7cm]{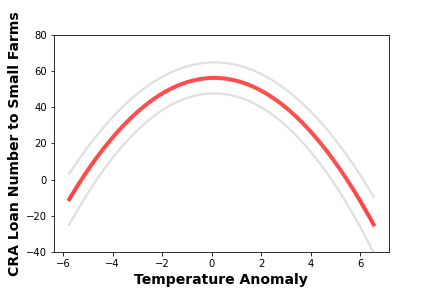}
		\caption{\textit{Frequencies to Small-Medium Farms}}
	\end{subfigure}%
	\begin{subfigure}{7cm}
		\centering\includegraphics[width=6.5cm]{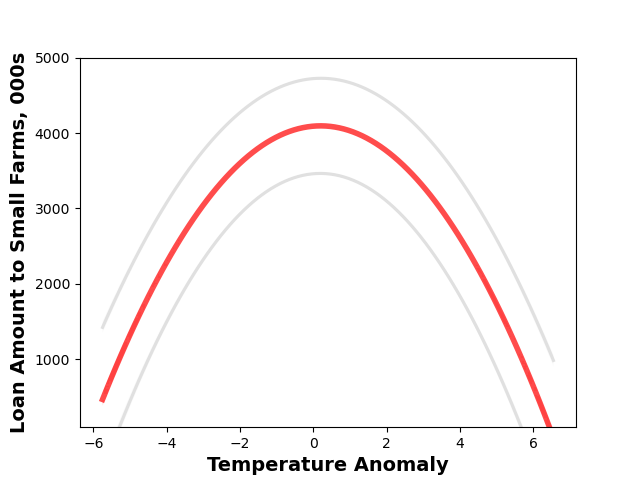}
		\caption{\textit{Amount to Small-Medium Farms}}
	\end{subfigure}
	\begin{subfigure}{7cm}
		\centering\includegraphics[width=7cm]{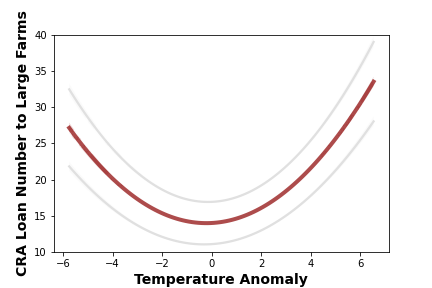}
		\caption{\textit{Frequencies to Large Farms}}
	\end{subfigure}
	\begin{subfigure}{7cm}
		\centering\includegraphics[width=6.5cm]{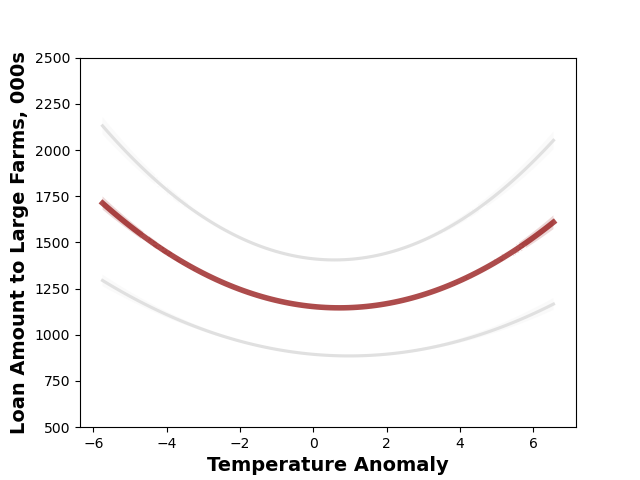}
		\caption{\textit{Amount to Large Farms}}
	\end{subfigure}\par
	{\tiny	\textit{Note}: Results based on regressions in Table (\ref{t5real});
		Gray lines indicate 95\% confidence intervals. Real values (2015 USD) for amount of lending variables. \par}
\end{figure}

The shapes of the effects can be visualized in Figure (\ref{f3}), where Panels (a) and (b) illustrate small-medium and large farms respectively. The shape of temperature anomaly effect for small-medium farms is concave. When temperature anomaly is negative (i.e., below normal), the level of CRA lending is increasing. But as the maximum temperature anomaly goes above normal, the level of lending goes down. This suggests that the overall effect of climate change, generally indicated by increasing temperature, is negative for lending to small-medium farms. In contrast, for large farms, the shape of the graph is convex. In particular, as temperature anomaly goes above normal, the CRA lending to this group actually rebounds. 

One possible explanation for this contrast is that the estimation here includes both direct and adaptation effect. Large farms, due to more available resources, may be better able to adapt to increasing temperature. Hence banks are more willing to lend to such farms that are more adaptive to climate change. Another related explanation is that banks consider large farms to be less risky, and replace their lending to small-medium farms with loans to large farms---in other words, the overall lending to a county may stay relatively constant, but it simply shifts from small to large farms (in fact, in Section \ref{banklevel} with bank level analysis, there is more evidence for this).

\subsubsection*{Marginal Effects by Climate Scenarios}
Since this paper focuses on the impact of climate change, it is useful to estimate the how such impact could materialize in the near future. The United Nations (UN) Intergovernmental Panel on Climate Change (IPCC) conducts regular assessments of the state of climate change.\footnote{For example, see a \href{http://web.archive.org/web/20210819125401/https://www.ipcc.ch/report/ar6/wg1/downloads/report/IPCC_AR6_WGI_SPM.pdf}{summary} of the most recent IPCC assessment} The IPCC assessments also provide projections for temperatures in different climate scenarios, known as the Shared Socioeconomic Pathways (SSPs), based on factors such as greenhouse gas emission, economic growth, and population growth.  

For example, if assuming no effective climate policy, under the scenario of `business as usual' (SSP5-8.5). the maximum temperature anomaly in continuous United States could be 2.8 Celsius (or 5.04 Fahrenheit). Using such projections,\footnote{The projected numbers are retrieved from IPCC WGI Atlas: \url{https://interactive-atlas.ipcc.ch/}} I thus estimate the marginal effects of temperature anomaly on CRA lending. It is important to note that only the measure of \textit{maximum} temperature anomaly, not \textit{mean} temperature, is used. From Equation (\ref{e2}), the overall marginal effect can be derived as
\begin{equation}\label{margin1}
	\frac{\partial y}{\partial T} = \beta_{1}  + 2 \beta_{2} T^*
\end{equation}

where $ T^*$ is a value of high temperature anomaly. For example, if the temperature anomaly is 1 degree Fahrenheit,  the overall marginal effects on number of loans are negative for both the small-medium and large farms, which can be calculated using coefficient estimates from Table (\ref{t5real}).

\begin{figure}[ht!]
	\centering
	\caption{Marginal Effect of Temperature Anomaly on CRA Lending, Climate Scenarios}\label{fmargin1} 
	\begin{subfigure}{7cm}
		\centering\includegraphics[width=7cm]{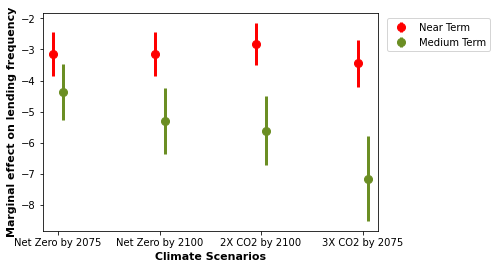}
		\caption{\tiny{Frequencies to Small-Medium Farms}}
	\end{subfigure}%
	\begin{subfigure}{7cm}
		\centering\includegraphics[width=6.5cm]{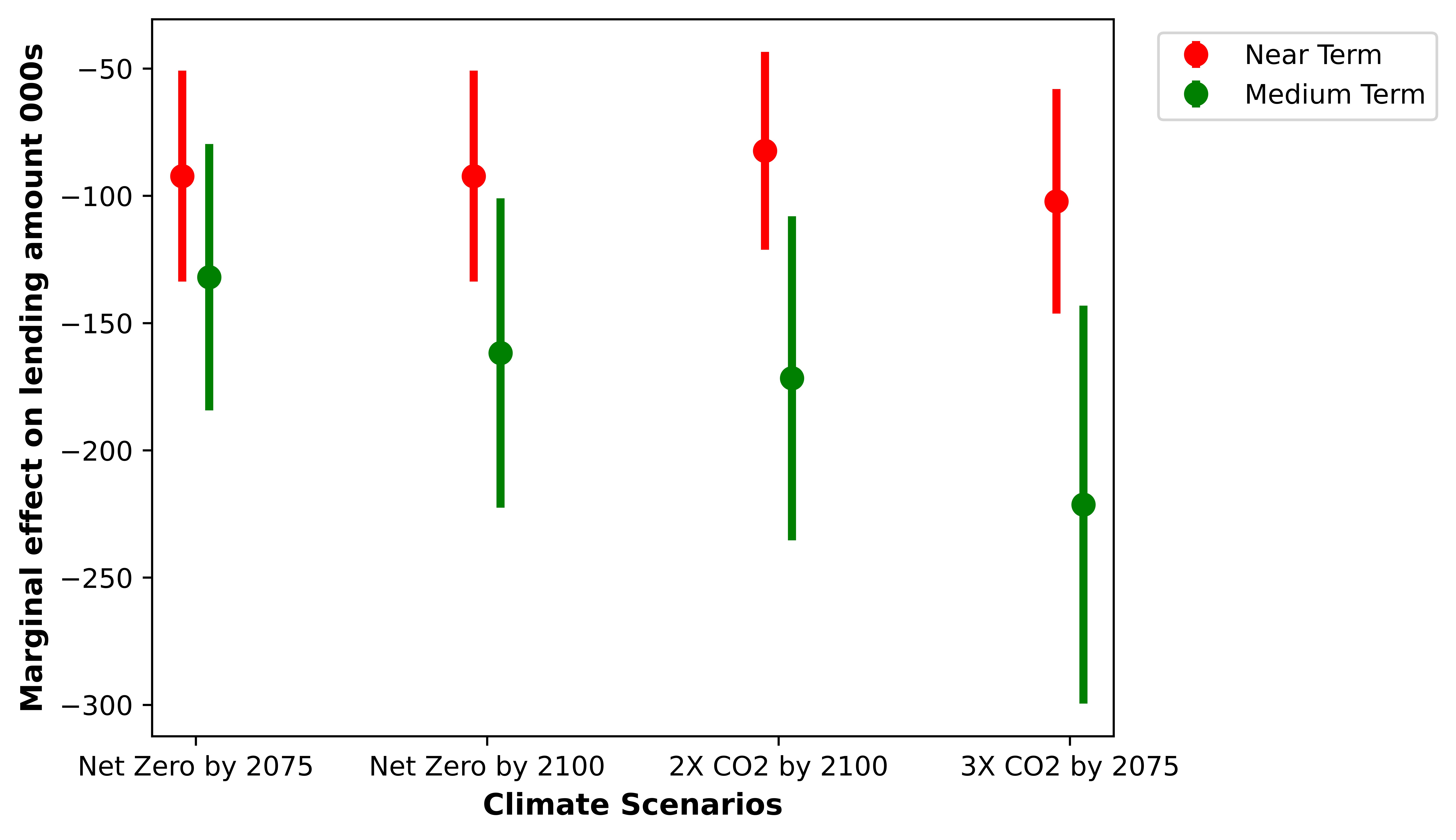}
		\caption{\tiny{Amount to Small-Medium Farms}}
	\end{subfigure}
	\begin{subfigure}{7cm}
		\centering\includegraphics[width=7cm]{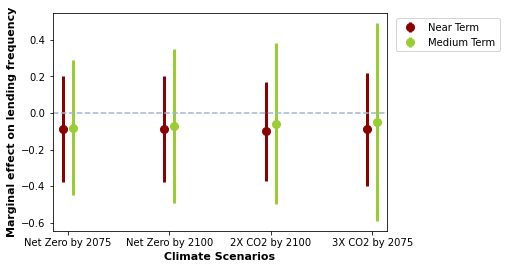}
		\caption{\tiny{Frequencies to Large Farms}}
	\end{subfigure}
	\begin{subfigure}{7cm}
		\centering\includegraphics[width=6.5cm]{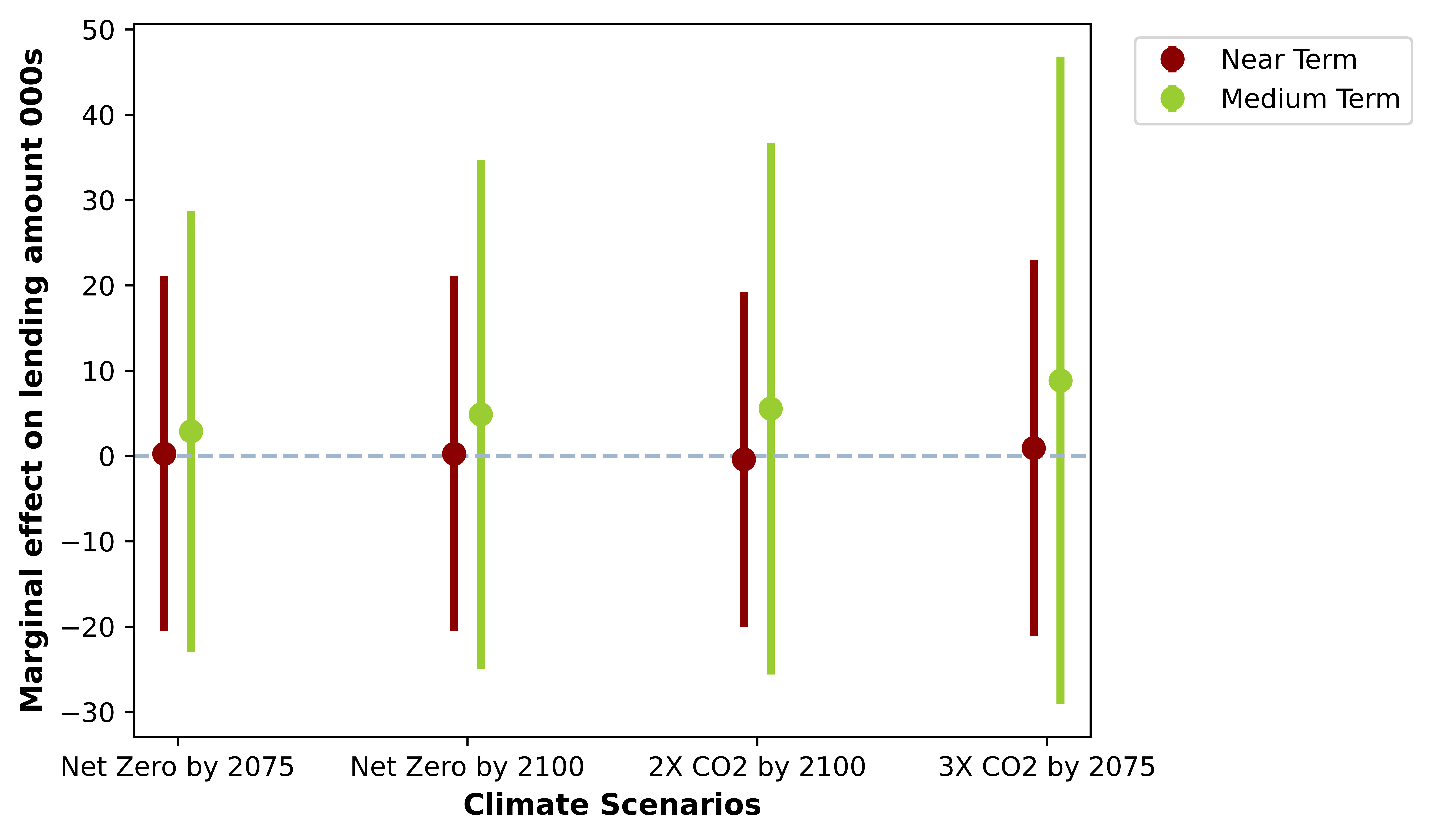}
		\caption{\tiny{Amount to Large Farms}}
	\end{subfigure}\par
	{\tiny \textit{Note}: Marginal effects with  95\% confidence intervals, calculated based on Equation (\ref{margin1}); The projected temperature estimates are from IPCC. The projections are specific to continuous United States, and refer to permanent increase of annual maximum temperature. The projections are based on Coupled Model Intercomparison Project Phase 6 (CMIP6) Model, taking into account emission uncertainty, and use 1986- 2005 as baseline. Climate scenarios are categorized into near term (now to 2040) and medium term (2041-2060)\par}
\end{figure}

Figure \ref{fmargin1} shows\footnote{
	Within the near term, \begin{inparaenum}[i)]
		\item `net zero by 2075' (SSP1-2.6) is equivalent to 1.5 ${}^{\circ}C$/ 2.7${}^{\circ}F$. `Net zero by 2100' (SSP2-4.5) is equivalent to 1.5 ${}^{\circ}C$/ 2.7${}^{\circ}F$. 
		\item `2X CO2 by 2100' (SSP5-8.5) is equivalent to  1.4 ${}^{\circ}C$/ 2.52${}^{\circ}F$. 
		\item `3X CO2 by 2100' (SSP3-7.0) is equivalent to  1.6 ${}^{\circ}C$/ 2.88${}^{\circ}F$. 
	\end{inparaenum}
	
	Within the medium term, \begin{inparaenum}[i)]
		\item `net zero by 2075' (SSP1-2.6) refers to 1.9 ${}^{\circ}C$/ 3.4${}^{\circ}F$. `net zero by 2100' (SSP2-4.5) is equivalent to 2.2 ${}^{\circ}C$/ 3.96${}^{\circ}F$. 
		\item `2X CO2 by 2100' (SSP5-8.5) is equivalent to  2.3 ${}^{\circ}C$/ 4.14${}^{\circ}F$. 
		\item `3X CO2 by 2100'  (SSP5-8.5) is equivalent to  2.8 ${}^{\circ}C$/ 5.04${}^{\circ}F$. 
\end{inparaenum}}
the estimated results under a range of climate scenarios. From left to right, the horizontal axes denote the climate scenarios, from the most optimistic to the most pessimistic. Within each graph, marginal effects are estimated for two different horizons: near term (now to 2040) and medium term (2041-2060). In general, the more pessimistic the scenario, the higher projected maximum temperature anomaly is. The longer the time horizon, the higher the projected temperature is. 

Three general patterns emerge from observing the figure. First, both small-medium and large farms suffer some loss of bank lending. Second, the marginal effects are large in magnitudes for small-medium farms. Third, the negative impact is minimal in terms of loan frequencies for large farms, and the impact is in fact positive in terms of loan amount. The figure here further suggests that farms fare differently, depending on their size, resulting in the lending approval region for large farms significantly wider. 

\paragraph{\textit{Robustness Checks}}

All the results so far focus on the level effect of temperature anomaly, and additional analysis is conducted on the growth effect: i.e., the dependent variable is the growth of CRA lending. The nonlinear effect still retains some significance, though the directions of impact are the same across farm groups. The results are reported in Tables (\ref{t20}) and (\ref{t21})  in Section \ref{d4} of Appendix. 

To assess robustness of the baseline results, I have conducted additional estimations: 	\begin{inparaenum}[1)]
	\item inclusion of precipitation as an additional control;
	\item distinguishing between growing and non-growing season.
\end{inparaenum}
The results are shown in \ref{robustnesscheck1} and \ref{robustnesscheck2}. In short, including precipitation does not significantly alter the baseline results. When focusing on growing season, the negative effects on small-medium farms become amplified. When focusing on non-growing season, the effects on large farms become statistically insignificant.

\subsubsection*{Extension I: Census Tract Income Areas}

So far the analysis has focused on county aggregates, with consideration of farm size. In this section, I expand the estimation by considering another dimension: the income areas where the farms are located. Whether farms are located in high- or low-income areas may matter for bank lending decisions, as CRA-qualifying loans in theory should go to low-income or moderate-income (LMI) communities.\footnote{For example, see \url{https://www.federalreserve.gov/consumerscommunities/cra_about.htm}}

Using the definition of an income area from FFIEC,  I conduct analysis at the level of county-income group pair. More specifically,  a county may have one or more income groups (defined at Census tract level), hence resulting in multiple county-income group pairs within the same county. This dimension of analysis is possible due to the fact that CRA lending data is also available at the Census tract level. The unit of analysis is essentially the income group aggregate \textit{within} each county, and there are 12,290 such county-income group pairs.\footnote{FFIEC categorizes Census tracts into the following income groups based on what the Median Family Income (MFI) of a Census tract compared to that of the Metropolitan Statistical Areas (MSA): 
	\begin{inparaenum}[1)]
		\item Low Income, less than 50\% of MFI of the MSA;
		\item Moderate Income, 50\% to 80\% of MFI of the MSA;
		\item Middle Income,  80\% to 120\% of MFI of the MSA;
		\item Upper Income,  greater than or equal to 120\% of MFI of the MSA;
	\end{inparaenum}
} 

Thus the regression adds the income group dimension, denoted by subscript $c$, and the specification is at the county-income group pair level, with income group fixed effect $ \xi_{c}$  added. 
\begin{equation}\label{e2cen} 
	y_{ict}=\beta_{1} T_{i t} + \beta_{2} T^2_{i t}+ u_{i}+ \xi_{c} + \eta_{t} +\lambda_{rt} +\gamma_{i} trend + \gamma_{i2} trend^2+e_{i ct}
\end{equation}

Table (\ref{t5realincomeg0}) shows the estimation results. For the entire sample and controlling for income area fixed effects, Table (\ref{t5realincomeg0}) shows very similar results to Table (\ref{t5real}), albeit with higher degrees of significance. Thus for all income areas, the impact of climate risks is significant for lending to large and smaller farms, and the signs of impact differ by the farm size. But this is not to say that income area has no relationship with the degrees of impact. To uncover this relationship, I repeat the analysis by four main income areas: low, moderate, middle, and high in Tables (\ref{t5realincomeg1}-\ref{t5realincomeg4}) in Appendix \ref{census-tables}. 

\begin{figure}[h]
	\centering
	\caption{Marginal Effect of Temperature Anomaly on CRA Lending, Climate Scenarios}\label{census-margins} 
	\begin{subfigure}{7cm}
		\centering\includegraphics[width=7cm]{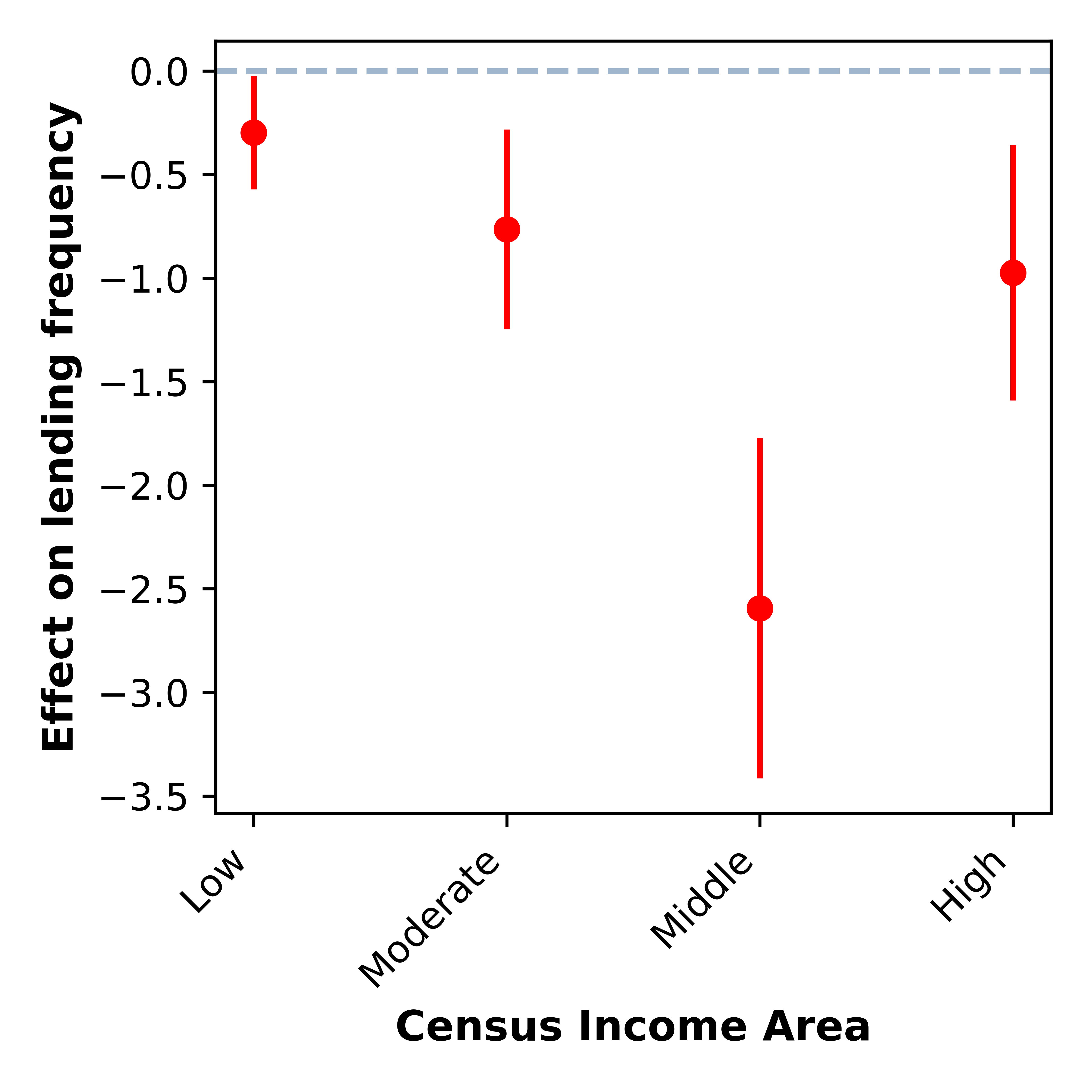}
		\caption{\tiny{Frequencies to Small-Medium Farms}}
	\end{subfigure}%
	\begin{subfigure}{7cm}
		\centering\includegraphics[width=7cm]{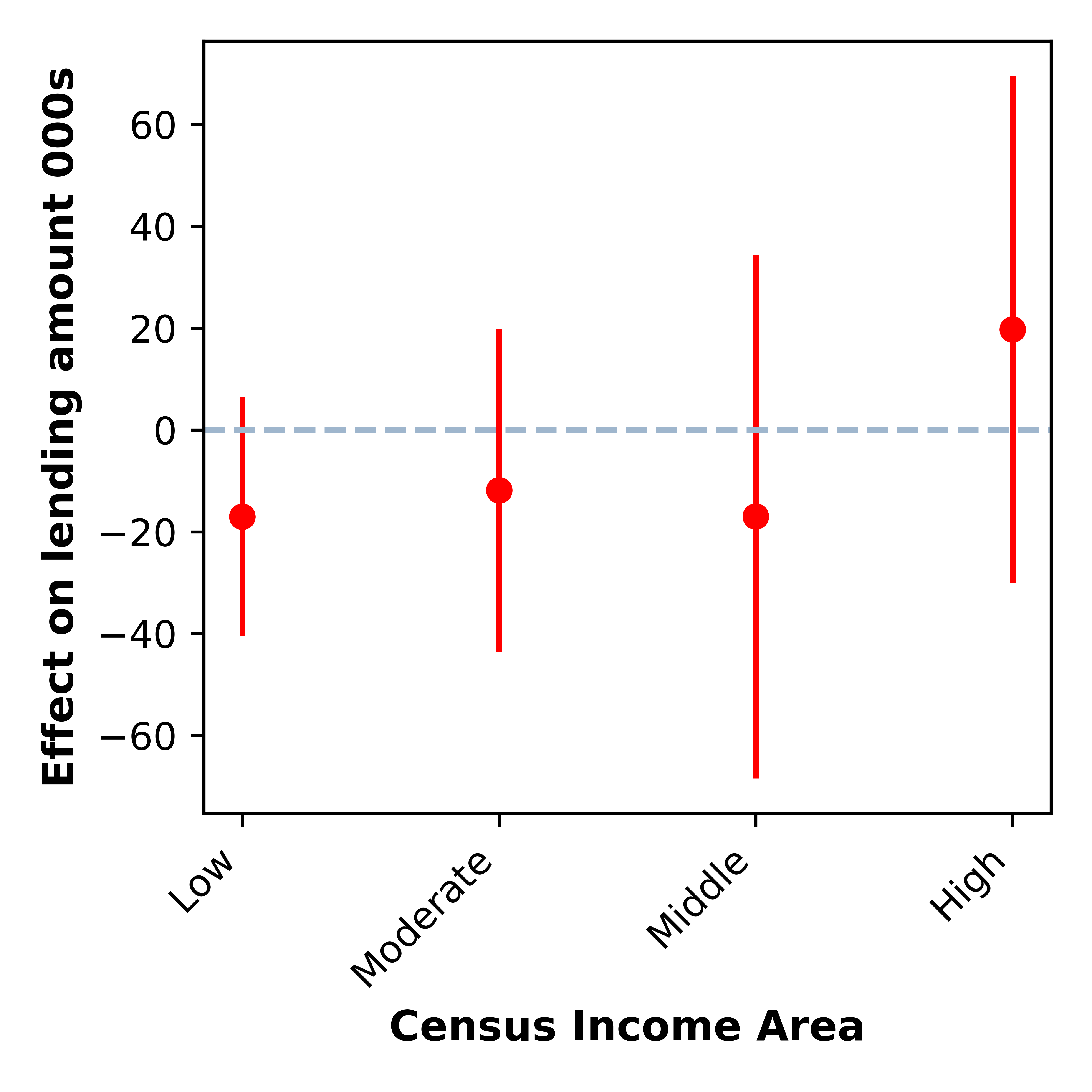}
		\caption{\tiny{Amount to Small-Medium Farms}}
	\end{subfigure}
	\begin{subfigure}{7cm}
		\centering\includegraphics[width=7cm]{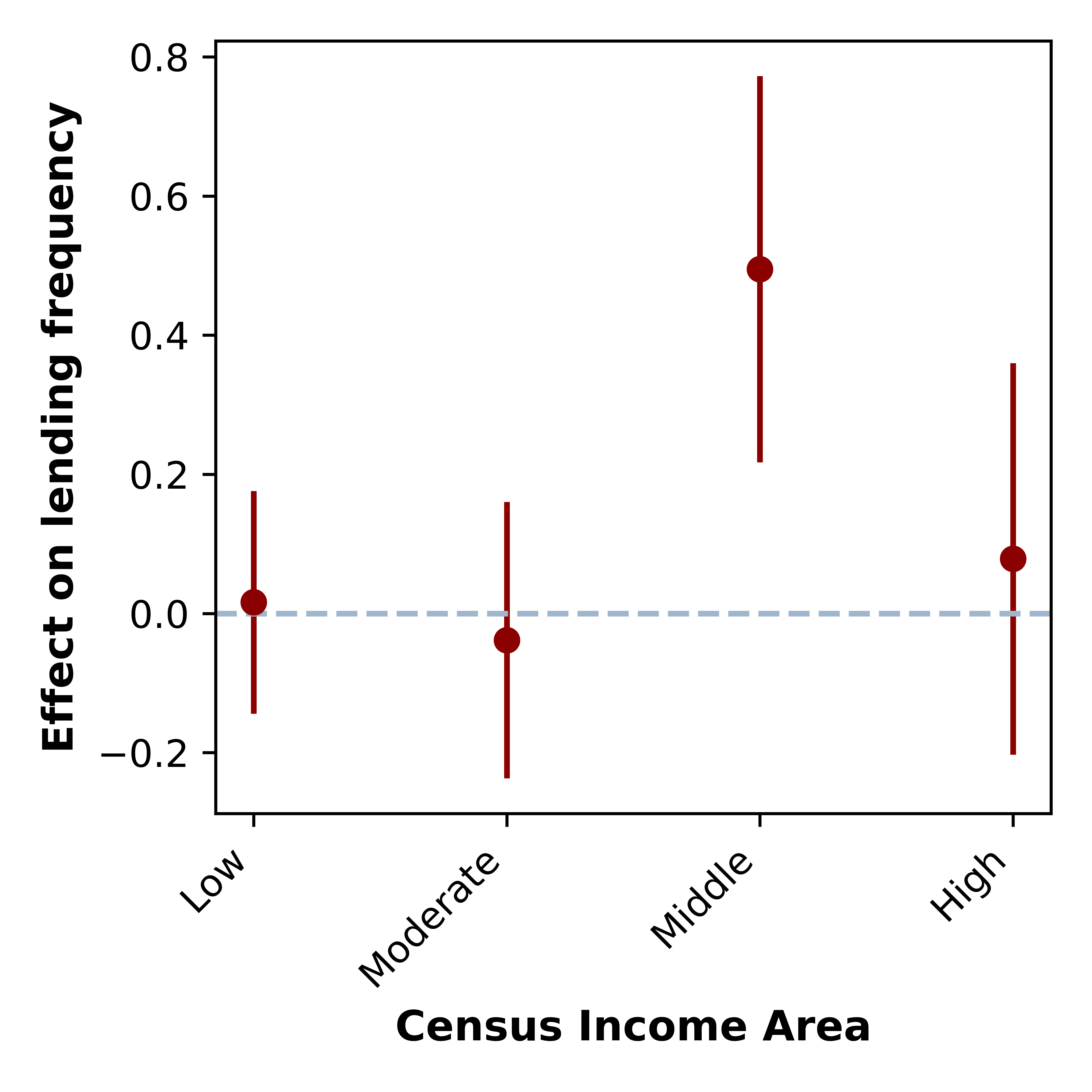}
		\caption{\tiny{Frequencies to Large Farms}}
	\end{subfigure}
	\begin{subfigure}{7cm}
		\centering\includegraphics[width=7cm]{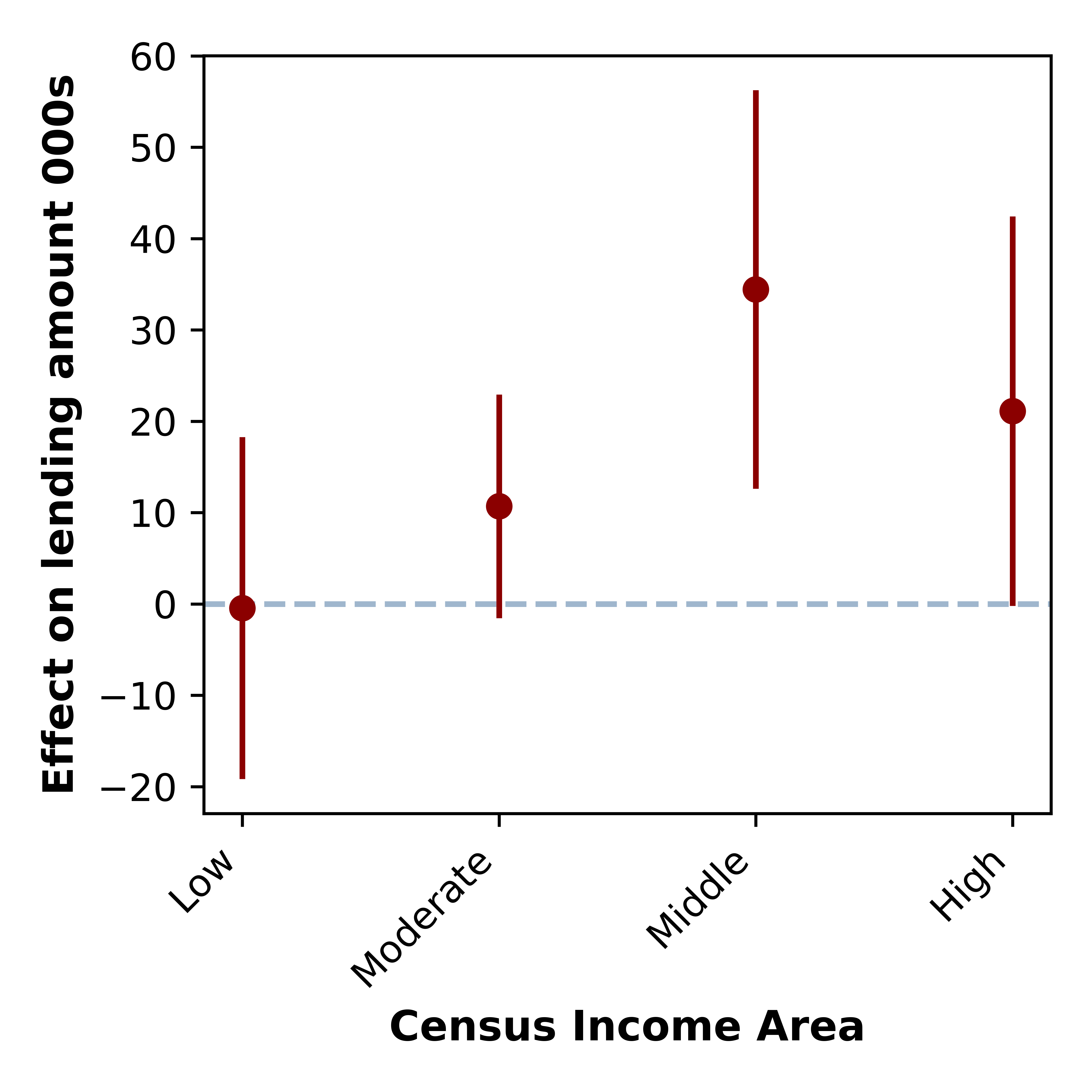}
		\caption{\tiny{Amount to Large Farms}}
	\end{subfigure}\par
	{\tiny \textit{Note}: Marginal effects with  95\% confidence intervals, calculated based on Equation (\ref{margin1}); The projected temperature estimates are from IPCC. The projections are specific to continuous United States, and refer to permanent increase of annual maximum temperature. The projections are based on Coupled Model Intercomparison Project Phase 6 (CMIP6) Model, taking into account emission uncertainty, and use 1986- 2005 as baseline. Climate scenarios are categorized into near term (now to 2040) and medium term (2041-2060)\par}
\end{figure}

To help with interpretation, I have computed the marginal effects for each of the Census tract income area and the results are shown in Figure (\ref{census-margins}). By comparing the figures, it becomes clear that the impact of climate vulnerability is heterogeneous across income groups. In terms of the frequency of lending to small-medium farms, the impact is consistent in terms of the sign of impact across income areas. Moreover, in general, conditional on being located in high income area, farms generally experience limited or insignificant impact of climate vulnerability on lending, with the exception of lending frequency to smaller farms. This makes intuitive sense in that such high income areas may have more resources available that improve farms' financial resilience to adverse shocks. 

When looking at the farms in low and moderate income areas, however, the results in the figures are seemingly surprising.  Conditional on being in these areas, large farms do not experience significant impact. Moreover, even for smaller farms, higher climate vulnerability largely do not ncessarily contribute to substantially lower amount of loans.  One plausible explanation is that to meet CRA requirements, banks need to ensure that they provide funding to farms located in low- and moderate-income communities. Therefore, such lending activities are less sensitive to changing climate vulnerabilities. 

In contrast, it seems that small farms located in middle income areas are most affected by climate vulnerabilities, especially in terms of the frequencies of loans. Compared with farms in high-income areas, they may not have as much financial resource to respond to climate change. Moreover, since banks are not mandated to maintain certain lending activities in middle income areas, they make lending decisions more purely based on their assessments of farms' terminal values in relation to climate vulnerabilities. In this case and consistent with Table  (\ref{t5realincomeg0}), smaller farms are less likely to receive financing when climate risks increase. Put another way, the variations within the middle income areas are driving the results in Table (\ref{t5realincomeg0}). 

\begin{table}[!]
	\caption {CRA Farm Loans and Climate Vulnerability, All Income Areas} \label{t5realincomeg0} 
	\begin{adjustbox}{width=\columnwidth,center}
		\begin{tabular}{lcccc} \hline
 & (1) & (2) & (3) & (4) \\
VARIABLES & Num. to large farms & Amount to large farms & Num. to small-mid farms & Amount to small-mid farms \\ \hline
 &  &  &  &  \\
Temp. anomaly & -0.20*** & -13.99*** & 1.10*** & 18.43*** \\
 & (0.04) & (2.91) & (0.10) & (5.87) \\
Temp. anomaly (square) & 0.08*** & 6.27*** & -0.52*** & -7.04*** \\
 & (0.02) & (1.12) & (0.04) & (2.43) \\
 &  &  &  &  \\
Observations & 160,137 & 160,137 & 160,137 & 160,137 \\
R-squared & 0.111 & 0.045 & 0.075 & 0.027 \\
County FE & Yes & Yes & Yes & Yes \\
Year FE & Yes & Yes & Yes & Yes \\
Income Area FE & Yes & Yes & Yes & Yes \\
Region x Year FE & Yes & Yes & Yes & Yes \\
 Robust SE & Cluster & Cluster & Cluster & Cluster \\ \hline
\multicolumn{5}{c}{ Robust standard errors in parentheses} \\
\multicolumn{5}{c}{ *** p$<$0.01, ** p$<$0.05, * p$<$0.1} \\
\end{tabular}

	\end{adjustbox}
	\begin{tablenotes}
		\vspace{5pt}
		\scriptsize
		\item \textit{Note}: Regression specification for this table is Equation (\ref{e2cen}), 
		$
		y_{ict}=\beta_{1} T_{i t} + \beta_{2} T^2_{i t}+ u_{i}+ \xi_{c} + \eta_{t} +\lambda_{rt} +\gamma_{i} trend + \gamma_{i2} trend^2+e_{i ct}
		$.
		Table (\ref{t5realincomeg0}) has greater number of observations than the sum of Tables (\ref{t5realincomeg1})-(\ref{t5realincomeg4}), as there are Census tracts that are uncategorizable by income
	\end{tablenotes}
\end{table}

\newpage
\subsubsection*{Extension II: Distributed Lag Model}
In the paper so far I focus on the contemporaneous relationship between temperature anomaly and CRA lending. However, the impact of climate change may play out over time rather than materializing instantaneously. In this section, I use a distributed lag specification to examine such longer-term effect. 
\begin{equation}\label{e5} 
	y_{it}=\sum_{l=0}^{2}\left({\beta}_{l} T_{i, t-l} \right) +\sum_{m=0}^{2}\left({\theta}_{m} T^2_{i, t-m} \right)+ u_{i}+ \eta_{t} +\lambda_{rt} +\gamma_{i} trend + \gamma_{i2} trend^2+e_{i t}
\end{equation}
where the lagged terms (1 and 2 years) of temperature anomaly are added to the baseline specification. 

If we assume at time $t$ there is a \textit{permanent} increase of temperature anomaly evaluated at $T^*$, the effect will materialize this period but also last into the next two periods. 
Additionally, the cumulative marginal effect of temperature anomaly over time is 
\begin{equation}\label{e6} 
	\sum_{l=0}^{2}{\beta}_{l} + \sum_{m=0}^{2}{\theta}_{m}= \beta_{0}  + 2 \theta_{0} T^* +  \beta_{1}  + 2 \theta_{1} T^*  +  \beta_{2}  + 2 \theta_{2} T^* 
\end{equation}

In other words, the cumulative effect of a temperature anomaly $T^*$ is a weighted sum of linear and nonlinear effects, both in the present and in two previous periods. 

The results from estimating Equation (\ref{e5}) are shown is Table (\ref{t7}). Within each of the columns, it is clear that the contemporaneous effects are still generally significant. However, it is interesting that the coefficient for the amount of loans to small-medium farm is now negative. Besides, for all the CRA variables, especially the lending to small-medium farms, the lagged effects are significant. This provides evidence showing that climate change impact accumulates over time. 

Using Equation (\ref{e6}), we can calculate the cumulative effect of a permanent increase of 1 high temperature anomaly---1 degree in Fahrenheit, or about 0.56 Celsius. For large farms, the effect is 0.36 for number of loans, and is 4.83 for the amount of loans. For small-medium farms, the effect is -0.92 for number of loans, and is -31.71 for the amount of loans. 

Based on the results, it is clear that small-medium farms are more vulnerable to bank lending cutback due to climate change. With 1 degree in Fahrenheit of temperature anomaly, lending to small-medium in one county would decrease by about 1 loan and \$31,710. However, if the anomaly is 3 degrees Fahrenheit (1.5 Celsius), the number of loans cut would be 14 and the amount reduced would be \$381,670. While the data here are at the county aggregate level, not farm-level, the magnitudes of effect are not trivial. In contrast, with temperature increasing, large farms will likely experience increase in both number and amount of lending. As the results in Appendix Tables (\ref{t9}) and (\ref{t10}) suggest, at least in terms of amount of loan, the different financial flows that large and small-medium farms experience are likely to due to banks' shift of lending between these types of farms. It is beyond the scope of the paper to explicit account for what may explain the differential outcome, but one possible explanation is that large farms are better able to adapt to climate change. Banks see less risk in such farms and are thus more willing to lend to them. 

\begin{table}[!]
	\caption {CRA Farm Loans and Climate Vulnerability, Distributed Lag} \label{t7} 
	\begin{adjustbox}{width=\columnwidth,center}
		\begin{tabular}{lcccc} \hline
 & (1) & (2) & (3) & (4) \\
VARIABLES & Num. to large farms & Amount to large farms & Num. to small-mid farms & Amount to small-mid farms \\ \hline
 &  &  &  &  \\
High temperature anomaly & -0.20** & -24.35*** & 1.93*** & 8.75 \\
 & (0.10) & (6.71) & (0.21) & (13.53) \\
High temperature anomaly (lag) & 0.11 & -19.28*** & 1.93*** & 52.55*** \\
 & (0.07) & (5.22) & (0.17) & (10.82) \\
High temperature anomaly (two lags) & 0.04 & -5.80 & 1.87*** & 81.97*** \\
 & (0.07) & (5.19) & (0.19) & (11.31) \\
High temperature anomaly (square) & 0.08** & 11.00*** & -1.31*** & -21.50*** \\
 & (0.04) & (2.48) & (0.10) & (5.21) \\
High temperature anomaly (square, lag) & 0.09** & 9.56*** & -1.28*** & -39.09*** \\
 & (0.04) & (2.48) & (0.10) & (5.13) \\
High temperature anomaly (square, two lags) & 0.03 & 6.57*** & -0.74*** & -26.90*** \\
 & (0.03) & (2.37) & (0.09) & (5.00) \\
 &  &  &  &  \\
Observations & 66,622 & 66,622 & 66,622 & 66,622 \\
R-squared & 0.201 & 0.091 & 0.161 & 0.063 \\
County FE & Yes & Yes & Yes & Yes \\
Year FE & Yes & Yes & Yes & Yes \\
Region x Year FE & Yes & Yes & Yes & Yes \\
 Robust SE & Cluster & Cluster & Cluster & Cluster \\ \hline
\multicolumn{5}{c}{ Robust standard errors in parentheses} \\
\multicolumn{5}{c}{ *** p$<$0.01, ** p$<$0.05, * p$<$0.1} \\
\end{tabular}

	\end{adjustbox}
	\begin{tablenotes}
		\vspace{5pt}
		\scriptsize
		\item \textit{Note}: Regression specification for this table is Equation (\ref{e5}), 
		$
		y_{it}=\sum_{l=0}^{2}\left({\beta}_{l} T_{i, t-l} \right) +\sum_{m=0}^{2}\left({\theta}_{m} T^2_{i, t-m} \right)+ u_{i}+ \eta_{t} +\lambda_{rt} +\gamma_{i} trend + \gamma_{i2} trend^2+e_{i t}
		$
	\end{tablenotes}
\end{table}

\subsection{Climate Change Measured by Temperature Anomaly: Bank-County Level}\label{banklevel}

Previous sections focus on the county-level aggregate frequencies and amounts of bank lending. The results in this section further answer the research question from a different perspective: bank-county level pair.\footnote{The bank level here refers to the bank entity level, not bank branch level. The bank-county level pair is not necessarily equivalent to bank branch in that county.} In other words, what is being estimated here is more granular: for a banking institution, whether there is difference in their lending to farms according to not only farms' exposure to climate risks, but also to bank-specific operations and market shares. The point of this bank-county pair section is to demonstrate that banks' own characteristics such as size, play a role in the impact, complementing the main results of the aggregate estimates. 

More specifically, I classify banks by their sizes based on CRA definitions\footnote{The categories are based on a 2014 CRA \href{https://www.occ.gov/publications-and-resources/publications/community-affairs/community-developments-fact-sheets/pub-fact-sheet-cra-reinvestment-act-mar-2014.pdf}{fact sheet}. As of now, the bank asset data in the estimates are in nominal values}: \begin{inparaenum}[1)]
	\item very small banks: asset value less than \$300 million;
	\item small-mid banks: asset value between \$300 million and \$1.2 billion;
	\item large banks: asset value over \$1.2 billion. 
\end{inparaenum}
It is important to point out the composition of the lending data: the vast majority of banks in the CRA dataset are large banks, and very small banks make up the smallest share. 

The loan size also matters in understanding the banks' lending behavior. The CRA dataset categorize loans into three brackets based on their sizes: \begin{inparaenum}[1)]
	\item smaller loans (`100k loans'): less than \$100 thousand per origination;
	\item medium loans (`250k loans'): between \$100 and 250 thousand;
	\item larger loans (`500k loans'): between \$250 and 500 thousand. 
\end{inparaenum} 
Given their market share in general, it is reasonable to assume the majority of the CRA loans originate from large banks. However, it is somewhat surprising that large banks are also dominant in lending out loans of less than \$100 thousand. For such small loans, almost 73\% of the total number of loan originations is made by large banks, whereas the shares for small and medium banks are 5\% and 22\% respectively.\footnote{The dominance of large banks in providing small loans mirrors the findings by \cite{DiSalvo} that examines the patterns of small business loans in metropolitan areas. Existing studies such as \cite{MKHAIBER2021} suggest that large banks tend to lend to large firms. But the interaction between large banks and small farms/firms is worth further investigation}

The intersection of large banks and smaller farms loans (size of less than \$100k) provides an important clue in understanding why different types of farms fare differently in terms of financial access, as these smaller loans make up the biggest share of total loan origination frequencies in years 1996-2019. The CRA dataset is not granular enough to decompose loans by farm type \textit{and} simultaneously by loan size brackets. In other words, it is difficult to say how much of the smaller loans go to a certain type of farms. Yet it is plausible that small farms are most likely the recipients of these smaller loans. Thus the effects of climate risks on loans of smaller sizes are especially relevant to these small farms, both qualitatively and quantitatively. 

To uncover the lending patterns at the bank-county level, I first set the scene through regressions by loan sizes. These results provide important contexts in understanding the ensuing estimations of loans to small and large farms, and for connecting the bank-county level results with the county-aggregate results.

Taking into account the interaction between bank size and temperature anomaly, the econometric specification in this section is
\begin{equation}\label{e2bankinteract} 
	\begin{aligned}
		y_{ibt}= & \beta_{1} T_{i t} + \beta_{2} T^2_{i t}+  (a_1+ a_2 T_{i t} + a_2 T^2_{i t}) \textit{small}+ (b_1+ b_2 T_{i t} + b_3 T^2_{i t}) \textit{medium} \\ 
		& + (c_1+ c_2 T_{i t} + c_3 T^2_{i t}) \textit{large} + u_{i}+ \psi_{b} + \eta_{t} +\lambda_{rt} +\gamma_{i} trend + \gamma_{i2} trend^2+e_{ibt}
	\end{aligned}
\end{equation}

Compared with Equation (\ref{e2}), the main difference of Equation (\ref{e2bankinteract}) is that the lending variables are at the bank entity level (with subscript $b$) in a specific county, with the additional $\psi_{b}$ as bank-level fixed effects. In short, what is being estimated here is that given the county and year, how a bank makes lending decisions, controlling for bank characteristics. Moreover, bank-size dummy variables are included: \textit{small} (asset less than \$300 million), \textit{medium} (asset between \$300 million and \$1.2 billion), and \textit{large} (asset over \$1.2 billion)). Additionally, these bank-size dummies interact with the linear and quadratic temperature anomaly terms. 

\begin{figure}[!]
	\centering
	\caption{Marginal Effect of Temperature Anomaly on CRA Lending Frequency, by Loan Size}\label{loansizemargin1} 
	\begin{subfigure}{3cm}
		\centering\includegraphics[width=3cm]{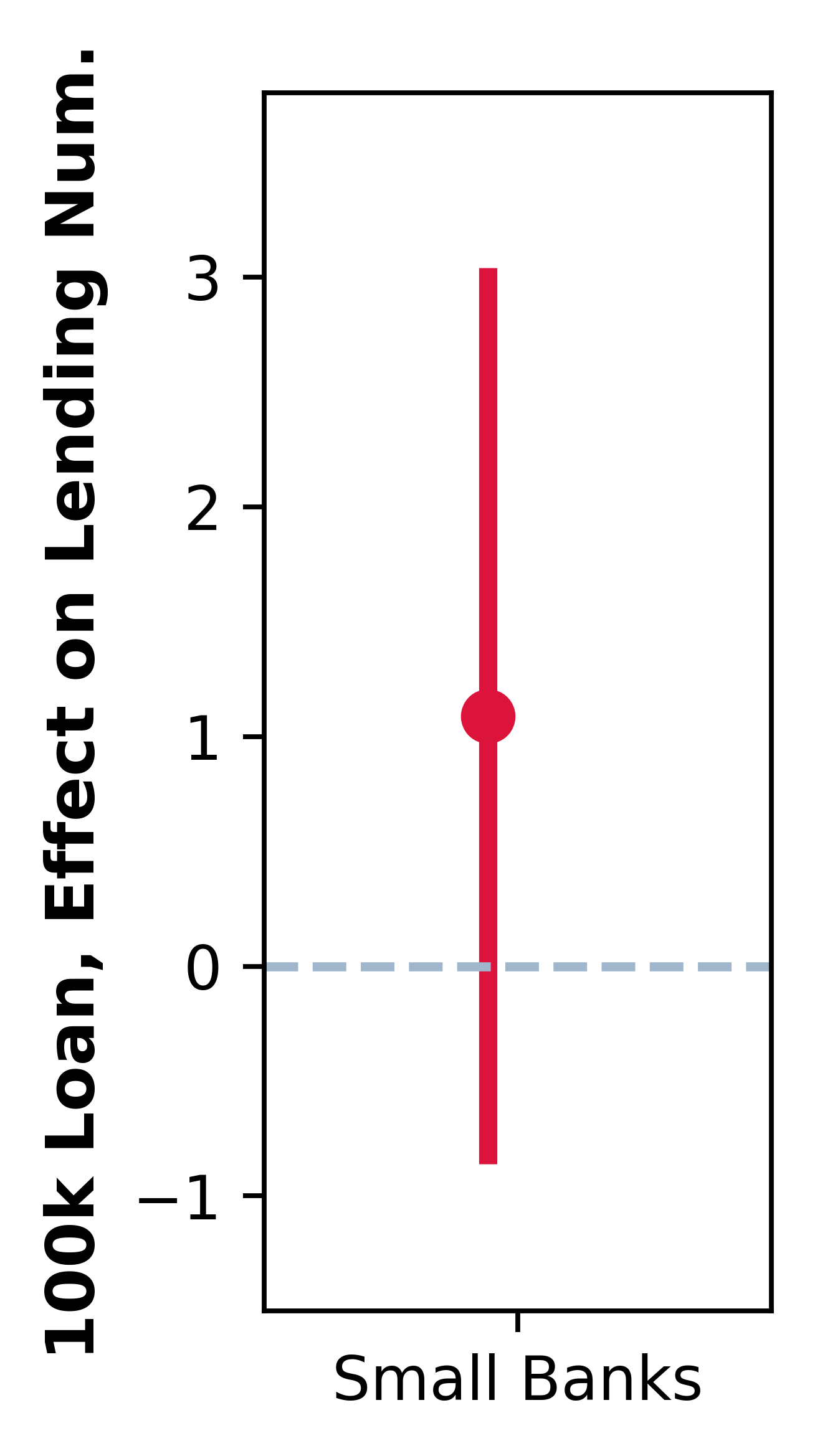}

	\end{subfigure}%
	\begin{subfigure}{3cm}
		\centering\includegraphics[width=3cm]{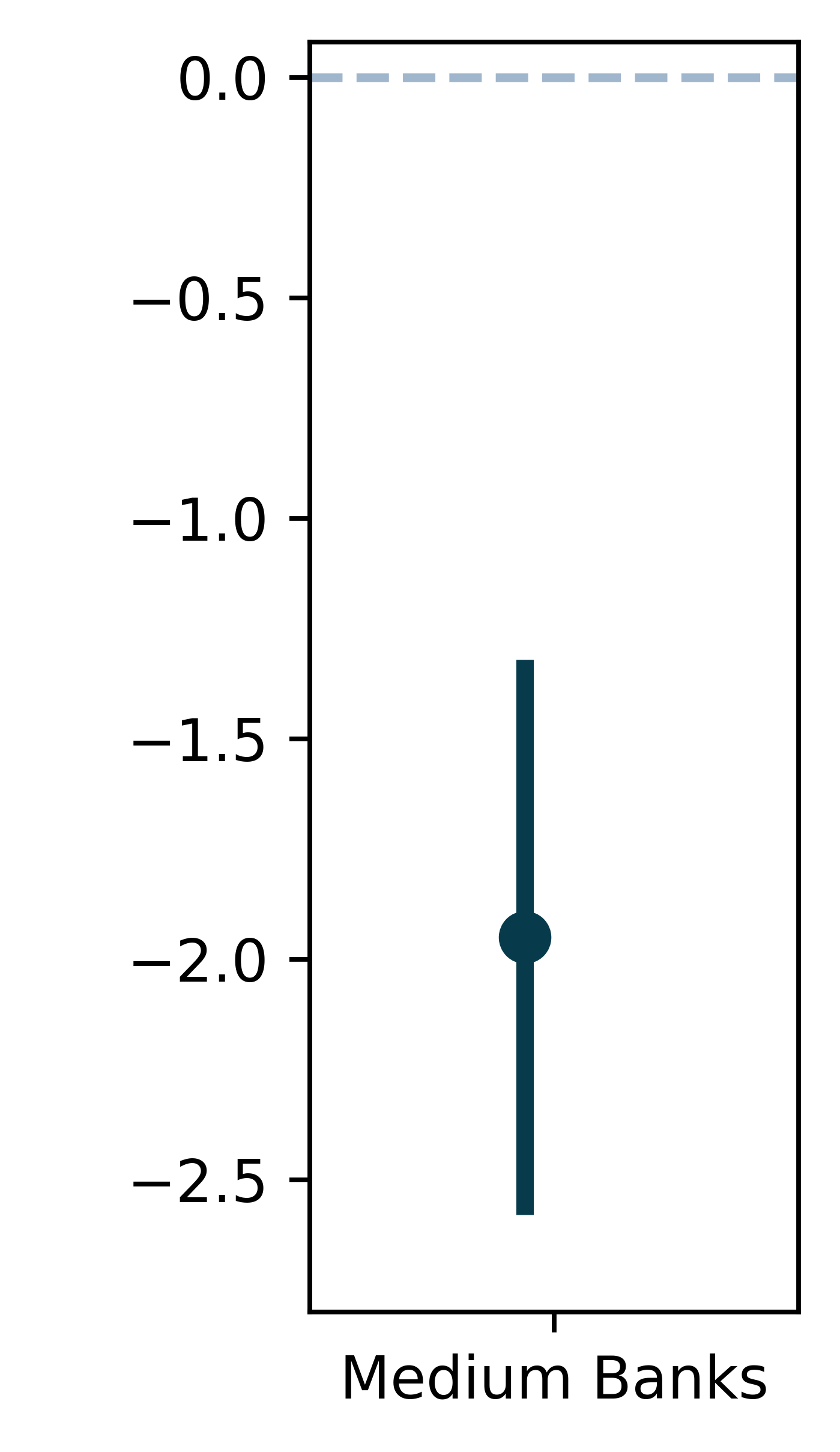}
	\end{subfigure}
	\begin{subfigure}{3cm}
		\centering\includegraphics[width=3cm]{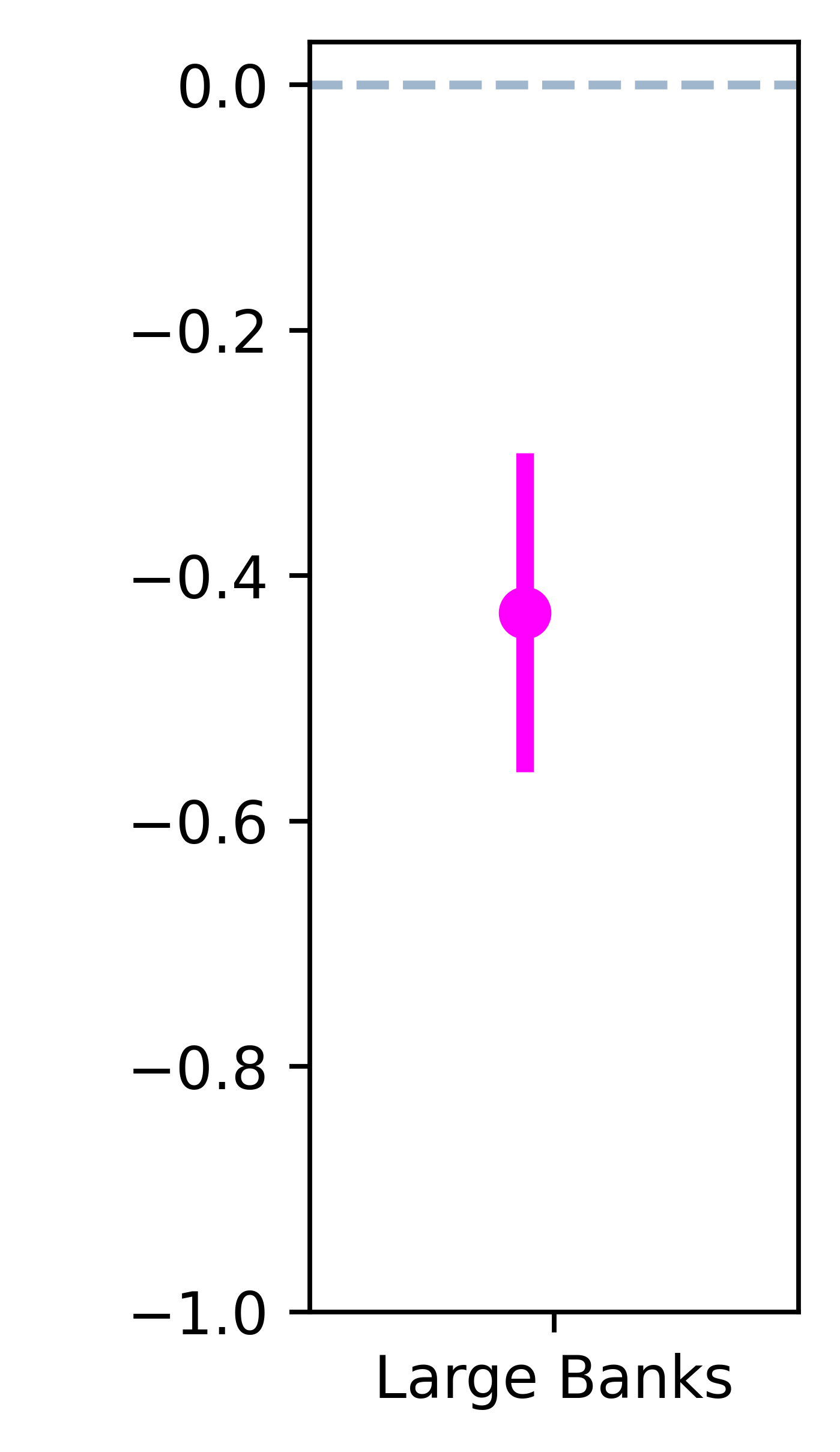}
	\end{subfigure}
	
	\begin{subfigure}{3cm}
		\centering\includegraphics[width=3cm]{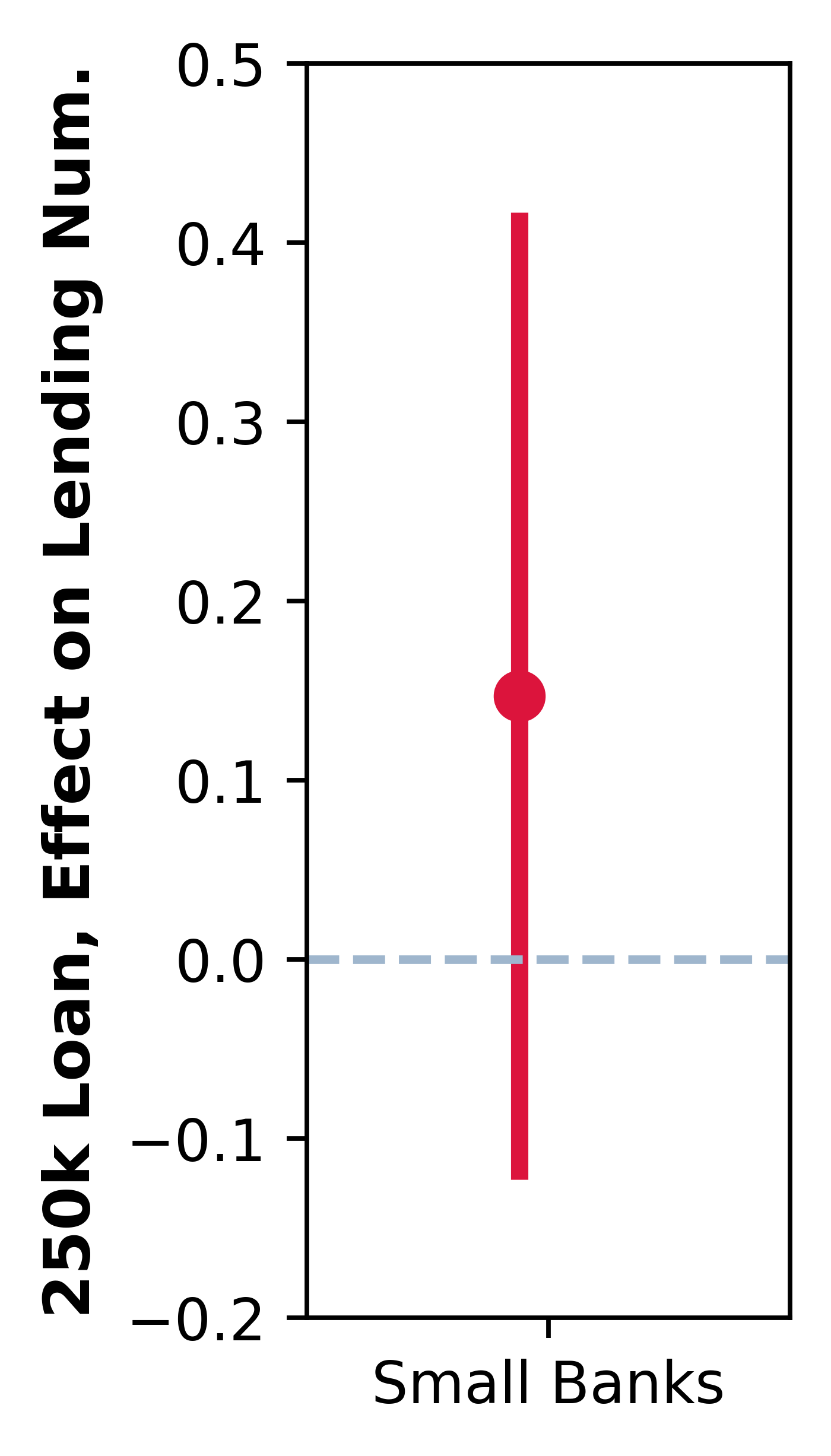}
	\end{subfigure}%
	\begin{subfigure}{3cm}
		\centering\includegraphics[width=3cm]{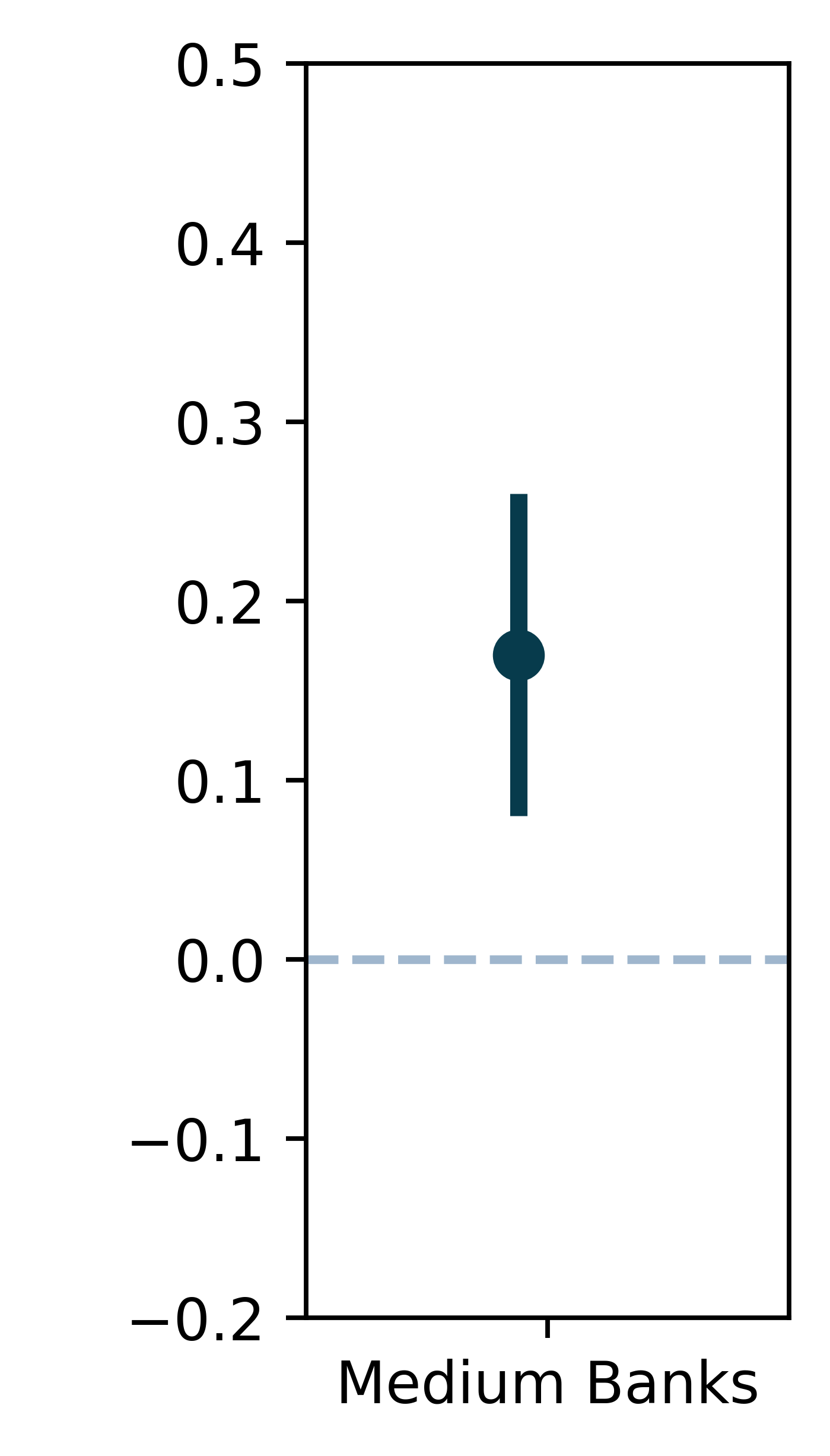}
	\end{subfigure}
	\begin{subfigure}{3cm}
		\centering\includegraphics[width=3cm]{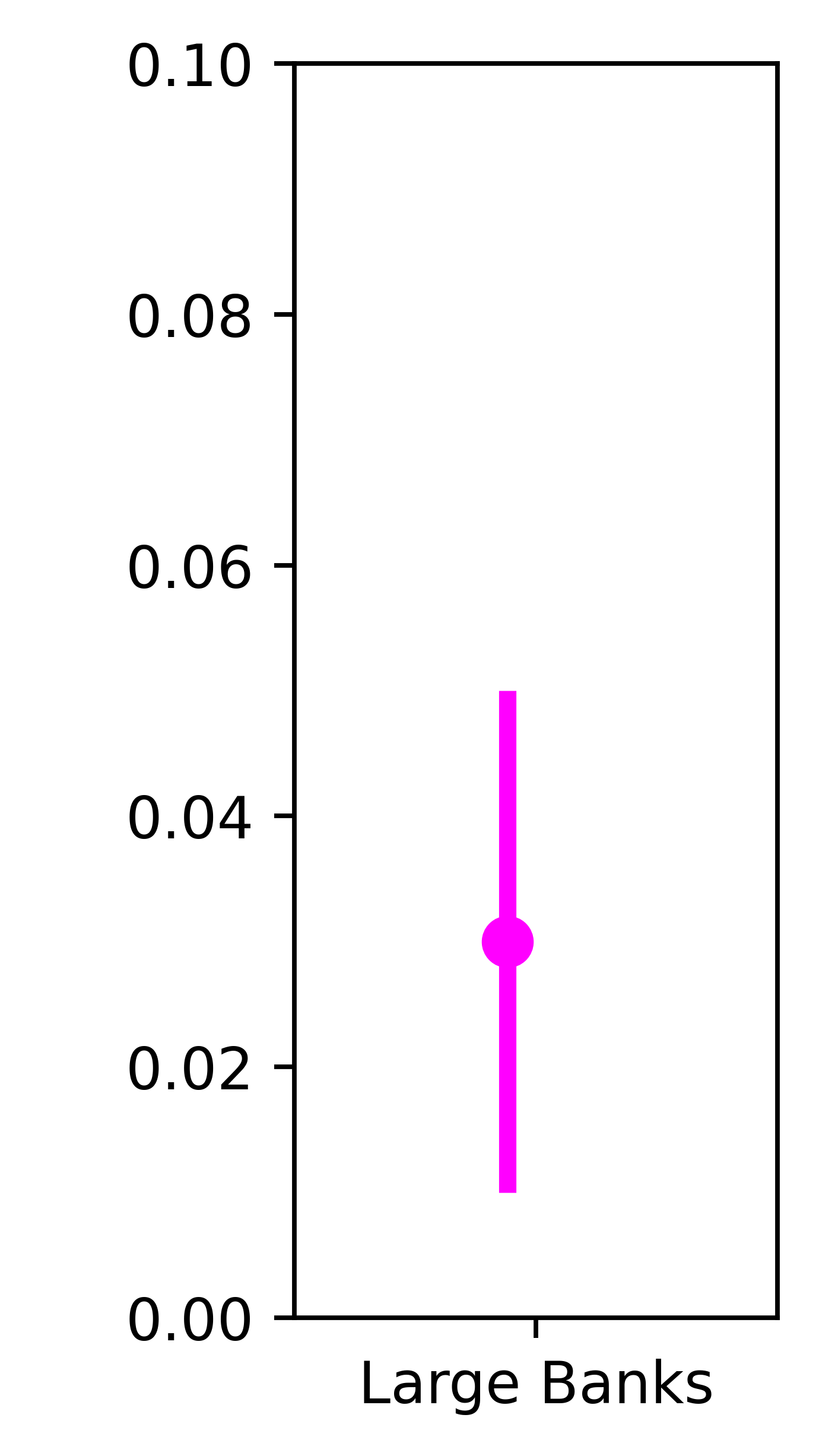}
	\end{subfigure}

	\begin{subfigure}{3cm}
		\centering\includegraphics[width=3cm]{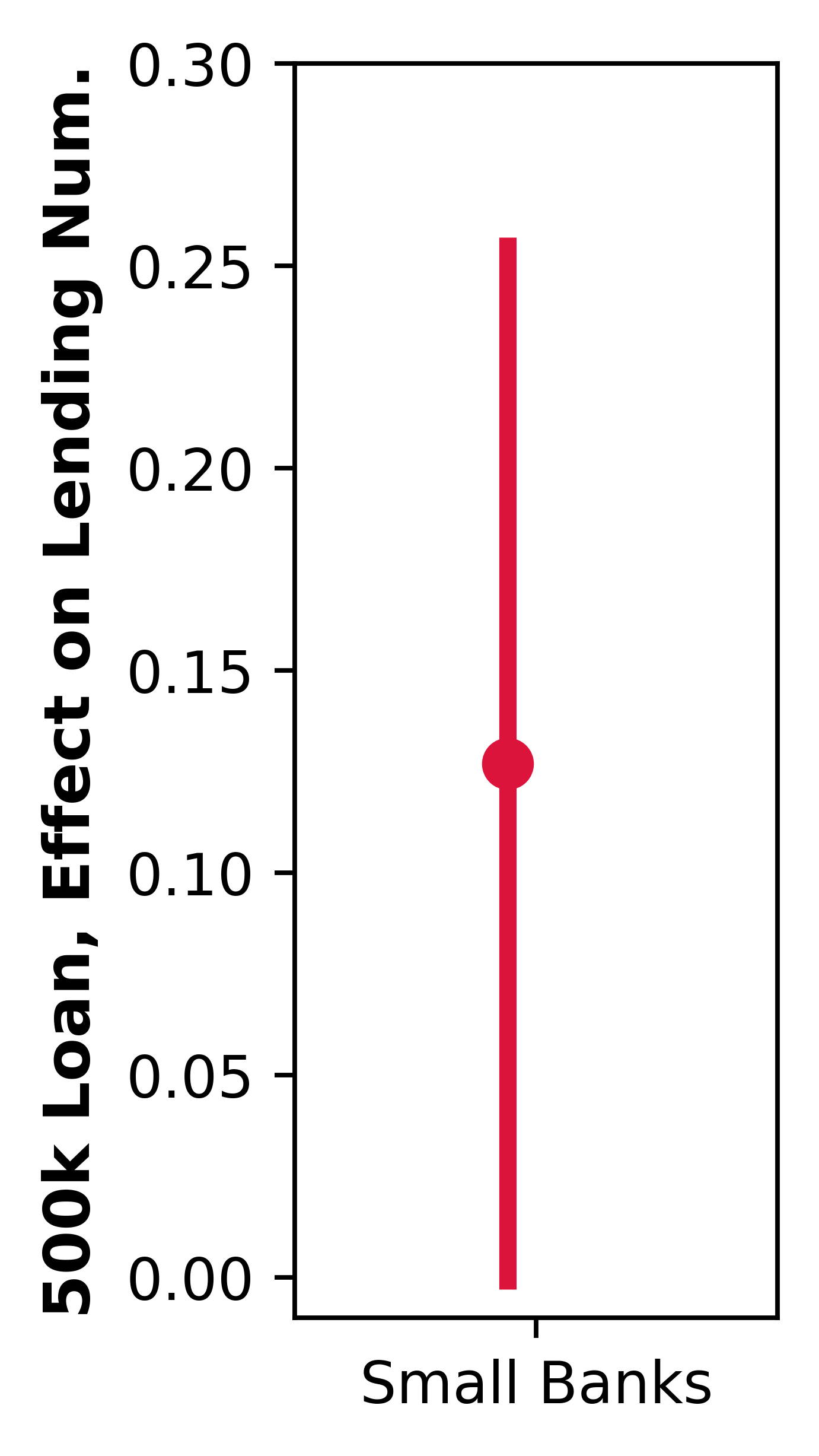}
	\end{subfigure}%
	\begin{subfigure}{3cm}
		\centering\includegraphics[width=3cm]{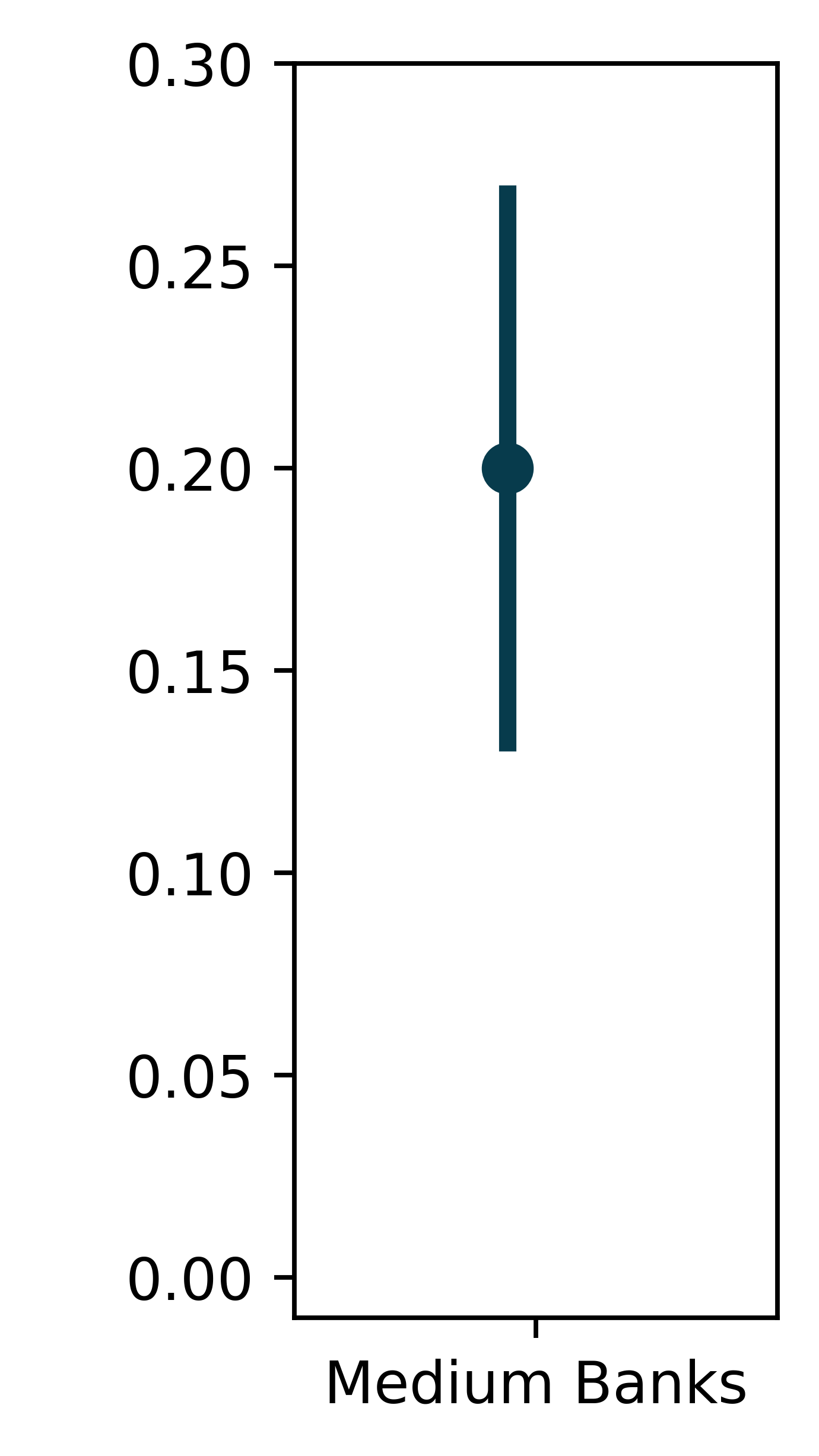}
	\end{subfigure}
	\begin{subfigure}{3cm}
		\centering\includegraphics[width=3cm]{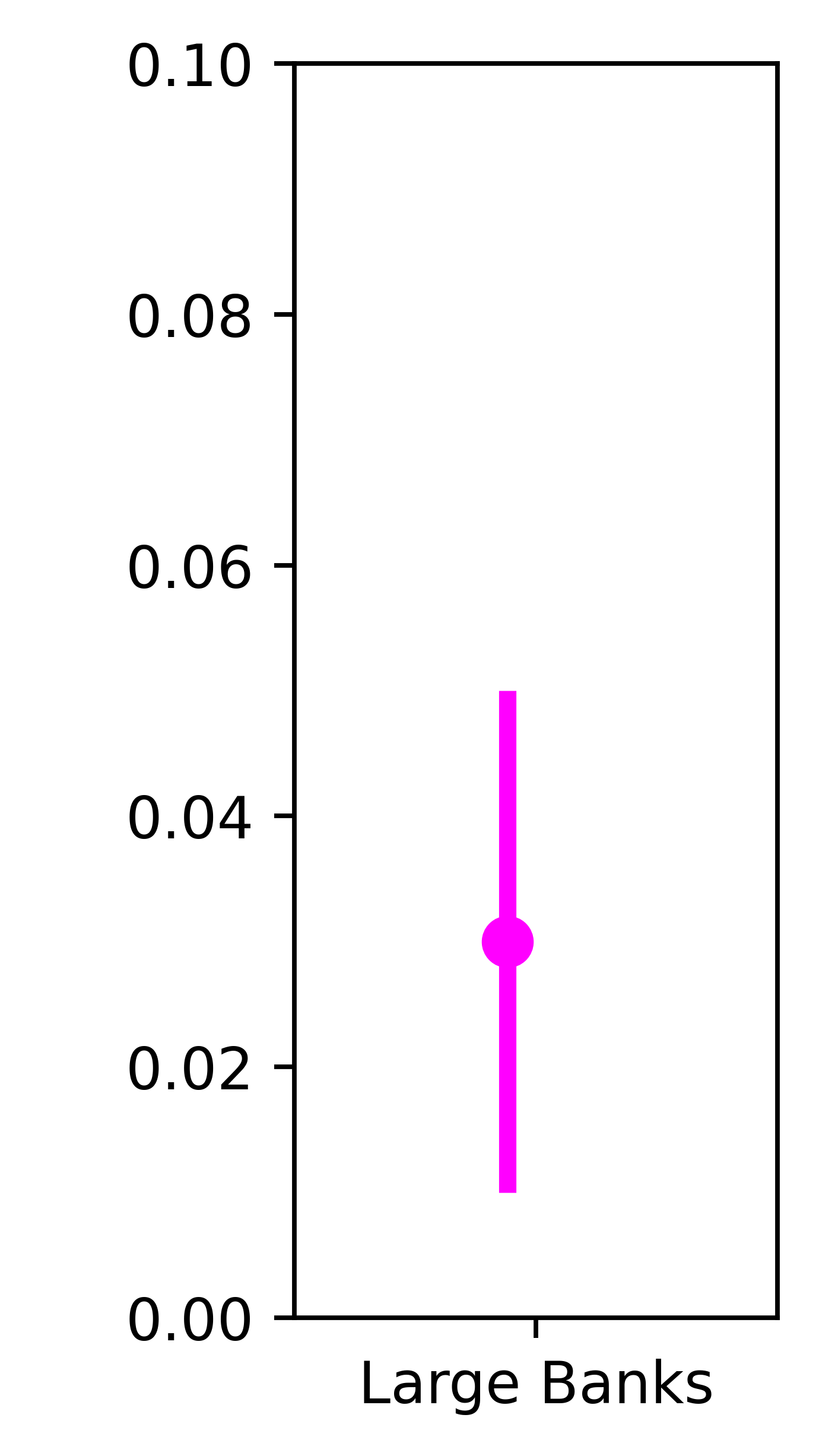}
	\end{subfigure}\par
	{\tiny	\textit{Note}: Average marginal effects with  95\% confidence intervals, assuming a temperature anomaly of  2.8 ${}^{\circ}C$/ 5.04${}^{\circ}F$. This is consistent with an adverse climate scenario, or`3X CO2 by 2100'  (SSP5-8.5). The regression results corresponding to this graph are in Table (\ref{loansize1}) in Appendix  \par}
\end{figure}

The first set of regression using Equation (\ref{e2bankinteract}) is at the loan bracket level, and the results for the effects on loan origination frequencies are illustrated in Figure (\ref{loansizemargin1}). Due to the interpretation challenge posed by the multitude of interaction terms, the full results of the regressions are reported in Table (\ref{loansize1}) in Appendix, where the vast majority of main and interaction terms are highly significant. To facilitate interpretation, I estimate the average marginal effects, reported in Figure (\ref{loansizemargin1}).\footnote{The calculation procedure is similar to that in Equation (\ref{margin1}). Note due to the peculiarity of how Stata calculates marginal effects, there may be slight differences between the results reported in the graph and those calculated by hand.}

In response to climate risks, the contrast between how banks make 100k loans and the other two sizes is stark. As shown by the first row of graphs of Figure (\ref{loansizemargin1}), holding all else constant, small banks will make more 100k loans to farms. In comparison, it is much more likely for medium and large banks to reject such loan applications. In terms of magnitude, the mean marginal effect of medium banks is much bigger than that of large banks. It is worth emphasizing the contrast of lending behavior here. As discussed previously, small farms are most likely the recipients of these smaller, 100k loans. With respect to the increase of climate risks, small banks actually want to provide more of such loans, likely due to small farms being important clients for them. However, medium substantially reduce funding access to the 100k loans. Most importantly, large banks also respond by lending less, and their dominance in the smaller loan market likely add up. Put another way, while small banks want to support (mostly small) farms who apply for 100k loans, their market share is not substantial enough to compensate for the withdraw of funding from large banks. 

The second and third rows of Figure (\ref{loansizemargin1}) provide further evidence of heterogeneous effects by loan size. For the 250k and 500k loans, all banks, including medium and large banks respond to increased temperature anomaly by lending more, and such larger loans are more likely to go to large farms. The lending behavior by bank type can also be viewed vertically. In general, small banks do not decrease funding access. In comparison, larger banks deny loans only of smaller sizes, but will approve larger size loans. 

Additional estimations are conducted in terms of loan amount, and the full results are reported in Table (\ref{tbankcountyinteract}) in Appendix. Figure (\ref{loansizemargin2}) illustrates the estimates of average marginal effects, and the results are consistent with Figure (\ref{loansizemargin1}) in terms of the directions of impact. 

\begin{figure}[!]
	\centering
	\caption{Marginal Effect of Temperature Anomaly on CRA Lending Amount, by Loan Size}\label{loansizemargin2} 
	\begin{subfigure}{3cm}
	\centering\includegraphics[width=3cm]{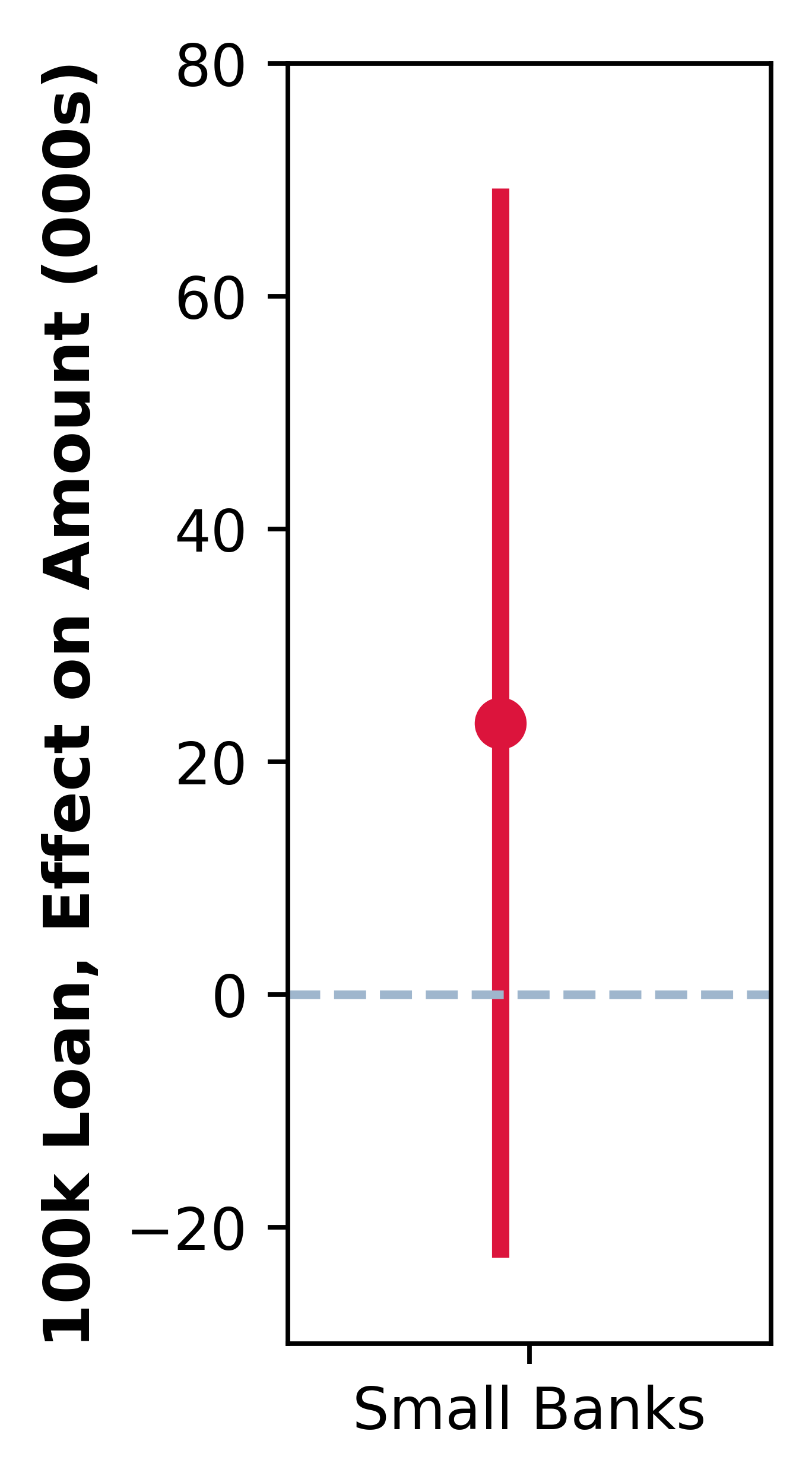}
	\end{subfigure}%
	\begin{subfigure}{3cm}
		\centering\includegraphics[width=3cm]{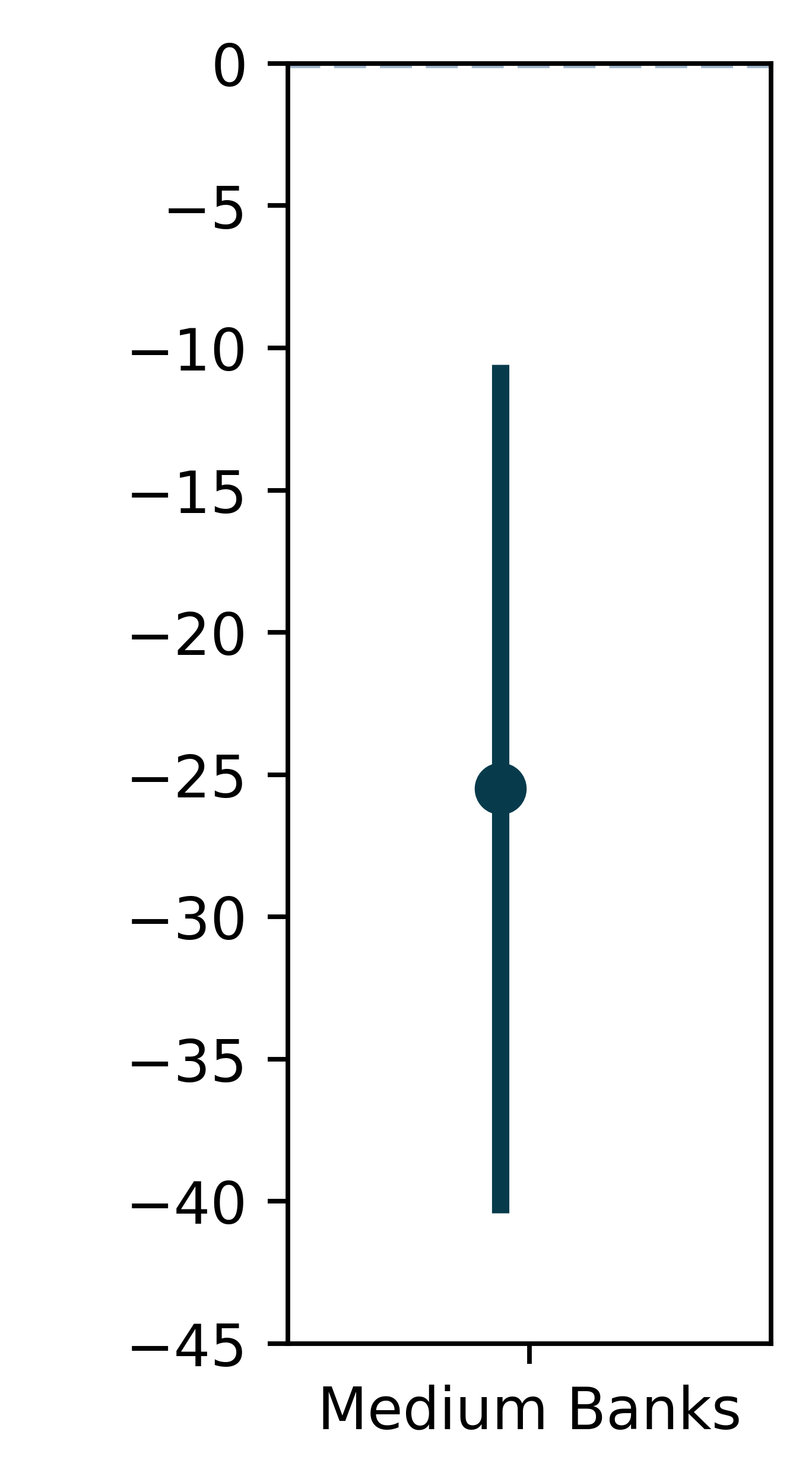}
	\end{subfigure}
	\begin{subfigure}{3cm}
		\centering\includegraphics[width=3cm]{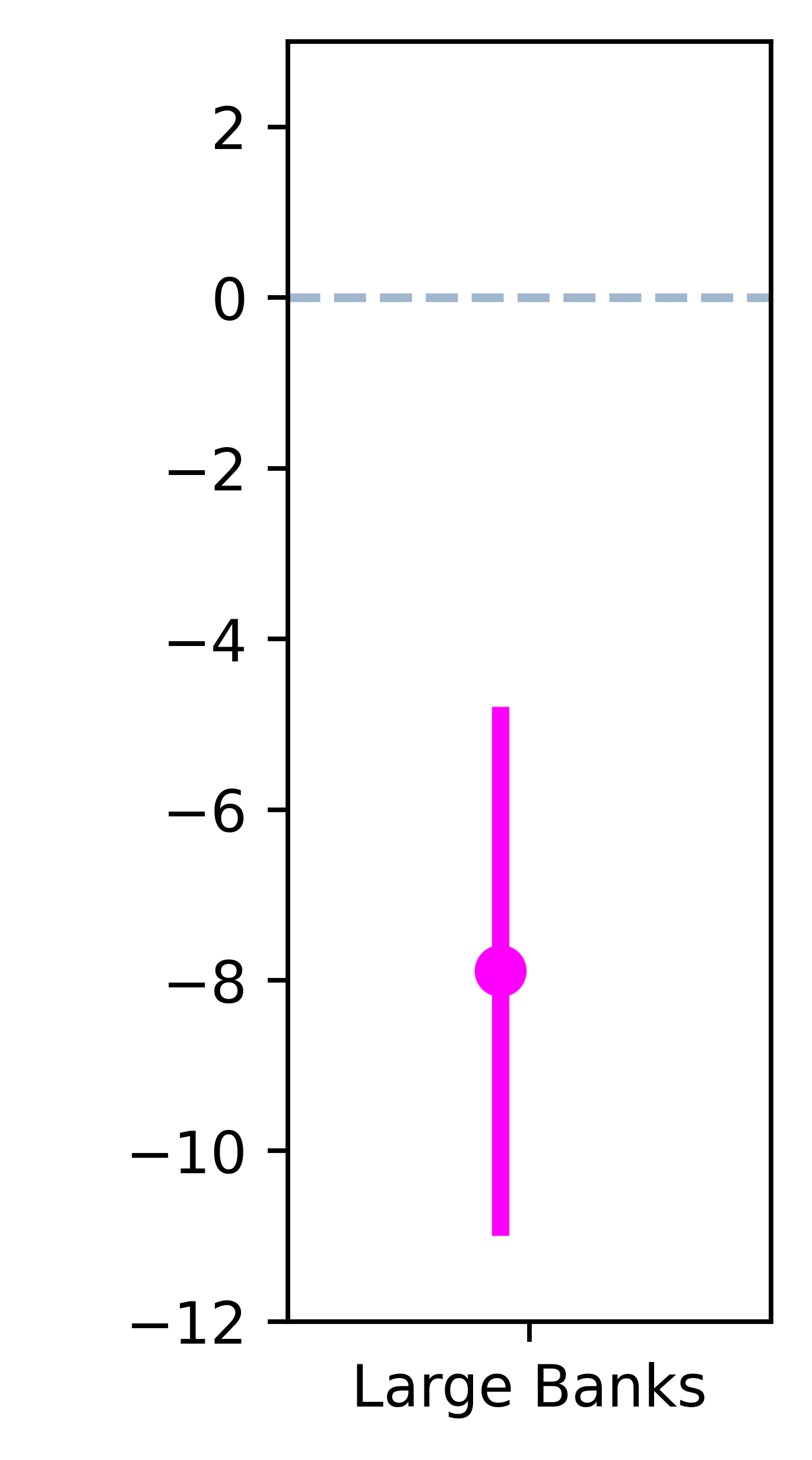}
	\end{subfigure}
	
	\begin{subfigure}{3cm}
		\centering\includegraphics[width=3cm]{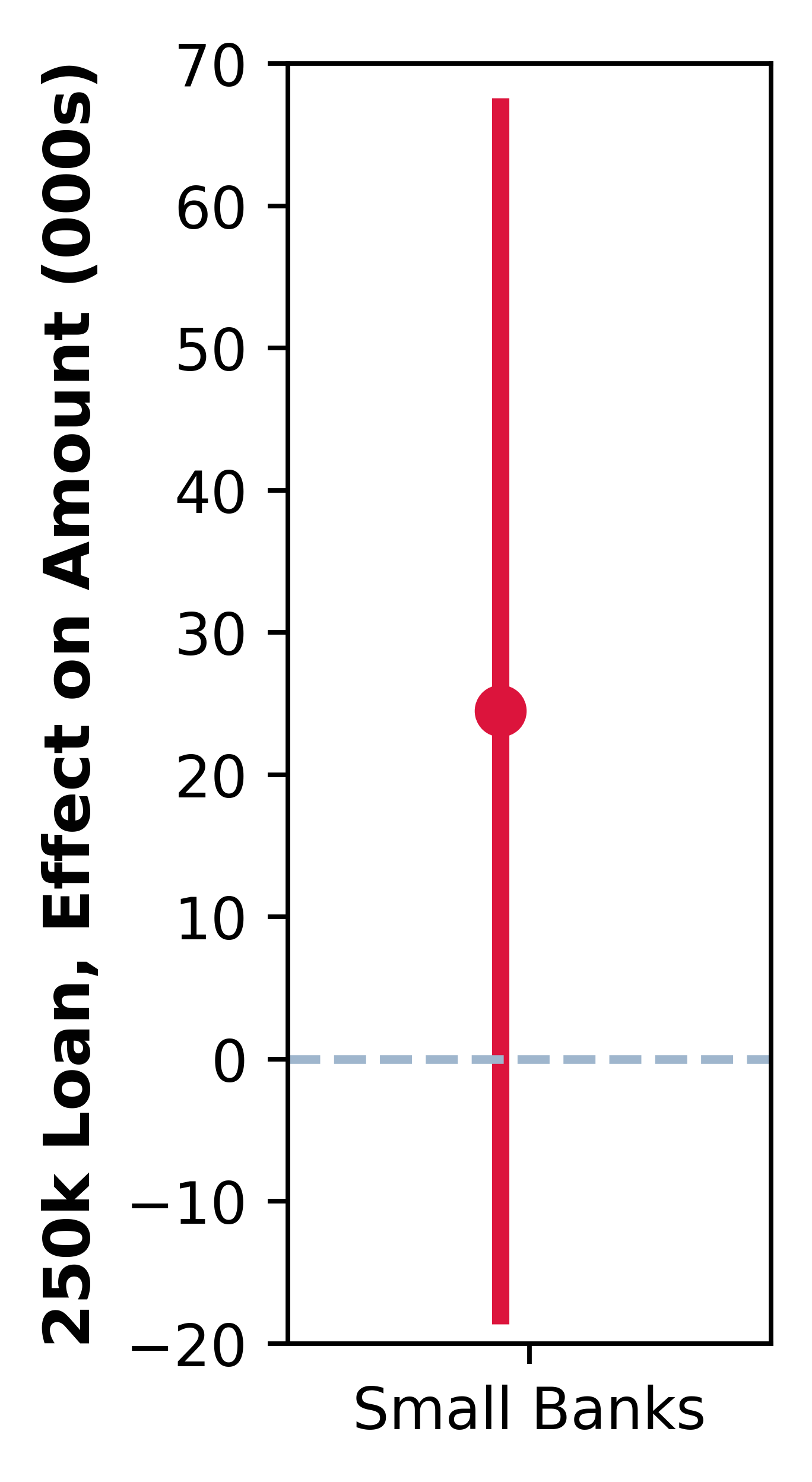}
	\end{subfigure}%
	\begin{subfigure}{3cm}
		\centering\includegraphics[width=3cm]{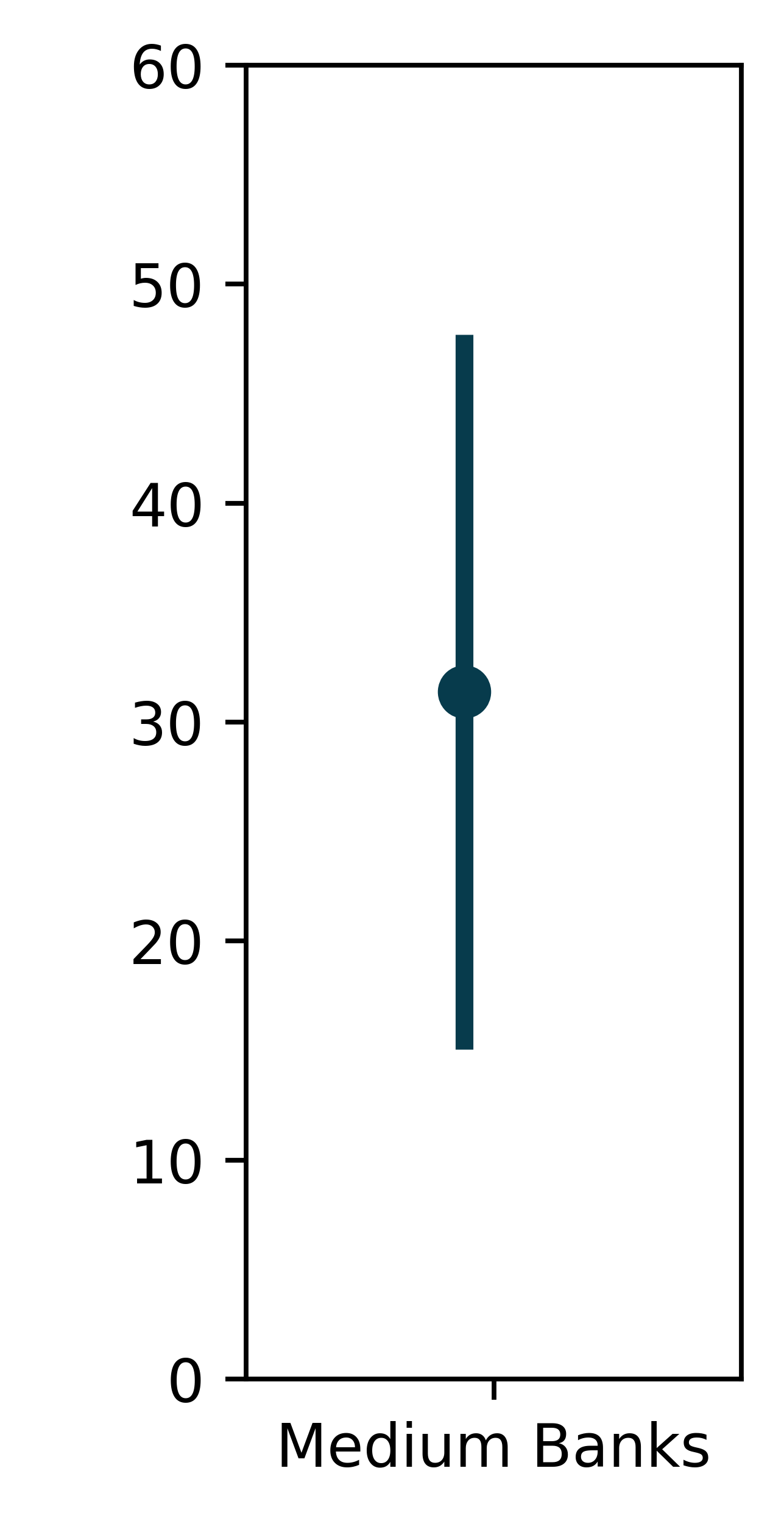}
	\end{subfigure}
	\begin{subfigure}{3cm}
		\centering\includegraphics[width=3cm]{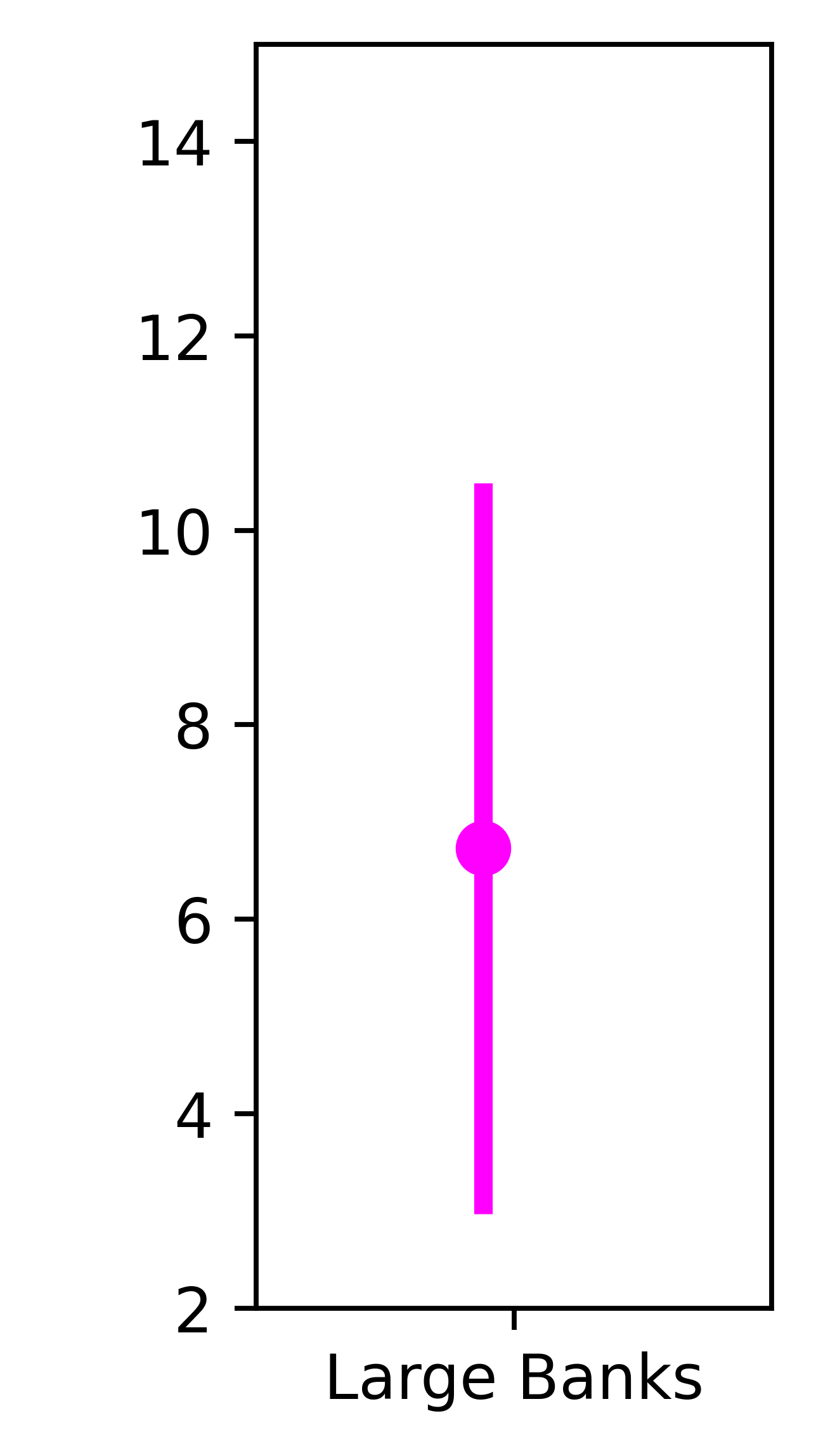}
	\end{subfigure}

	\begin{subfigure}{3cm}
		\centering\includegraphics[width=3cm]{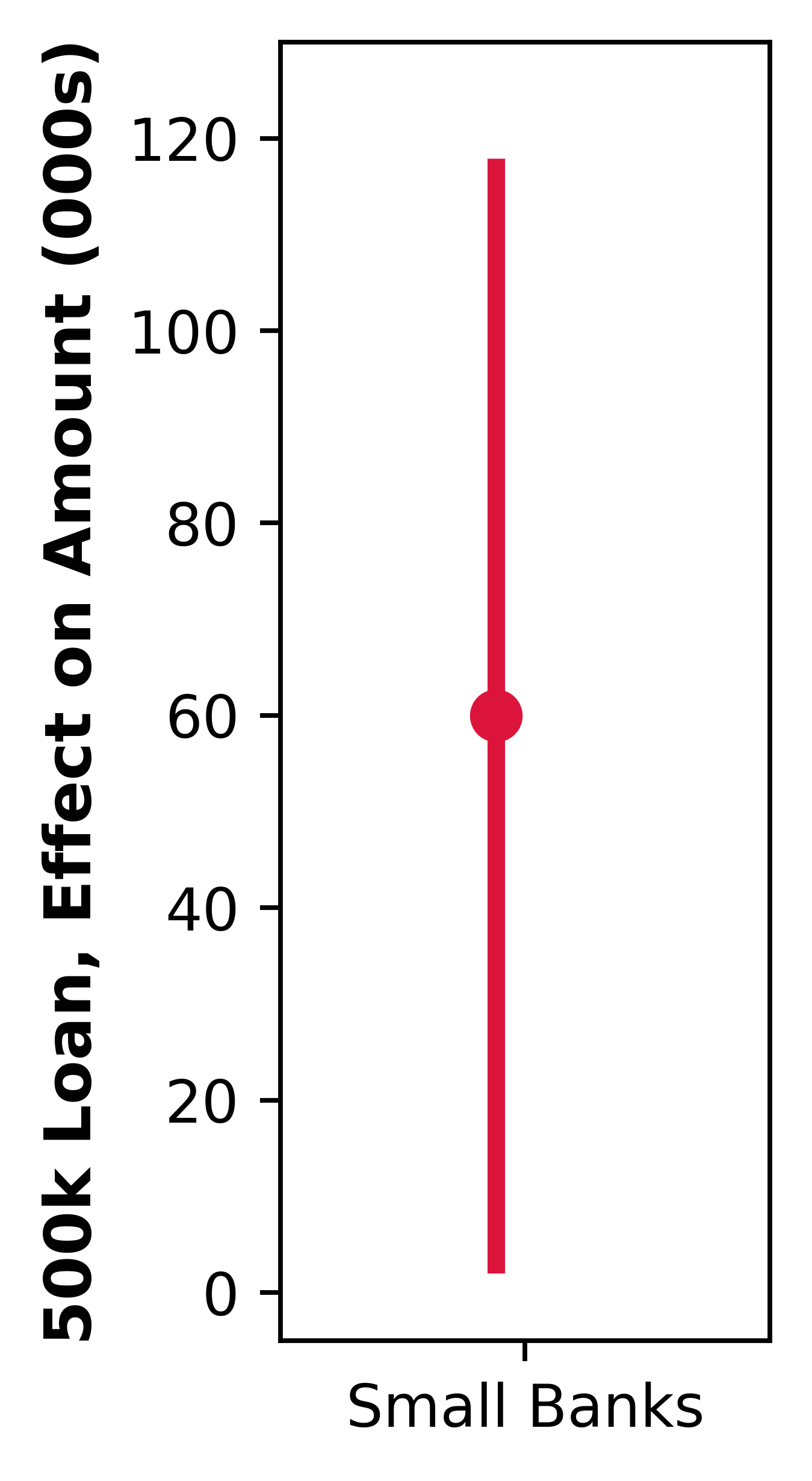}
	\end{subfigure}%
	\begin{subfigure}{3cm}
		\centering\includegraphics[width=3cm]{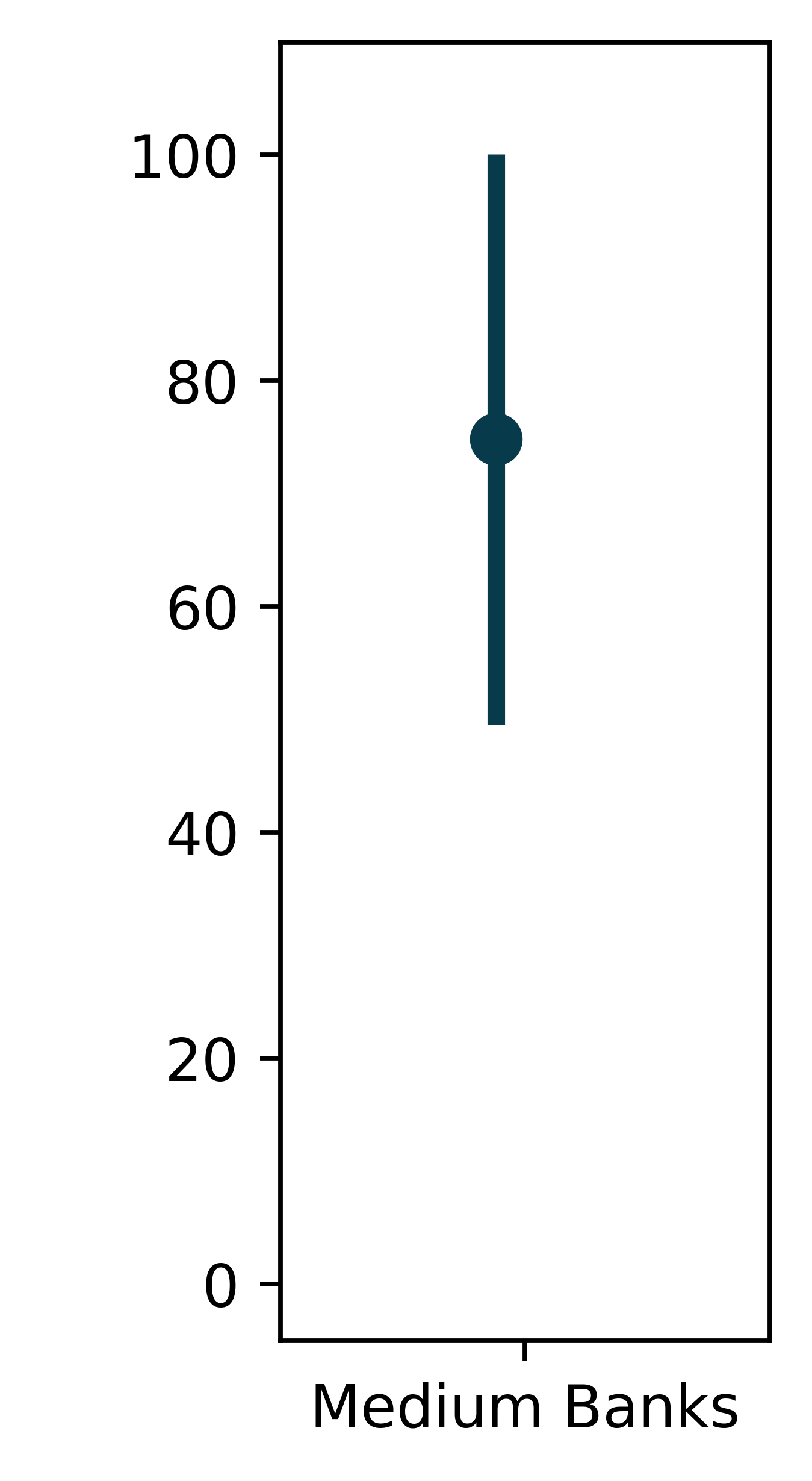}
	\end{subfigure}
	\begin{subfigure}{3cm}
		\centering\includegraphics[width=3cm]{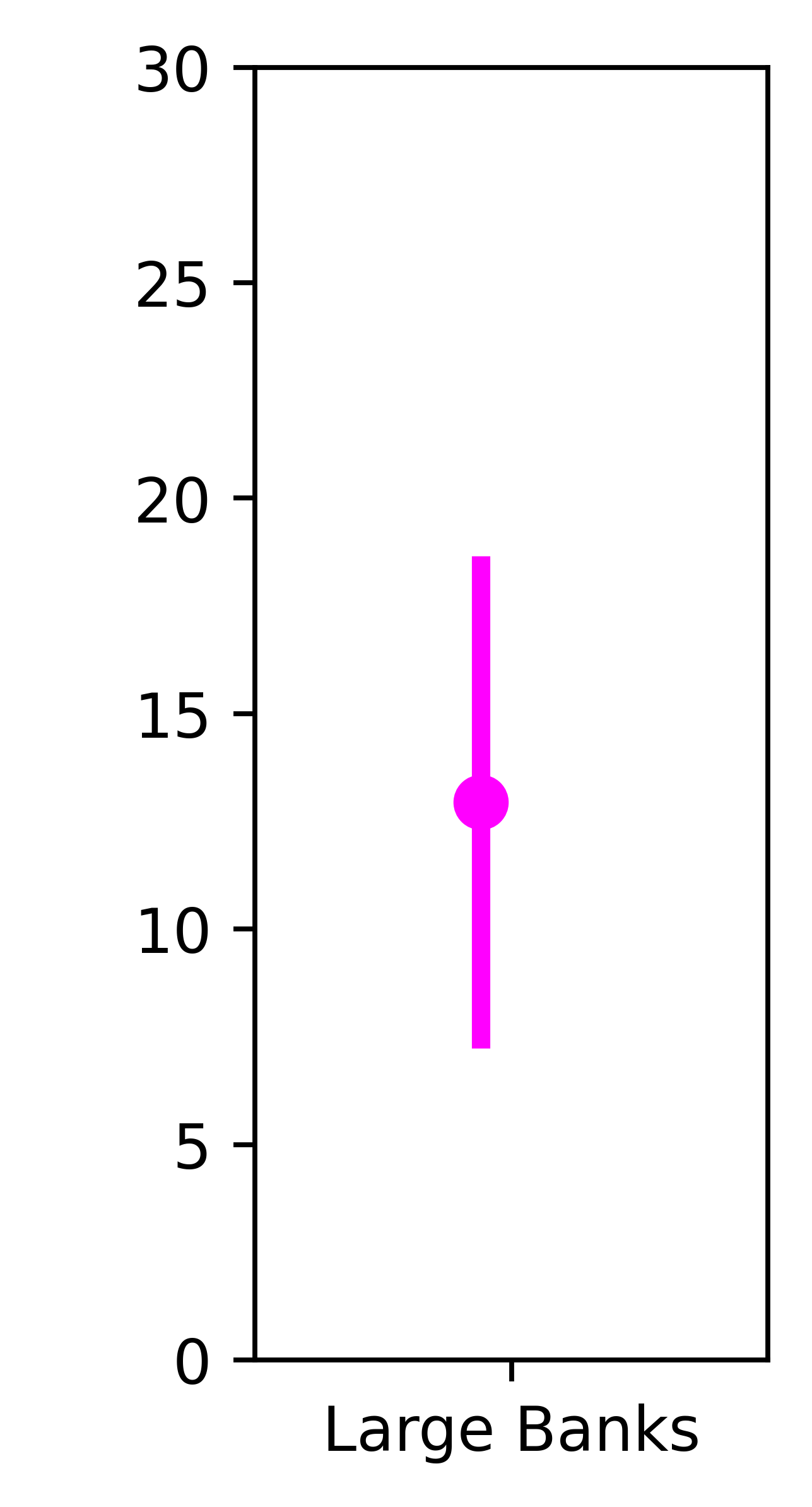}
	\end{subfigure}\par
	{\tiny	\textit{Note}: Average marginal effects with  95\% confidence intervals, assuming a temperature anomaly of  2.8 ${}^{\circ}C$/ 5.04${}^{\circ}F$. This is consistent with an adverse climate scenario, or`3X CO2 by 2100'  (SSP5-8.5). The regression results corresponding to this graph are in Table (\ref{loansize2}) in Appendix  \par}
\end{figure}

\begin{figure}[!]
	\centering
	\caption{Marginal Effect of Temperature Anomaly on CRA Lending Frequencies, by Farm Size}\label{bankmargin1} 
	\begin{subfigure}{3cm}
		\centering\includegraphics[width=3cm]{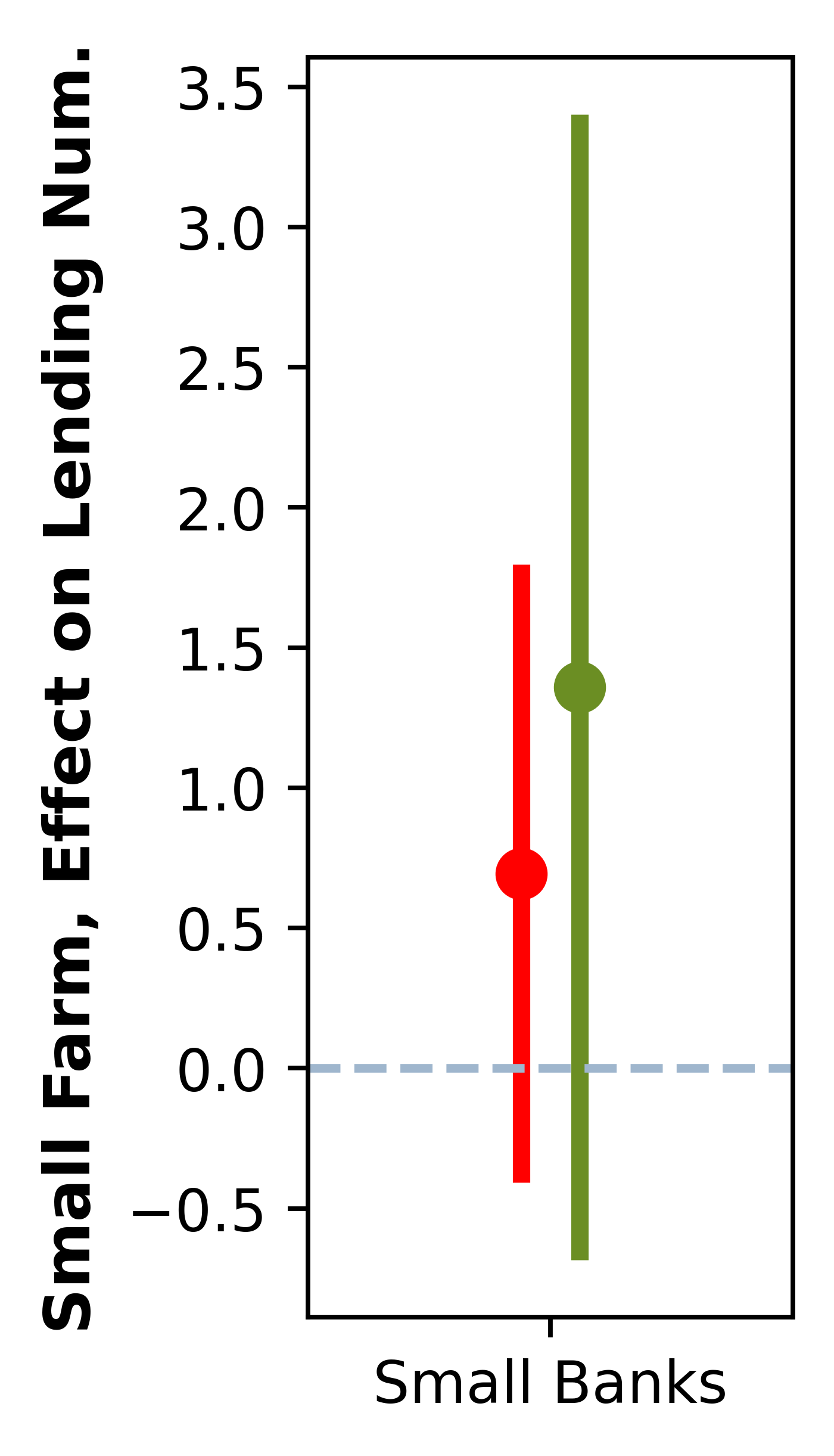}
	\end{subfigure}%
	\begin{subfigure}{3cm}
		\centering\includegraphics[width=3cm]{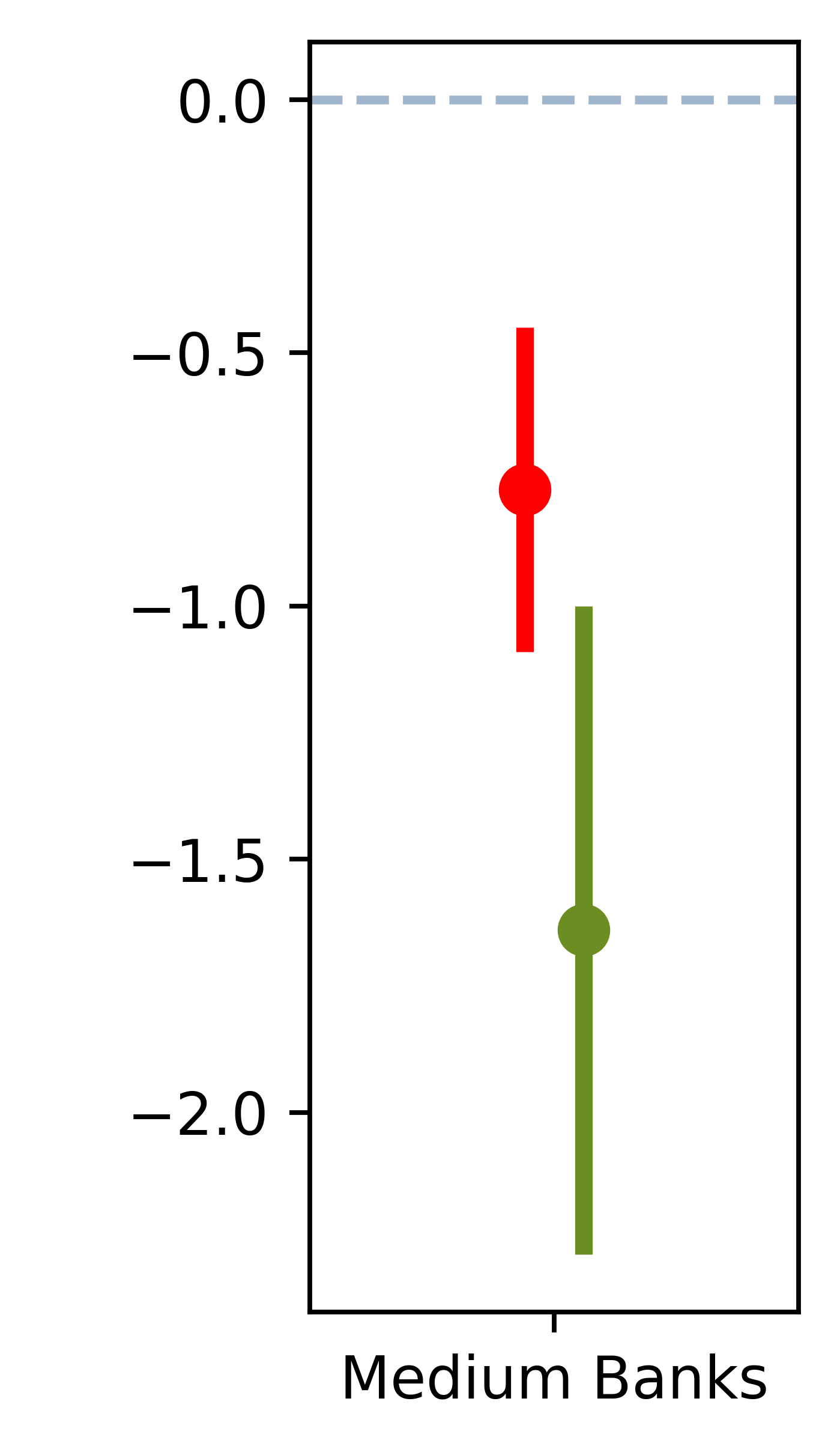}
	\end{subfigure}
	\begin{subfigure}{3cm}
		\centering\includegraphics[width=5.95cm]{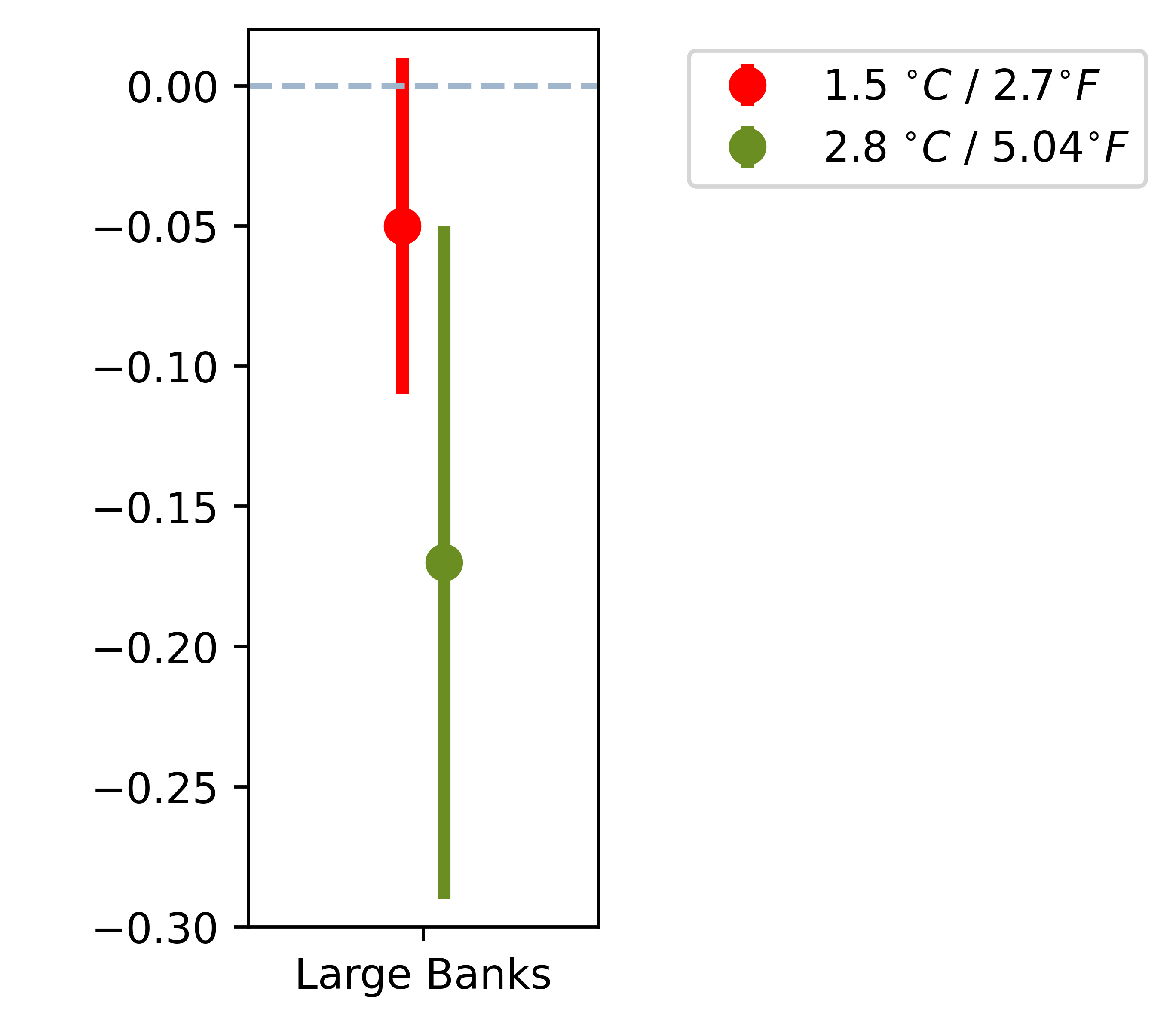}
	\end{subfigure}
	
	\begin{subfigure}{3cm}
		\centering\includegraphics[width=3cm]{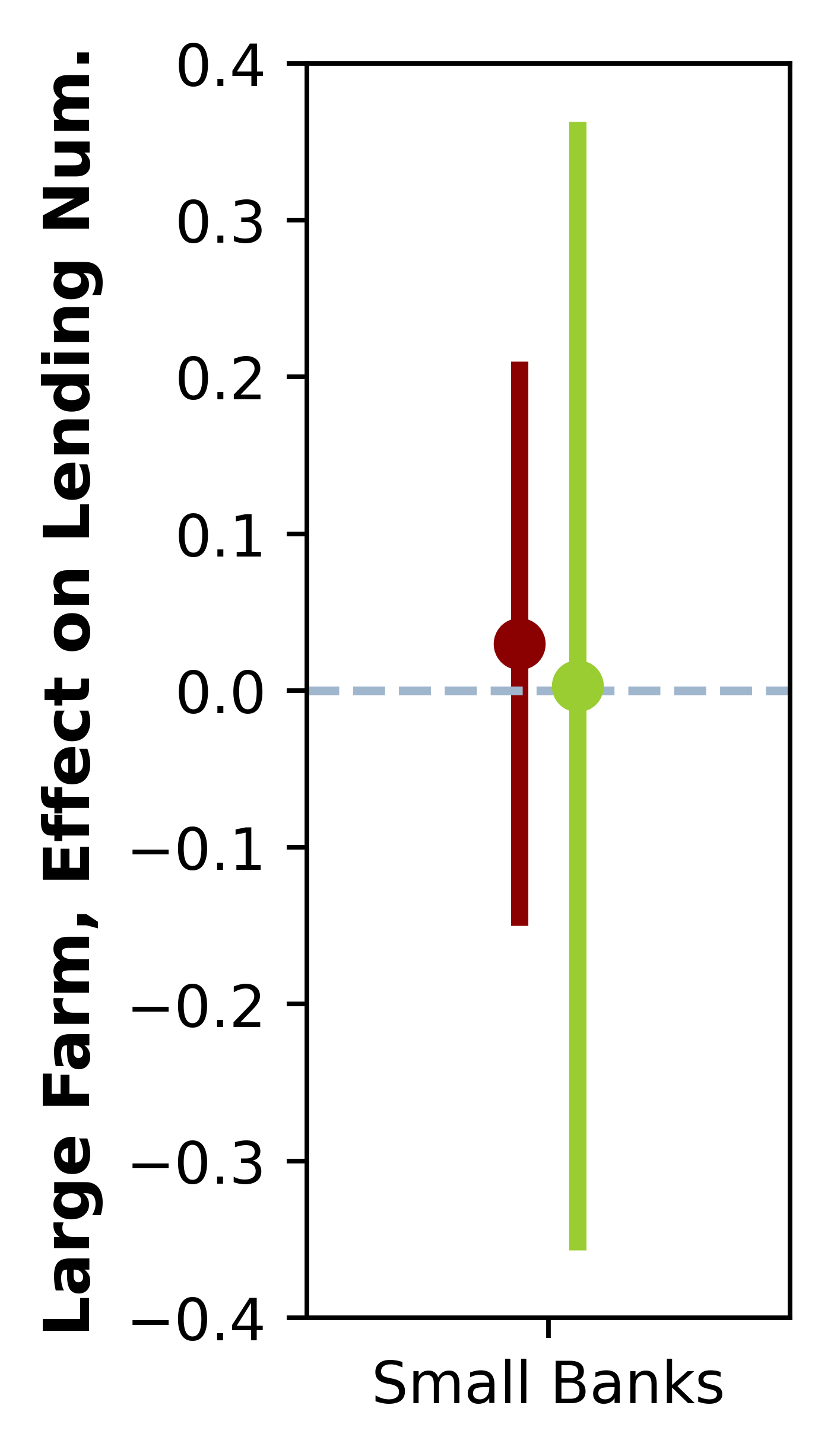}
	\end{subfigure}%
	\begin{subfigure}{3cm}
		\centering\includegraphics[width=3cm]{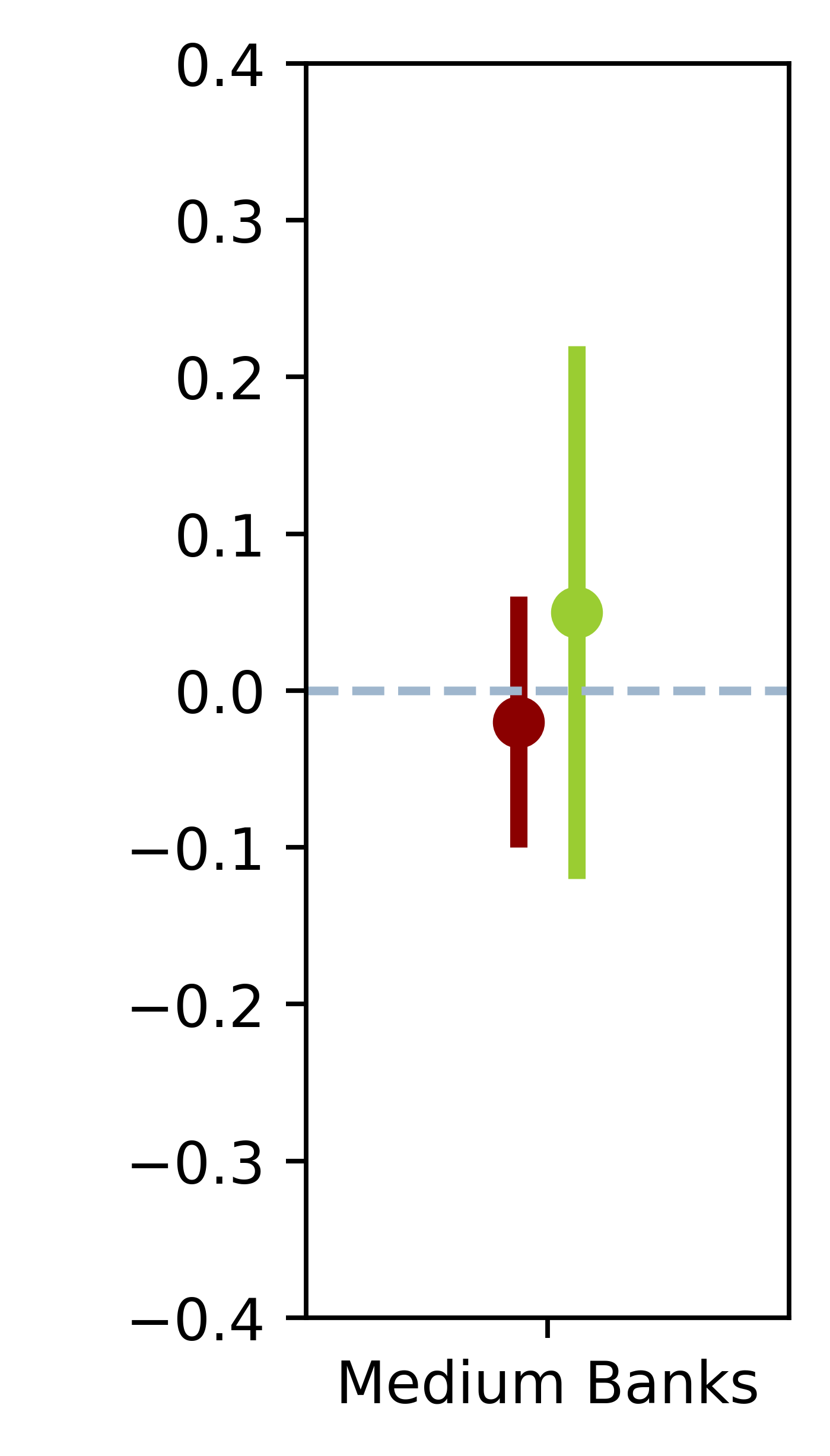}
	\end{subfigure}
	\begin{subfigure}{3cm}
		\centering\includegraphics[width=5.95cm]{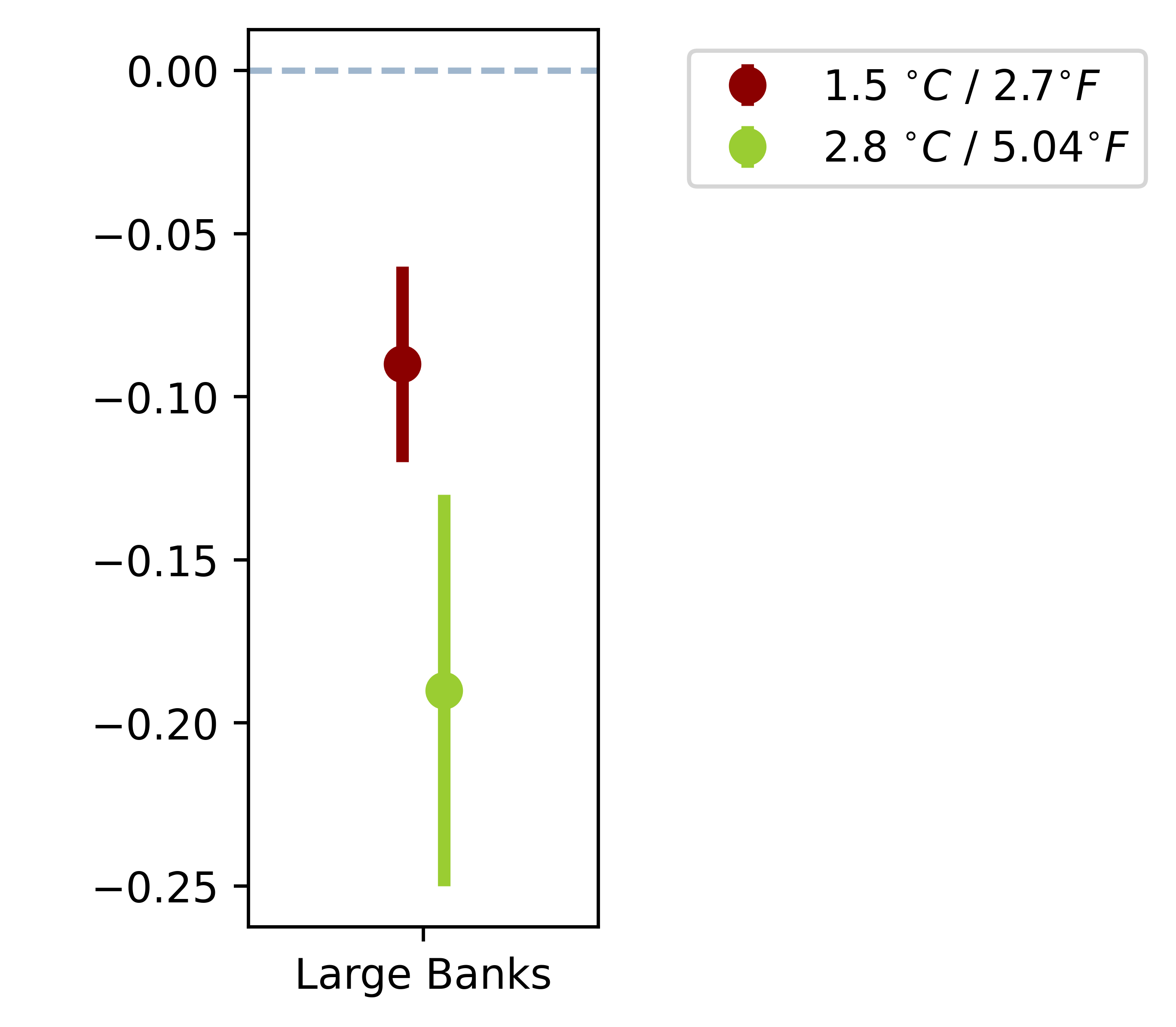}
	\end{subfigure}\par
	{\tiny	\textit{Note}: Average marginal effects with  95\% confidence intervals, with two scenarios: i). temperature anomaly of 1.5 ${}^{\circ}C$/ 2.7${}^{\circ}F$; ii). temperature anomaly of  2.8 ${}^{\circ}C$/ 5.04${}^{\circ}F$;  The first is consistent with a moderate climate scenario, or `net zero by 2075' (SSP1-2.6). The second is consistent with an adverse climate scenario, or`3X CO2 by 2100'  (SSP5-8.5). The regression results corresponding to this graph are in Table (\ref{tbankcountyinteract}) in Appendix.  \par}
\end{figure}

With the results of loan size providing contexts, I use Equation (\ref{e2bankinteract}) to run regressions on the loans  to small and large farms. The results of loan frequencies are illustrated in Figure (\ref{bankmargin1}), and those of loan magnitudes shown in Figure (\ref{bankmargin2}) (regression results shown in Table (\ref{tbankcountyinteract}) in Appendix; most of the main and interaction coefficients are significant). Additionally, the results are estimated for two climate scenarios: moderate and severe. 

The first row of Figure (\ref{bankmargin1}) illustrates the results for small farms, while the second row shows the results for large farms. For small farms, the contraction of lending frequencies from medium and large banks are noticeable, and the magnitude of impact from medium banks is particularly large. Small banks do increase their lending frequencies to small farms. But such increased support is unlikely to overturn the overall loss of lending access that small farms experience, simply due to small banks' very limited market share. 

In contrast, for large farms, the impact of lending frequencies from small and medium banks are negligible or even positive at times. Only large banks noticeably reduce lending frequencies to large farms. One way to interpret these results is that as farms' climate vulnerability increase, small farms do not seem to drastically reduce their perceived risk exposure. Instead, they increase lending to small farms likely due to that they rely on this type of farms as clients. Medium banks try to reduce their exposure to small farms by reducing loan frequencies, while maintaining relatively similar level of lending activities to large farms. Thus medium banks de-risk by lending less to small farms. Other than that, they largely maintain their lending activities within the same county. Moreover, since large banks reduce their lending frequencies to both small and large farms, it is plausible that these large banks de-risk by moving their farm lending to other less-risky counties (the only other valid interpretation is that large banks reduce their farm lending altogether). Thus, for large banks, there may be leakage of lending across counties. This is plausible as such banks have more extensive branches and operations than smaller-sized banks. 

Figure (\ref{bankmargin2}) shows the results for the estimation of lending magnitudes, which is based on the bank-county level estimation showing the effect \textit{conditional on} the loans already being approved. The qualitative model of the paper predicts that as climate risks increase, farms' financing needs also increase. Therefore it is reasonable to expect that the actual amount of lending will increase. This is exactly what is shown in Figure (\ref{bankmargin2}). Conditional on already being approved for loans, both small and large farms obtain increased amount of lending. It is interesting that the magnitudes of impact are more pronounced for small farms. But this is not surprising, if one assumes that small farms want to access more funding to improve their climate resilience.  It is important to first point out the difference of the results here versus county-level estimations of lending amount. For the county-level regression results shown earlier in the paper, the results are underpinned by the composition effect: there is a multitude of banks operating within a county, and on aggregation the effect could be negative for a county. 

\begin{figure}[!]
	\centering
	\caption{Marginal Effect of Temperature Anomaly on CRA Lending Amount, by Farm Size}\label{bankmargin2} 
	\begin{subfigure}{3cm}
		\centering\includegraphics[width=3cm]{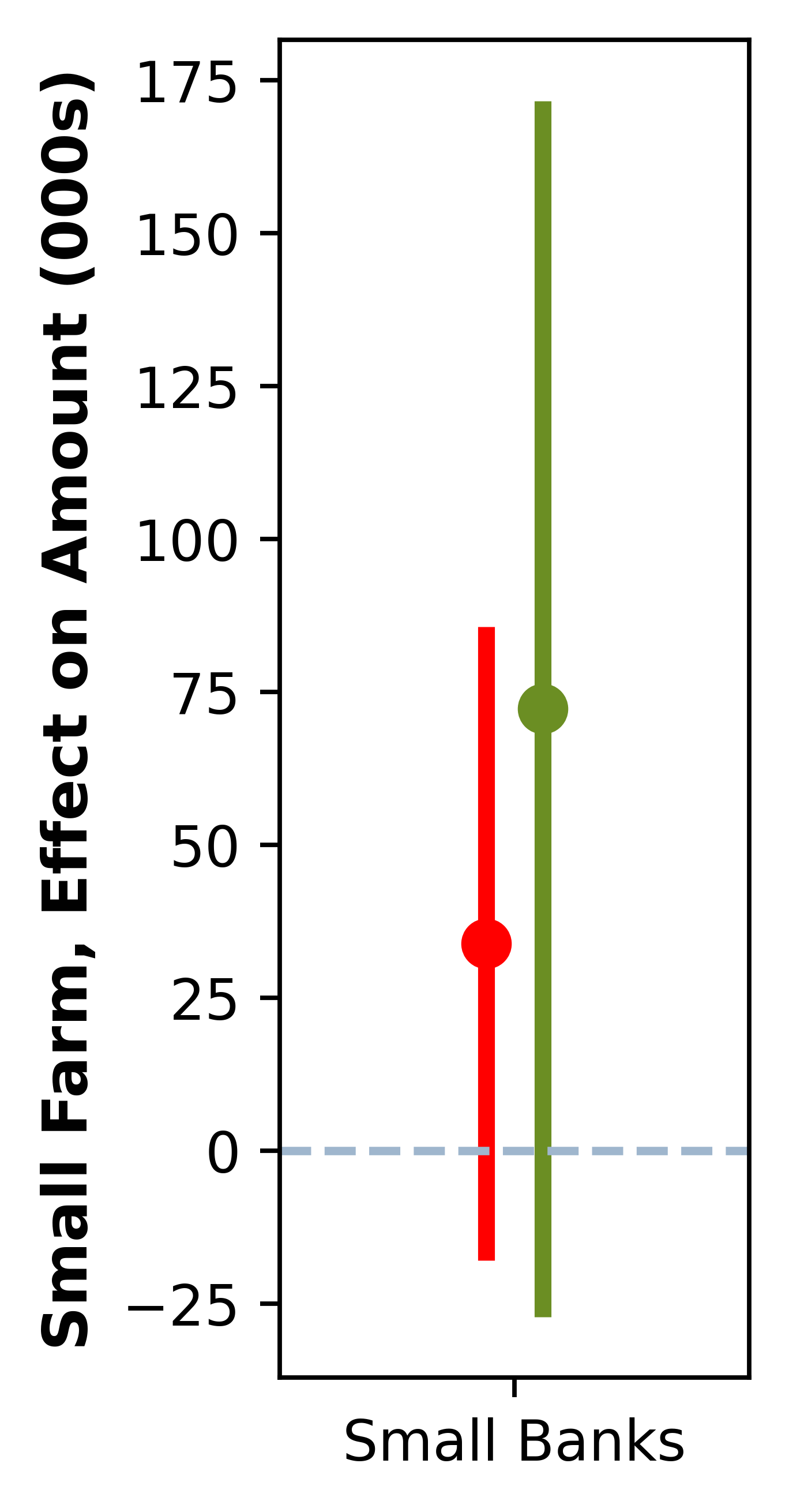}
	\end{subfigure}%
	\begin{subfigure}{3cm}
		\centering\includegraphics[width=3cm]{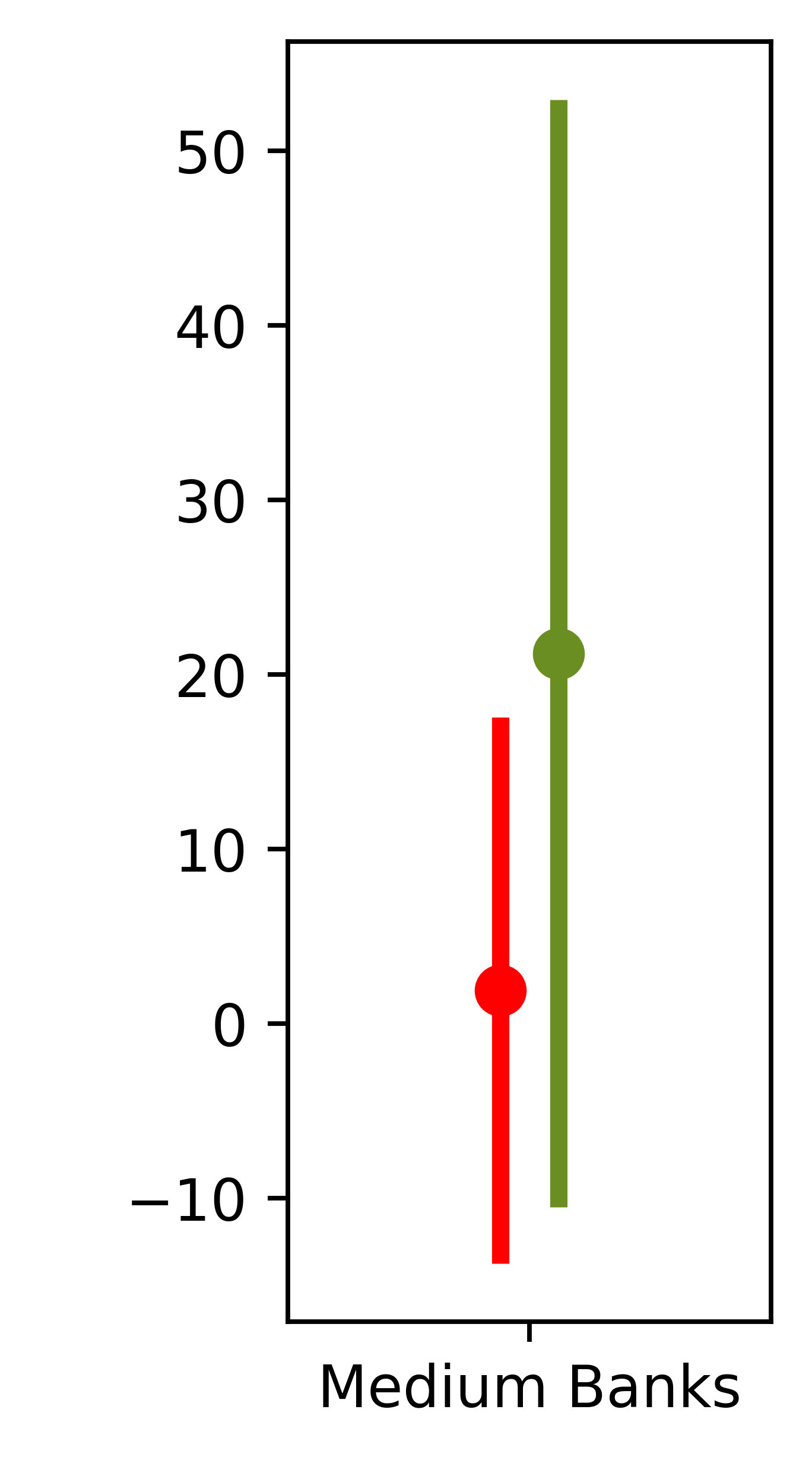}
	\end{subfigure}
	\begin{subfigure}{3cm}
		\centering\includegraphics[width=5.95cm]{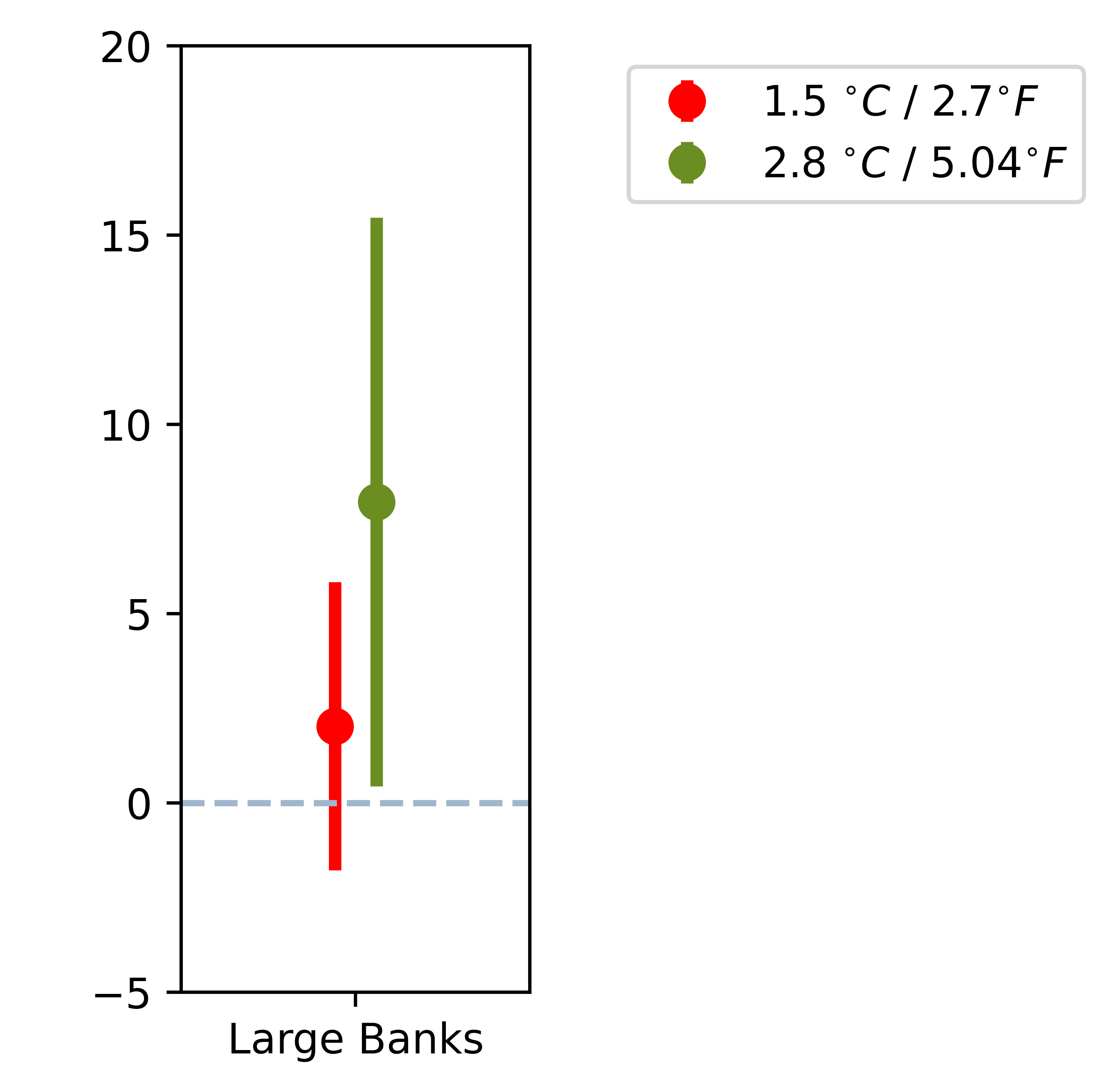}
	\end{subfigure}
	
	\begin{subfigure}{3cm}
		\centering\includegraphics[width=3cm]{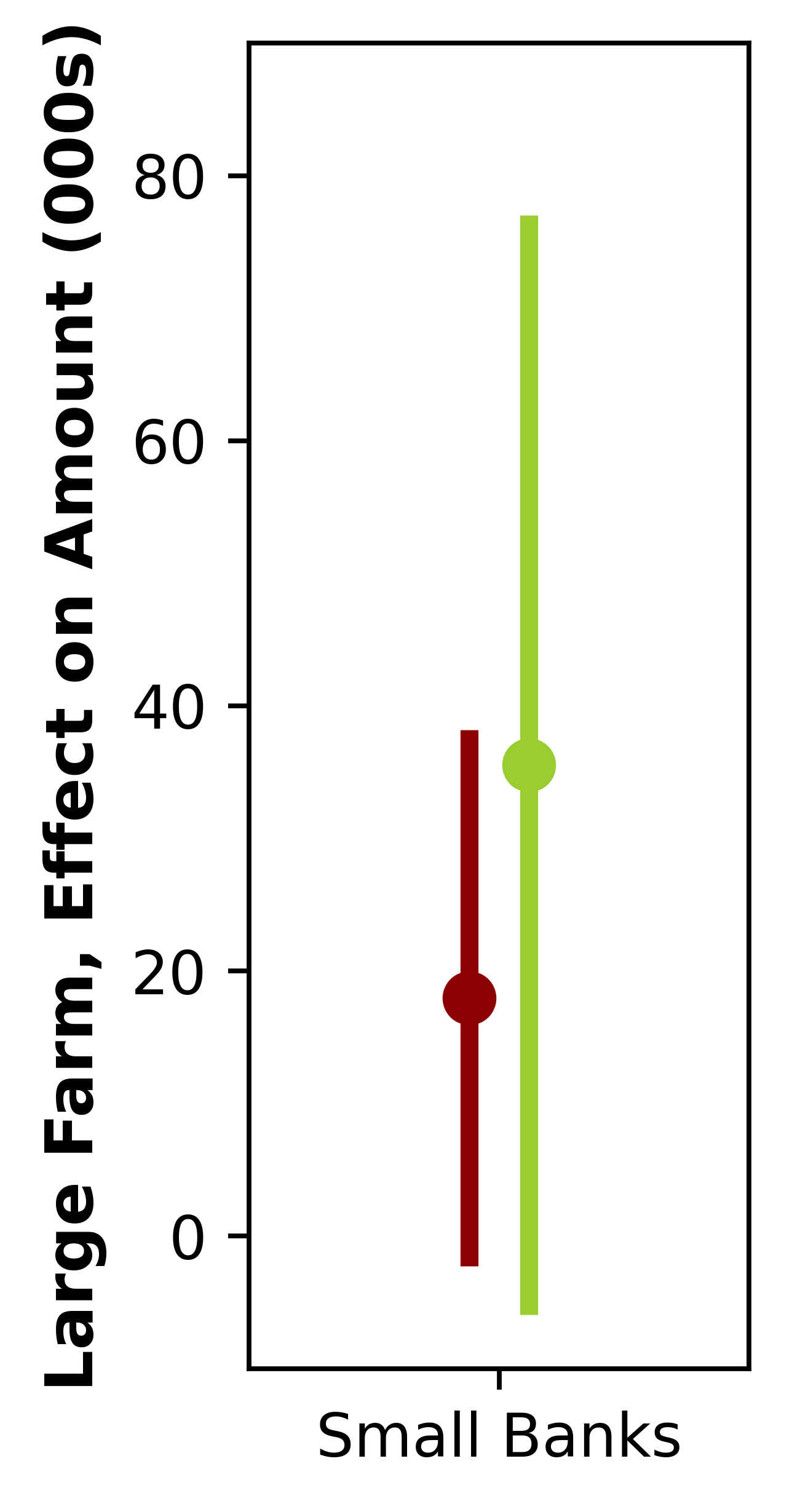}
	\end{subfigure}%
	\begin{subfigure}{3cm}
		\centering\includegraphics[width=3cm]{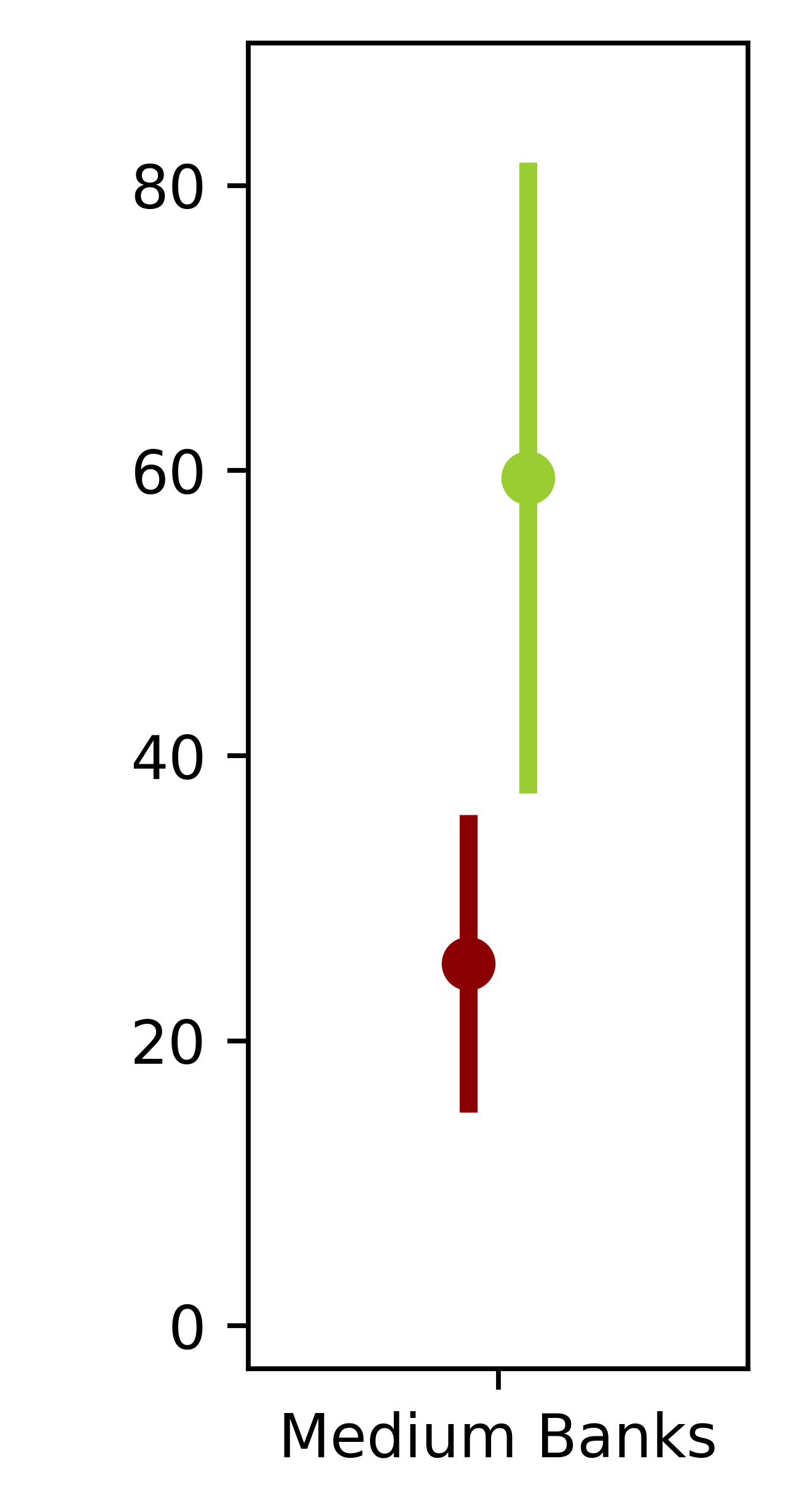}
	\end{subfigure}
	\begin{subfigure}{3cm}
		\centering\includegraphics[width=5.95cm]{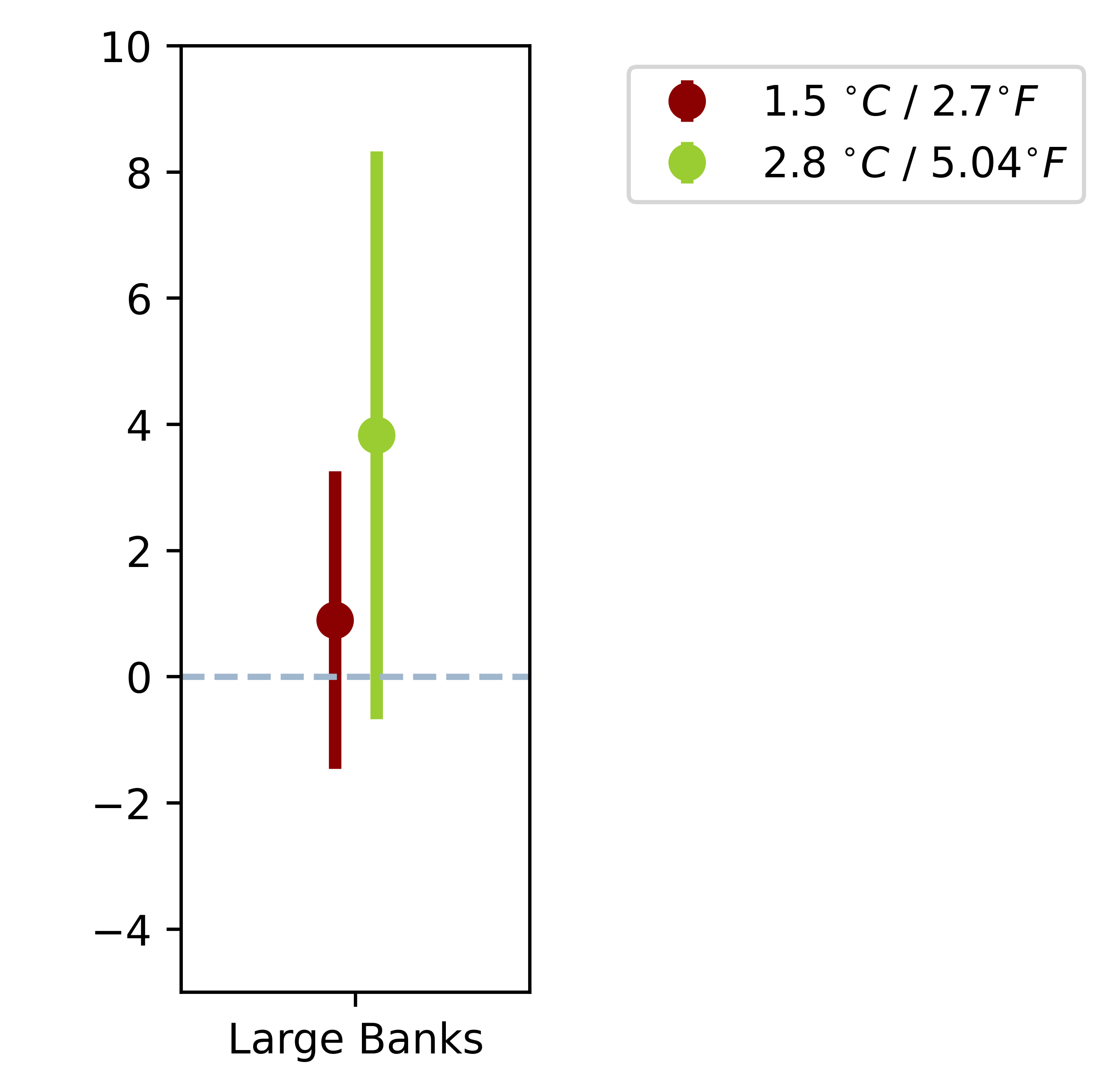}
	\end{subfigure}\par
	{\tiny	\textit{Note}: Average marginal effects with  95\% confidence intervals, with two scenarios: i). temperature anomaly of 1.5 ${}^{\circ}C$/ 2.7${}^{\circ}F$; ii). temperature anomaly of  2.8 ${}^{\circ}C$/ 5.04${}^{\circ}F$;  The first is consistent with a moderate climate scenario, or `net zero by 2075' (SSP1-2.6). The second is consistent with an adverse climate scenario, or`3X CO2 by 2100'  (SSP5-8.5). The regression results corresponding to this graph are in Table (\ref{tbankcountyinteract}) in Appendix.  \par}
\end{figure}

Besides the aforementioned results, I have estimated additional regressions separately for three groups of banks: very small, small-mid, and large. For the sake of brevity, Tables (\ref{t5realincomeg0bankvs}-\ref{t5realincomeg0bankl}) appear in the Appendix. The results are largely the same as in the estimations with bank dummies and interaction effects.\footnote{Tables (\ref{t5realincomeg0bankvs}-\ref{t5realincomeg0bankl}) show results from the estimation, with bank-level fixed effects included. As seen in Table (\ref{t5realincomeg0bankvs}), very small banks' lending do not change in a significant way in response to farms' climate risks. What is more interesting is that as such risks increase, the lending to smaller farms actually increase, both in terms of frequency and amount. In contrast, as Table (\ref{t5realincomeg0bankm}) shows, for small to midsize banks, their lending to small farms shrinks, while the amount of lending to large farms increases. As farms' exposure to climate risks increase, large banks in general are less willing to lend to farms, regardless of farm size, the magnitude of the effect seem slightly larger for small-medium farms.}

While more analyses are needed, one potential mechanism at work is a bank's ability to relocate their operations or businesses. The large banks in the sample are generally national entities with operations in many counties. Thus they have more leeway to move their businesses elsewhere, should the lending activities in a few locations are anticipated to be less profitable due to climate risks. In comparison, smaller banks, and especially those very small banks are much more localized. For some of them, their entire operations are constrained to one county, and do not have the flexibility as large banks for geographic risk-sharing or arbitrage. For small-mid banks, it is possible that they anticipate large farms to be more climate-resilient, thus continue lending to them. For very small banks, their lending decision could be more dictated by their established relations with small-medium farms, or they are liquidity-constrained such that they cannot meet the financing needs of large farms.  

In summary, results in this section provide richer insights into how lending decisions are made at the bank level with each county. Even though the analysis is conducted at a more granular dimension, many results here are highly consistent with those in the county level results. More specifically, there continues to be a wide gap of how large farms and smaller farms fare in terms of financial access---smaller farms generally have a harder time getting a loan approved in response to increased climate risks. Conditional on getting a loan, however, both smaller and large farms do receive financial support. The bank-county level analysis also reveal additional insights in that banks bear the different levels of burden of financing: large (likely national, multi-branch) banks tend to de-risk by withdrawing funding, as is the often the case for medium size banks. It is more likely for the localized, small banks to continue lending to small farms. While this is not a paper about the financial risks arising from climate change, the results here do point to concerning signs that the sharing of risks among banks is not distributed equally.


\newpage
\section{Conclusion and Discussions}\label{conclusion}
In this paper, I answer the question of how exposure to climate change risks affect farms' financial access. There is causal effect because extreme temperature and disasters reduce farms' output and revenue, and therefore increase their likelihood of defaulting on bank loans. By designing a two-period model, I show that farm size matters in modulating such impact: it is more likely for smaller farms to lose financial access. Using data from the Community Reinvestment Act (CRA), the empirical estimation then shows that vulnerability to climate change indeed has negative and significant impact on bank lending to farms, and such effects are nonlinear. Moreover, the financial impact on large farms is negligible or at times positive.  In contrast,  small to medium farms generally suffer from loss of credit access. 

In addition to the overall patterns by farm size, I also present additional granular results based on bank type and income area. Banks' own size also acts as a mechanism in determining the frequencies and amount of lending. Large banks tend to lend less frequently altogether in risky counties, and likely move their operations elsewhere. Medium banks are less willing to lend to small farms, and are in fact more wiling to lend to large farms as climate risks increase. Due to their highly localized operations, very small banks maintain and even increase lending to small farms. 
Moreover, the income areas where farms are located also matter, and the impact is more pronounced in middle income areas.

While it is difficult to directly test whether there is diversion of lending from smaller farms to large farms, the results suggest there is such evidence, particularly observed by the lending behavior of medium banks. In short, it is not necessarily the case that all banks reduce lending completely as climate risks increase. Rather, banks reassess and readjust their lending strategies to minimize potential loss and maximize profits. Consequently, with the advantages of size and higher productivity, large farms are less vulnerable to the adverse financial impact, while smaller farms are not. Though focusing on farm lending, the results of the paper suggest the financial impact of climate change may hit smaller stakeholders the hardest. This calls for further research not only in bank lending but in other financial issues such as insurance premium. Deeper understandings of such inequity are necessary to broaden communities' financial access to improve their climate resilience.

\clearpage

\begin{appendix}
\section{APPENDIX---Additional Results}

\subsection{Additional Figure}\label{addfigure}
\begin{figure}[h]
	\centering\includegraphics[width=10cm]{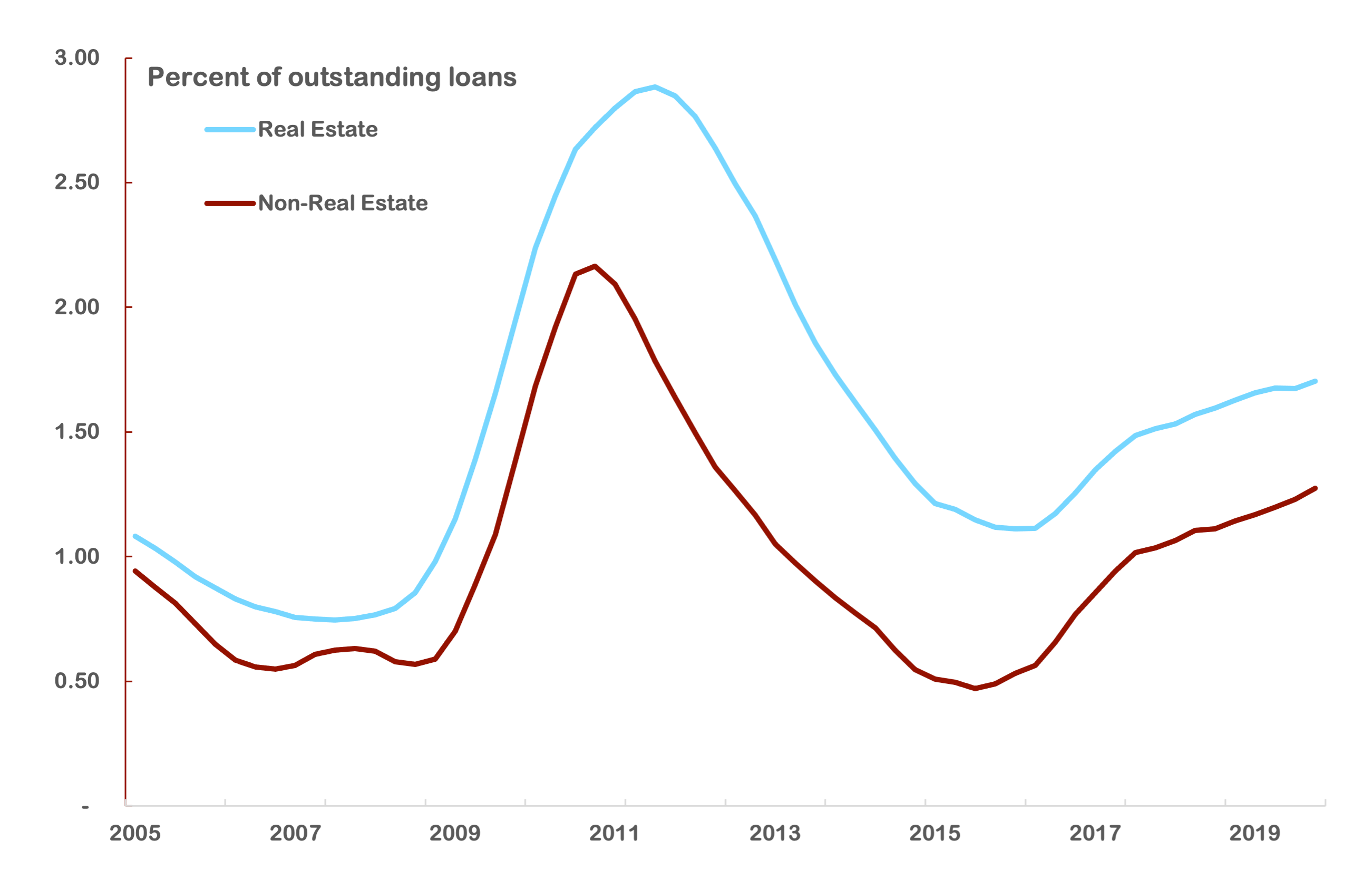}
	\caption{Non-Performing Farm Loans at U.S. Commercial Banks}\label{f1}
	{\tiny \textit{Note}: 4-quarter moving average; accruing and non-accruing loans past due 90 or more days} \\
	{\tiny Sources: Bank call reports and \cite{Kansasfed1}}
\end{figure}

\newpage
\subsection{Summary Statistics}\label{tsum}

\begin{table}[htbp]
	\centering
	\caption{Summary Statistics of Number of CRA Loans, County Level}\label{t1b} 
	\begin{adjustbox}{width=\textwidth}
		\begin{tabular}{rlrrrrl}
			\toprule
			\multicolumn{1}{l}{\textbf{Variable}} &       & \multicolumn{1}{l}{\textbf{Mean}} & \multicolumn{1}{l}{\textbf{Std. Dev.}} & \multicolumn{1}{l}{\textbf{Min}} & \multicolumn{1}{l}{\textbf{Max}} & \textbf{Observations} \\
			\midrule
			&       &       &       &       &       &  \\
			\multicolumn{1}{l}{\textit{\textbf{Number of loans smaller than \$100k}}} & overall & 54.32 & 77.09 & 0.00  & 1518.00 & N =   72834 \\
			& between &       & 62.23 & 0.25  & 940.21 & n =    3106 \\
			& within &       & 45.18 & -438.10 & 1058.90 & T-bar = 23.45 \\
			\midrule
			&       &       &       &       &       &  \\
			\multicolumn{1}{l}{\textit{\textbf{Number of loans \$100k to \$250k}}} & overall & 8.24  & 14.43 & 0.00  & 349.00 & N =   72834 \\
			& between &       & 12.55 & 0.00  & 204.96 & n =    3106 \\
			& within &       & 6.98  & -103.72 & 152.28 & T-bar = 23.45 \\
			\midrule
			&       &       &       &       &       &  \\
			\multicolumn{1}{l}{\textit{\textbf{Number of loans \$250k to \$500k}}} & overall & 3.97  & 8.12  & 0.00  & 199.00 & N =   72834 \\
			& between &       & 7.03  & 0.00  & 124.13 & n =    3106 \\
			& within &       & 3.96  & -85.16 & 78.84 & T-bar = 23.45 \\
			\bottomrule
		\end{tabular}%
	\end{adjustbox}
\end{table}%

\begin{table}[htbp]
	\centering
	\caption{Summary Statistics of Amount of CRA Loans, County Level, in thousand 2015 \$}\label{t2b} 
	\begin{adjustbox}{width=\textwidth}
		\begin{tabular}{rlrrrrl}
			\toprule
			\multicolumn{1}{l}{\textbf{Variable}} &       & \multicolumn{1}{l}{\textbf{Mean}} & \multicolumn{1}{l}{\textbf{Std. Dev.}} & \multicolumn{1}{l}{\textbf{Min}} & \multicolumn{1}{l}{\textbf{Max}} & \textbf{Observations} \\
			\midrule
			&       &       &       &       &       &  \\
			\multicolumn{1}{l}{\textit{\textbf{Amount of loans smaller than \$100k}}} & overall & 1230.26 & 1844.59 & 0.00  & 47299.75 & N =   72834 \\
			& between &       & 1604.50 & 0.43  & 28959.89 & n =    3106 \\
			& within &       & 894.52 & -13151.50 & 19570.12 & T-bar = 23.45 \\
			\midrule
			&       &       &       &       &       &  \\
			\multicolumn{1}{l}{\textit{\textbf{Amount of loans \$100k to \$250k}}} & overall & 1218.87 & 2178.50 & 0.00  & 56579.62 & N =   72834 \\
			& between &       & 1882.13 & 0.00  & 31524.24 & n =    3106 \\
			& within &       & 1074.01 & -20163.81 & 26274.25 & T-bar = 23.45 \\
			\midrule
			&       &       &       &       &       &  \\
			\multicolumn{1}{l}{\textit{\textbf{Amount of loans \$250k to \$500k}}} & overall & 1307.94 & 2783.98 & 0.00  & 72390.09 & N =   72834 \\
			& between &       & 2348.64 & 0.00  & 42286.68 & n =    3106 \\
			& within &       & 1464.12 & -31752.53 & 31411.35 & T-bar = 23.45 \\
			\bottomrule
		\end{tabular}%
	\end{adjustbox}
\end{table}%

\newpage
\subsection{Estimation of Effects of Climate Disasters}\label{reg-climate-disasters}
Climate disasters are the realizations of climate change risks. Thus, examining the effect of climate-related disasters on bank lending establishes helpful baseline understandings. 

The econometric specification for the county level analysis is 
\begin{equation}\label{e1} 
	y_{it}=\beta_k \textbf{disaster}_{i t}+\gamma_{i} trend + \gamma_{i2} trend^2+ u_{i}+ \eta_{t} +\lambda_{rt} + e_{i t}
\end{equation}
where $y_{it}$ refers to CRA lending variables, with subscript $i$ as county index and subscript $t$ as year. Additionally, each of the county $i$ corresponds to one climate region indexed by $r$. $\textbf{disaster}_{i t}$ refers to a vector of climate-related disasters reported by FEMA. 

Since the evolution of bank lending in relation to climate change may be nonlinear, the linear and quadratic county-specific year trends, denoted as $trend$ and $trend^2$, are both included, to allow for more flexibility in estimation. 
$u_{i}$ and $\eta_{t}$ are county and year fixed effects.  $u_{i}$ controls for the time-invariant heterogeneities of counties. The year fixed effect controls for global shocks such as commodity cycles in specific years. Moreover, it is likely that the effects of such shocks differ by climate region, thus the region by year (interaction) fixed effect $\lambda_{rt}$ is included to explicitly control for this. Finally, $e_{i t}$ is the error term. 

Moreover, there are four main CRA variables used in the analysis:   \begin{inparaenum}[1)]
	\item Number of loans to large farms
	\item Amount of loans to large farms
	\item Number of loans to small-medium farms
	\item Amount of loans to small-medium farms.
\end{inparaenum}

To measure climate disasters, I count the number of FEMA-declared disasters that are related to climate in each county in each year. It is important to note that such events are all extreme/anomaly events. FEMA, a federal agency, only declares an event as a disaster when local and state governments are unable to cope with it (\cite{FEMA}). The main types of FEMA-declared disasters include:  \begin{inparaenum}[1)]
	\item coastal storm,
	\item flood,
	\item freezing,
	\item hurricane,
	\item landslide,
	\item ice storm, 
	\item storm,
	\item snow,
	\item and tornado.
\end{inparaenum}
The disaster events are orthogonal to each other (i.e., no double counting) due to how FEMA categorizes such events. For example, while snow and ice storm may seem related, they are generally distinct events. Additionally, for disaster events such as freezing, ice storm, and snow that occur in winter, it is likely that they occur at the end of a year and thus their effects are lagged compared with other disasters. Thus, in the regression these winter disasters are lagged by 1 year. 

Table (\ref{t4}) reports the results from estimation of Equation (\ref{e1}). With the exception of freezing, almost all the disasters have some significant effect on at least one of the CRA variables. In particular, the coefficients for hurricane and storm are negative and significant across all the lending variables. Similarly, the relationship between CRA lending and flood, landslide, ice storm, and snow is generally negative. In other words, for the aforementioned events, the more they occur, the less the level of CRA lending there is. Section \ref{d1} reports the same estimations but grouped by the loan size thresholds. Additionally, the results are robust to dropping observations from years 2008-2009 (the Great Recession), reported in Section \ref{d2}.

The results of Table (\ref{t4}) suggest that there is a negative link between climate-related disasters and the level of bank lending. Moreover, this relationship holds for loans to both large and small-medium farms. However, one key limitation from the estimation of Equation (\ref{e1}) is that it is difficult to plausibly identify that such climate-related disasters are all occurring due to climate change. Put another way, weather events are not the same as climate change. Therefore in the remainder of the paper, I use temperature and precipitation data as more direct measures of climate change.

\begin{table}[h]
	\caption {CRA Farm Loans and Climate Disasters, County Total} \label{t4} 
	\begin{adjustbox}{width=\columnwidth,center}
		\begin{tabular}{lcccc} \hline
 & (1) & (2) & (3) & (4) \\
VARIABLES & Num. to large farms & Amount to large farms & Num. to small-mid farms & Amount to small-mid farms \\ \hline
 &  &  &  &  \\
Coastal Storm & 1.65** & 73.64 & 0.01 & -14.92 \\
 & (0.73) & (61.97) & (1.89) & (148.94) \\
Flood & -0.82*** & -55.33** & -1.71** & -152.44*** \\
 & (0.30) & (26.42) & (0.80) & (58.33) \\
Freezing (lagged) & 10.66 & 720.68** & 7.43 & 513.00 \\
 & (8.47) & (294.34) & (8.48) & (384.11) \\
Hurricane & -0.91*** & -27.51*** & -0.59* & -37.27* \\
 & (0.15) & (9.48) & (0.31) & (20.20) \\
Landslide & -8.12*** & -289.85 & 8.17** & 399.65 \\
 & (1.76) & (242.23) & (3.18) & (284.70) \\
Ice storm (lagged) & -0.78*** & -42.09** & 1.50 & 155.17*** \\
 & (0.22) & (18.19) & (0.95) & (59.47) \\
Storm & -0.32** & -20.91* & -1.48*** & -69.71** \\
 & (0.12) & (11.54) & (0.44) & (28.65) \\
Snow (lagged) & -0.01 & -77.10*** & -5.61*** & -268.72*** \\
 & (0.24) & (26.62) & (0.95) & (71.89) \\
Tornado & -1.12* & 33.28 & 3.39 & 382.39* \\
 & (0.67) & (73.53) & (2.77) & (206.06) \\
 &  &  &  &  \\
Observations & 69,728 & 69,728 & 69,728 & 69,728 \\
R-squared & 0.203 & 0.051 & 0.140 & 0.111 \\
County FE & Yes & Yes & Yes & Yes \\
Year FE & Yes & Yes & Yes & Yes \\
Region x Year FE & Yes & Yes & Yes & Yes \\
 Robust SE & Cluster & Cluster & Cluster & Cluster \\ \hline
\multicolumn{5}{c}{ Robust standard errors in parentheses} \\
\multicolumn{5}{c}{ *** p$<$0.01, ** p$<$0.05, * p$<$0.1} \\
\end{tabular}

	\end{adjustbox}
	\begin{tablenotes}
		\scriptsize
		\vspace{5pt}
		\item \textit{Note}: Regression specification for this table is Equation (\ref{e1})
		$
		y_{it}=\beta_k \textbf{disaster}_{i t}+\gamma_{i} trend + \gamma_{i2} trend^2+ u_{i}+ \eta_{t} +\lambda_{rt} + e_{i t}
		$. The lending amount variables in real terms (2015 USD)
	\end{tablenotes}
\end{table}

\newpage
\subsection{Robustness Check: Including Precipitation}\label{robustnesscheck1}

Besides temperature anomaly, precipitation anomaly is another measurement of climate change. In \cite{Burke2016}, both mean temperature and precipitation are included in the estimation. At the same time, in the climate process, temperature and precipitation closely interact with each other. Thus the inclusion of precipitation may be unnecessary or could lead to problem of multicollinearity. However, for completeness, I extend the baseline regression to examine how considering precipitation shapes the overall results. 
\begin{flushleft}
	
	\begin{equation}\label{e3} 
		y_{it}=\beta_{1} T_{i t} + \beta_{2} T^2_{i t}+\theta_{1} P_{i t} + \theta_{2} P^2_{i t} + u_{i}+ \eta_{t} +\lambda_{rt} +\gamma_{i} trend + \gamma_{i2} trend^2+e_{i t}
	\end{equation}
	
\end{flushleft}
	
	where $ P_{i t}$ refers to precipitation anomaly observed in each county in each year,  $P^2_{i t}$ is the quadratic form. Everything else is the same as in Equation (\ref{e2}).
	
	Table (\ref{t6}) shows the results: including precipitation does not significantly change the coefficient estimates of temperature, though it does make the coefficient of linear high temperature anomaly in Columns (1) and (2) insignificant. Moreover, precipitation anomaly seems to have positive effect on the amount of loans for both large and small-medium farms. In comparison, both the linear and nonlinear effects on the number of loans to small-medium farms are negative. In short, the patterns revealed by precipitation anomaly is less clear compared with temperature. One possibility is that the NOAA data on precipitation does not distinguish between high or low precipitation. Therefore, for remainder of the paper, I focus on using high temperature anomaly to measure climate change.

		\begin{table}[ht]
		\caption {CRA Farm Loans and Climate Vulnerability, County Total} \label{t6} 
		\begin{adjustbox}{width=\columnwidth,center}
			\begin{tabular}{lcccc} \hline
 & (1) & (2) & (3) & (4) \\
VARIABLES & Num. to large farms & Amount to large farms & Num. to small-mid farms & Amount to small-mid farms \\ \hline
 &  &  &  &  \\
High temperature anomaly & -0.09 & -5.15 & 1.34*** & 60.80*** \\
 & (0.09) & (6.32) & (0.20) & (13.00) \\
High temperature anomaly (square) & 0.02 & 1.88 & -0.85*** & -28.15*** \\
 & (0.03) & (2.05) & (0.08) & (4.30) \\
Precipitation anomaly & 0.24** & 21.92** & -0.76** & 24.23 \\
 & (0.12) & (8.72) & (0.34) & (19.16) \\
Precipitation anomaly (square) & -0.11 & 14.37** & -0.63*** & 38.85*** \\
 & (0.08) & (5.81) & (0.22) & (13.39) \\
 &  &  &  &  \\
Observations & 72,834 & 72,834 & 72,834 & 72,834 \\
R-squared & 0.193 & 0.093 & 0.133 & 0.056 \\
County FE & Yes & Yes & Yes & Yes \\
Year FE & Yes & Yes & Yes & Yes \\
Region x Year FE & Yes & Yes & Yes & Yes \\
 Robust SE & Cluster & Cluster & Cluster & Cluster \\ \hline
\multicolumn{5}{c}{ Robust standard errors in parentheses} \\
\multicolumn{5}{c}{ *** p$<$0.01, ** p$<$0.05, * p$<$0.1} \\
\end{tabular}

		\end{adjustbox}
	\end{table}

\newpage
	\subsection{Robustness Check: Seasonal Effects}\label{robustnesscheck2}
	While the estimation is conducted at annual frequency, it is possible to isolate the seasonal effects of high temperature anomaly. For example, \cite{Diffenbaugh2021} splits the estimation into growing season (April through October) and non-growing season. Thus I follow their approaching by calculating the temperature anomaly for the growing season and the rest of the year, using original, monthly observations of the NOAA NClimDiv dataset. Other than the new measurements of temperature, the specification follows that of Equation (\ref{e2}). Tables (\ref{t5grow}) and (\ref{t5nogrow}) show the results. For small-medium farms, a temperature anomaly shock seems costly during the growing season, compared to what larger farms experience. On the other hand, small-medium farms remain vulnerable during the non-growing season. Yet for large farms, during non-growing season, the impact of a climate shock is mostly statistically insignificant.

		\begin{table}[ht]
		\caption {CRA Farm Loans and Climate Vulnerability, County Total} \label{t5grow} 
		\begin{adjustbox}{width=\columnwidth,center}
			\begin{tabular}{lcccc} \hline
 & (1) & (2) & (3) & (4) \\
VARIABLES & Num. to large farms & Amount to large farms & Num. to small-mid farms & Amount to small-mid farms \\ \hline
 &  &  &  &  \\
High temp. anomaly (growing season) & -0.33*** & -16.30*** & 1.12*** & 11.16 \\
 & (0.07) & (5.25) & (0.17) & (10.60) \\
High temp. anomaly (square, growing season) & -0.01 & 4.00*** & -0.66*** & -18.87*** \\
 & (0.02) & (1.53) & (0.06) & (3.49) \\
 &  &  &  &  \\
Observations & 72,834 & 72,834 & 72,834 & 72,834 \\
R-squared & 0.195 & 0.094 & 0.133 & 0.056 \\
County FE & Yes & Yes & Yes & Yes \\
Year FE & Yes & Yes & Yes & Yes \\
Region x Year FE & Yes & Yes & Yes & Yes \\
 Robust SE & Cluster & Cluster & Cluster & Cluster \\ \hline
\multicolumn{5}{c}{ Robust standard errors in parentheses} \\
\multicolumn{5}{c}{ *** p$<$0.01, ** p$<$0.05, * p$<$0.1} \\
\end{tabular}

		\end{adjustbox}
	\end{table}
	
	\begin{table}[ht]
		\caption {CRA Farm Loans and Climate Vulnerability, County Total} \label{t5nogrow} 
		\begin{adjustbox}{width=\columnwidth,center}
			\begin{tabular}{lcccc} \hline
 & (1) & (2) & (3) & (4) \\
VARIABLES & Num. to large farms & Amount to large farms & Num. to small-mid farms & Amount to small-mid farms \\ \hline
 &  &  &  &  \\
High temp. anomaly (non-growing season) & 0.11*** & 0.03 & 0.67*** & 44.94*** \\
 & (0.04) & (3.30) & (0.12) & (6.42) \\
High temp. anomaly (square, non-growing season) & 0.01 & -0.25 & -0.22*** & -8.92*** \\
 & (0.01) & (0.84) & (0.03) & (1.86) \\
 &  &  &  &  \\
Observations & 72,834 & 72,834 & 72,834 & 72,834 \\
R-squared & 0.195 & 0.094 & 0.133 & 0.057 \\
County FE & Yes & Yes & Yes & Yes \\
Year FE & Yes & Yes & Yes & Yes \\
Region x Year FE & Yes & Yes & Yes & Yes \\
 Robust SE & Cluster & Cluster & Cluster & Cluster \\ \hline
\multicolumn{5}{c}{ Robust standard errors in parentheses} \\
\multicolumn{5}{c}{ *** p$<$0.01, ** p$<$0.05, * p$<$0.1} \\
\end{tabular}
			
		\end{adjustbox}
	\end{table}

\newpage
\subsection{Estimations by Loan Size Brackets}\label{d1}

To further analyze how bank branching plays a role, I conduct additional regressions using a different dimension of CRA lending---loan size bracket: \begin{inparaenum}[1)]
	\item loans less than \$100 thousand
	\item  loans between \$100 thousand and \$250 thousand
	\item  loans between \$250 thousand and \$500 thousand. 
\end{inparaenum}

\paragraph{\textit{Estimation by Loan Sizes}}
The CRA dataset does not provide information to compare loan size brackets and lending to large versus small farms. In other words, it is difficult to know whether loans in the smaller bracket are directed primarily towards to small-medium farms. However, it is plausible that in absolute terms, large farms have bigger financing needs and thus may be more likely to have larger loans. Moreover, independently of which type of farms the lending goes to, the sheer size of loans correlates with banks' exposure, and is indicative of how banks may want to manage their exposure in light of climate risks. 

Table (\ref{t9}) presents the estimation in terms of amount of loans. As shown by Column (2), the coefficients of temperature for loans less than \$100 thousand mirror those for overall loans to small-medium farms---suggesting a concave curve. In contrast, the coefficients for loans of size \$250 to \$500 thousand have the opposite signs. Moreover, as seen in Column (2), the interaction effect between temperature anomaly and bank branches is highly significant for small loans, suggesting that banks adjust the size of their exposure in relation to climate change. This adjustment may involve substitution between loan sizes, as suggested by Column (1), there is no significant impact of temperature on the total amount of loans. 

Similarly, Table (\ref{t10}) shows the results for the number of loans. The effects of temperature on loans smaller than \$100 thousand are similar to Column (2) of Table (\ref{t9}). However, temperature has almost no impact on the number of loans of size \$250 to \$500 thousand. Additionally, while statistically significant, the coefficients for loans of \$100 to \$250 thousand have the same signs as Column (2), as are total number of loans in Column (1). In contrast to Table (\ref{t9}), in terms of number of loans, Table (\ref{t10}) shows that there does not seem to be substitution between different sizes of lending. Rather, climate change, measured in terms of increasing high temperature anomaly, has overall negative impact on the total number of loans.

\begin{table}[!]
	\caption {CRA Farm Loan Amount and Climate Vulnerability, Loan Size Bracket} \label{t9} 
	\begin{adjustbox}{width=\columnwidth,center}
		\begin{tabular}{lcccc} \hline
 & (1) & (2) & (3) & (4) \\
VARIABLES & Total Amt. & Amt. (less 100k) & Amt. (100k to 250k) & Amt. (250k to 500k) \\ \hline
 &  &  &  &  \\
High temperature anomaly & 47.18*** & 37.51*** & 20.20*** & -10.53 \\
 & (14.00) & (4.97) & (5.50) & (6.61) \\
High temperature anomaly (square) & -25.90*** & -21.22*** & -8.37*** & 3.69 \\
 & (5.21) & (1.82) & (2.02) & (2.51) \\
Total number of bank branches & 2.61 & 6.23*** & -1.12 & -2.49 \\
 & (3.10) & (1.24) & (0.89) & (2.05) \\
Temp. anomaly x Bank Branches & -1.42* & -0.08 & -0.39 & -0.96** \\
 & (0.81) & (0.38) & (0.25) & (0.47) \\
Temp. anomaly (square) x Bank Branches & 0.49 & -0.73*** & 0.40** & 0.81 \\
 & (0.64) & (0.24) & (0.20) & (0.60) \\
 &  &  &  &  \\
Observations & 72,447 & 72,447 & 72,447 & 72,447 \\
R-squared & 0.049 & 0.082 & 0.046 & 0.076 \\
County FE & Yes & Yes & Yes & Yes \\
Year FE & Yes & Yes & Yes & Yes \\
Region x Year FE & Yes & Yes & Yes & Yes \\
 Robust SE & Cluster & Cluster & Cluster & Cluster \\ \hline
\multicolumn{5}{c}{ Robust standard errors in parentheses} \\
\multicolumn{5}{c}{ *** p$<$0.01, ** p$<$0.05, * p$<$0.1} \\
\end{tabular}

	\end{adjustbox}
	\begin{tablenotes}
		\vspace{5pt}
		\scriptsize
		\item \textit{Note}: Regression specification for this table is Equation (\ref{e4}), 
		$
		y_{it}=  \beta_{1} T_{i t} + \beta_{2} T^2_{i t}+\theta_{1} branch_{i t} +\theta_{2} T_{i t} \cdot branch_{i t} +\theta_{3}  T^2_{i t} \cdot branch_{i t}  + u_{i}+ \eta_{t} +\lambda_{rt} +\gamma_{i} trend + \gamma_{i2} trend^2+e_{i t}
		$, where $y_{it}$ is the loan variable based on loan size bracket.
	\end{tablenotes}
\end{table}

\begin{table}[!]
	\caption {CRA Farm Loan Frequency and Climate Vulnerability, Loan Size Bracket} \label{t10} 
	\begin{adjustbox}{width=\columnwidth,center}
		\begin{tabular}{lcccc} \hline
 & (1) & (2) & (3) & (4) \\
VARIABLES & Total Num. of Loans & Number of loan (less 100k) & Number of loan (100k to 250k) & Number of loan (250k to 500k) \\ \hline
 &  &  &  &  \\
High temperature anomaly & 1.40*** & 1.30*** & 0.13*** & -0.02 \\
 & (0.21) & (0.18) & (0.03) & (0.02) \\
High temperature anomaly (square) & -0.83*** & -0.78*** & -0.06*** & 0.01 \\
 & (0.08) & (0.07) & (0.01) & (0.01) \\
Total number of bank branches & 0.43*** & 0.44*** & -0.00 & -0.01 \\
 & (0.04) & (0.04) & (0.01) & (0.01) \\
Temp. anomaly x Bank Branches & -0.03 & -0.03 & -0.00 & -0.00** \\
 & (0.04) & (0.04) & (0.00) & (0.00) \\
Temp. anomaly (square) x Bank Branches & -0.04*** & -0.04*** & 0.00* & 0.00 \\
 & (0.01) & (0.01) & (0.00) & (0.00) \\
 &  &  &  &  \\
Observations & 72,447 & 72,447 & 72,447 & 72,447 \\
R-squared & 0.080 & 0.087 & 0.046 & 0.073 \\
County FE & Yes & Yes & Yes & Yes \\
Year FE & Yes & Yes & Yes & Yes \\
Region x Year FE & Yes & Yes & Yes & Yes \\
 Robust SE & Cluster & Cluster & Cluster & Cluster \\ \hline
\multicolumn{5}{c}{ Robust standard errors in parentheses} \\
\multicolumn{5}{c}{ *** p$<$0.01, ** p$<$0.05, * p$<$0.1} \\
\end{tabular}

	\end{adjustbox}
	\begin{tablenotes}
		\vspace{5pt}
		\scriptsize
		\item \textit{Note}: Regression specification for this table is Equation (\ref{e4}), 
		$
		y_{it}=  \beta_{1} T_{i t} + \beta_{2} T^2_{i t}+\theta_{1} branch_{i t} +\theta_{2} T_{i t} \cdot branch_{i t} +\theta_{3}  T^2_{i t} \cdot branch_{i t}  + u_{i}+ \eta_{t} +\lambda_{rt} +\gamma_{i} trend + \gamma_{i2} trend^2+e_{i t}
		$, where $y_{it}$ is the loan variable based on loan size bracket.
	\end{tablenotes}
\end{table}

\clearpage
\subsection{Alternative Measures of Temperature Anomaly}\label{d3}

\begin{table}[h]
	\caption {CRA Farm Loans and Climate Vulnerability, County Total} \label{t22} 
	\begin{adjustbox}{width=\columnwidth,center}
		\begin{tabular}{lcccc} \hline
 & (1) & (2) & (3) & (4) \\
VARIABLES & Num. to large farms & Amount to large farms & Num. to small-mid farms & Amount to small-mid farms \\ \hline
 &  &  &  &  \\
High temperature anomaly (50 years) & -0.14 & -10.59 & 2.08*** & 78.13*** \\
 & (0.09) & (6.91) & (0.22) & (13.58) \\
High temperature anomaly (square, 50 years) & 0.00 & 1.51 & -0.80*** & -27.84*** \\
 & (0.03) & (2.09) & (0.08) & (4.22) \\
 &  &  &  &  \\
Observations & 72,834 & 72,834 & 72,834 & 72,834 \\
R-squared & 0.193 & 0.093 & 0.132 & 0.056 \\
County FE & Yes & Yes & Yes & Yes \\
Year FE & Yes & Yes & Yes & Yes \\
Region x Year FE & Yes & Yes & Yes & Yes \\
 Robust SE & Cluster & Cluster & Cluster & Cluster \\ \hline
\multicolumn{5}{c}{ Robust standard errors in parentheses} \\
\multicolumn{5}{c}{ *** p$<$0.01, ** p$<$0.05, * p$<$0.1} \\
\end{tabular}

	\end{adjustbox}
\end{table}

\begin{table}[h]
	\caption {CRA Farm Loans and Climate Vulnerability, County Total} \label{t23} 
	\begin{adjustbox}{width=\columnwidth,center}
		\begin{tabular}{lcccc} \hline
 & (1) & (2) & (3) & (4) \\
VARIABLES & Num. to large farms & Amount to large farms & Num. to small-mid farms & Amount to small-mid farms \\ \hline
 &  &  &  &  \\
High temperature anomaly (70 years) & -0.16* & -11.29 & 2.20*** & 83.78*** \\
 & (0.09) & (7.14) & (0.23) & (14.01) \\
High temperature aneromaly (square, 70 years) & 0.01 & 1.81 & -0.76*** & -27.86*** \\
 & (0.03) & (2.11) & (0.07) & (4.18) \\
 &  &  &  &  \\
Observations & 72,834 & 72,834 & 72,834 & 72,834 \\
R-squared & 0.193 & 0.093 & 0.132 & 0.056 \\
County FE & Yes & Yes & Yes & Yes \\
Year FE & Yes & Yes & Yes & Yes \\
Region x Year FE & Yes & Yes & Yes & Yes \\
 Robust SE & Cluster & Cluster & Cluster & Cluster \\ \hline
\multicolumn{5}{c}{ Robust standard errors in parentheses} \\
\multicolumn{5}{c}{ *** p$<$0.01, ** p$<$0.05, * p$<$0.1} \\
\end{tabular}

	\end{adjustbox}
\end{table}

\begin{table}[h]
	\caption {CRA Farm Loans and Climate Vulnerability, County Total} \label{t24} 
	\begin{adjustbox}{width=\columnwidth,center}
		\begin{tabular}{lcccc} \hline
 & (1) & (2) & (3) & (4) \\
VARIABLES & Num. to large farms & Amount to large farms & Num. to small-mid farms & Amount to small-mid farms \\ \hline
 &  &  &  &  \\
High temperature anomaly (100 years) & -0.15 & -10.34 & 2.01*** & 79.20*** \\
 & (0.09) & (7.03) & (0.22) & (13.52) \\
High temperature anomaly (square, 100 years) & 0.01 & 1.17 & -0.66*** & -26.22*** \\
 & (0.03) & (2.08) & (0.07) & (3.94) \\
 &  &  &  &  \\
Observations & 72,834 & 72,834 & 72,834 & 72,834 \\
R-squared & 0.193 & 0.093 & 0.132 & 0.056 \\
County FE & Yes & Yes & Yes & Yes \\
Year FE & Yes & Yes & Yes & Yes \\
Region x Year FE & Yes & Yes & Yes & Yes \\
 Robust SE & Cluster & Cluster & Cluster & Cluster \\ \hline
\multicolumn{5}{c}{ Robust standard errors in parentheses} \\
\multicolumn{5}{c}{ *** p$<$0.01, ** p$<$0.05, * p$<$0.1} \\
\end{tabular}

	\end{adjustbox}
\end{table}

\begin{table}[H]
	\caption {CRA Farm Loans and Climate Vulnerability, County Total} \label{t25} 
	\begin{adjustbox}{width=\columnwidth,center}
		\begin{tabular}{lcccc} \hline
 & (1) & (2) & (3) & (4) \\
VARIABLES & Num. to large farms & Amount to large farms & Num. to small-mid farms & Amount to small-mid farms \\ \hline
 &  &  &  &  \\
High temperature anomaly (since 1895) & -0.17* & -10.85 & 2.13*** & 85.48*** \\
 & (0.10) & (7.38) & (0.23) & (14.11) \\
High temperature anomaly (square, since 1895) & 0.02 & 1.26 & -0.61*** & -25.01*** \\
 & (0.03) & (2.03) & (0.07) & (3.80) \\
 &  &  &  &  \\
Observations & 72,834 & 72,834 & 72,834 & 72,834 \\
R-squared & 0.193 & 0.093 & 0.132 & 0.056 \\
County FE & Yes & Yes & Yes & Yes \\
Year FE & Yes & Yes & Yes & Yes \\
Region x Year FE & Yes & Yes & Yes & Yes \\
 Robust SE & Cluster & Cluster & Cluster & Cluster \\ \hline
\multicolumn{5}{c}{ Robust standard errors in parentheses} \\
\multicolumn{5}{c}{ *** p$<$0.01, ** p$<$0.05, * p$<$0.1} \\
\end{tabular}

	\end{adjustbox}
\end{table}
 
\clearpage
\subsection{Dropping Years 2008 and 2009}\label{d2}

%
%
%

\begin{table}[ht]
	\caption {Number of CRA Farm Loans and Climate Vulnerability, County Total} \label{t17} 
	\begin{adjustbox}{width=\columnwidth,center}
		\begin{tabular}{lcccc} \hline
 & (1) & (2) & (3) & (4) \\
VARIABLES & Num. to large farms & Amount to large farms & Num. to small-mid farms & Amount to small-mid farms \\ \hline
 &  &  &  &  \\
High temperature anomaly & -0.18** & -3.10 & 1.20*** & 57.99*** \\
 & (0.08) & (6.74) & (0.20) & (12.76) \\
High temperature anomaly (square) & 0.00 & 0.79 & -0.84*** & -28.54*** \\
 & (0.03) & (2.06) & (0.08) & (4.26) \\
 &  &  &  &  \\
Observations & 66,748 & 66,748 & 66,748 & 66,748 \\
R-squared & 0.204 & 0.099 & 0.135 & 0.059 \\
County FE & Yes & Yes & Yes & Yes \\
Year FE & Yes & Yes & Yes & Yes \\
Region x Year FE & Yes & Yes & Yes & Yes \\
 Robust SE & Cluster & Cluster & Cluster & Cluster \\ \hline
\multicolumn{5}{c}{ Robust standard errors in parentheses} \\
\multicolumn{5}{c}{ *** p$<$0.01, ** p$<$0.05, * p$<$0.1} \\
\end{tabular}

	\end{adjustbox}
\end{table}

\clearpage
\subsection{Lagged and Growth Effect}\label{d4}

\begin{table}[ht]
	\caption {CRA Farm Loans and Climate Vulnerability (Lagged), County Total} \label{t19} 
	\begin{adjustbox}{width=\columnwidth,center}
		\begin{tabular}{lcccc} \hline
 & (1) & (2) & (3) & (4) \\
VARIABLES & Num. to large farms & Amount to large farms & Num. to small-mid farms & Amount to small-mid farms \\ \hline
 &  &  &  &  \\
High temperature anomaly (lag) & 0.26*** & -0.19 & 1.65*** & 82.98*** \\
 & (0.07) & (5.44) & (0.18) & (11.03) \\
High temperature anomaly (square, lag) & -0.00 & -0.91 & -0.94*** & -43.24*** \\
 & (0.03) & (1.98) & (0.08) & (4.21) \\
 &  &  &  &  \\
Observations & 69,728 & 69,728 & 69,728 & 69,728 \\
R-squared & 0.202 & 0.092 & 0.142 & 0.059 \\
County FE & Yes & Yes & Yes & Yes \\
Year FE & Yes & Yes & Yes & Yes \\
Region x Year FE & Yes & Yes & Yes & Yes \\
 Robust SE & Cluster & Cluster & Cluster & Cluster \\ \hline
\multicolumn{5}{c}{ Robust standard errors in parentheses} \\
\multicolumn{5}{c}{ *** p$<$0.01, ** p$<$0.05, * p$<$0.1} \\
\end{tabular}

	\end{adjustbox}
\end{table}

\begin{table}[ht]
	\caption {CRA Farm Loans (growth) and Climate Vulnerability, County Total} \label{t20} 
	\begin{adjustbox}{width=\columnwidth,center}             
		\begin{tabular}{lcccc} \hline
 & (1) & (2) & (3) & (4) \\
VARIABLES & Num. of loans to large farms (\%) & Amount of loans to large farms (\%) & Num. of loans to small-medium farms (\%) & Amount of loans to small-medium farms (\%) \\ \hline
 &  &  &  &  \\
High temperature anomaly & 0.30*** & 7.12 & -0.54*** & -8.04 \\
 & (0.06) & (4.66) & (0.13) & (9.38) \\
High temperature anomaly (square) & 0.07** & -3.79** & 0.01 & -1.14 \\
 & (0.03) & (1.84) & (0.04) & (3.07) \\
 &  &  &  &  \\
Observations & 69,728 & 69,728 & 69,728 & 69,728 \\
R-squared & 0.053 & 0.031 & 0.049 & 0.045 \\
County FE & Yes & Yes & Yes & Yes \\
Year FE & Yes & Yes & Yes & Yes \\
Region x Year FE & Yes & Yes & Yes & Yes \\
 Robust SE & Cluster & Cluster & Cluster & Cluster \\ \hline
\multicolumn{5}{c}{ Robust standard errors in parentheses} \\
\multicolumn{5}{c}{ *** p$<$0.01, ** p$<$0.05, * p$<$0.1} \\
\end{tabular}

	\end{adjustbox}
\end{table}

\begin{table}[ht]
	\caption {CRA Farm Loans (growth) and Climate Vulnerability (lag), County Total} \label{t21} 
	\begin{adjustbox}{width=\columnwidth,center}
		\begin{tabular}{lcccc} \hline
 & (1) & (2) & (3) & (4) \\
VARIABLES & Num. of loans to large farms (\%) & Amount of loans to large farms (\%) & Num. of loans to small-medium farms (\%) & Amount of loans to small-medium farms (\%) \\ \hline
 &  &  &  &  \\
High temperature anomaly (lag) & 0.26*** & 6.52 & 0.13 & -1.35 \\
 & (0.07) & (5.42) & (0.14) & (10.02) \\
High temperature anomaly (square, lag) & -0.04 & -3.95** & -0.03 & -6.42* \\
 & (0.03) & (1.91) & (0.05) & (3.38) \\
 &  &  &  &  \\
Observations & 69,728 & 69,728 & 69,728 & 69,728 \\
R-squared & 0.053 & 0.031 & 0.049 & 0.045 \\
County FE & Yes & Yes & Yes & Yes \\
Year FE & Yes & Yes & Yes & Yes \\
Region x Year FE & Yes & Yes & Yes & Yes \\
 Robust SE & Cluster & Cluster & Cluster & Cluster \\ \hline
\multicolumn{5}{c}{ Robust standard errors in parentheses} \\
\multicolumn{5}{c}{ *** p$<$0.01, ** p$<$0.05, * p$<$0.1} \\
\end{tabular}

	\end{adjustbox}
\end{table}

%
%
%
%

\clearpage
\subsection{Results by Census Regions}\label{census-tables}
\begin{table}[!]
	\caption {CRA Farm Loans and Climate Vulnerability, Low Income Areas} \label{t5realincomeg1} 
	\begin{adjustbox}{width=\columnwidth,center}
		\begin{tabular}{lcccc} \hline
 & (1) & (2) & (3) & (4) \\
VARIABLES & Num. to large farms & Amount to large farms & Num. to small-mid farms & Amount to small-mid farms \\ \hline
 &  &  &  &  \\
Temp. anomaly & -0.05 & -1.98 & 0.17** & -0.73 \\
 & (0.04) & (3.54) & (0.07) & (6.53) \\
Temp. anomaly (square) & 0.01 & 0.28 & -0.09*** & -3.01 \\
 & (0.02) & (1.71) & (0.03) & (2.91) \\
 &  &  &  &  \\
Observations & 5,129 & 5,129 & 5,129 & 5,129 \\
R-squared & 0.045 & 0.033 & 0.067 & 0.052 \\
County FE & Yes & Yes & Yes & Yes \\
Year FE & Yes & Yes & Yes & Yes \\
Region x Year FE & Yes & Yes & Yes & Yes \\
 Robust SE & Cluster & Cluster & Cluster & Cluster \\ \hline
\multicolumn{5}{c}{ Robust standard errors in parentheses} \\
\multicolumn{5}{c}{ *** p$<$0.01, ** p$<$0.05, * p$<$0.1} \\
\end{tabular}

	\end{adjustbox}
	\begin{tablenotes}
		\vspace{5pt}
		\scriptsize
		\item \textit{Note}: Regression specification for this table is Equation (\ref{e2cen}), 
		$
		y_{ict}=\beta_{1} T_{i t} + \beta_{2} T^2_{i t}+ u_{i}+ \xi_{c} + \eta_{t} +\lambda_{rt} +\gamma_{i} trend + \gamma_{i2} trend^2+e_{i ct}
		$, but only for the sample of low income areas.
	\end{tablenotes}
\end{table}

\begin{table}[h]
	\caption {CRA Farm Loans and Climate Vulnerability, Moderate Income Areas} \label{t5realincomeg2} 
	\begin{adjustbox}{width=\columnwidth,center}
		\begin{tabular}{lcccc} \hline
 & (1) & (2) & (3) & (4) \\
VARIABLES & Num. to large farms & Amount to large farms & Num. to small-mid farms & Amount to small-mid farms \\ \hline
 &  &  &  &  \\
Temp. anomaly & -0.07 & -0.38 & 0.47*** & 12.34* \\
 & (0.04) & (3.87) & (0.11) & (7.23) \\
Temp. anomaly (square) & 0.01 & 2.05 & -0.23*** & -4.48 \\
 & (0.02) & (1.31) & (0.06) & (3.57) \\
 &  &  &  &  \\
Observations & 31,702 & 31,702 & 31,702 & 31,702 \\
R-squared & 0.060 & 0.014 & 0.069 & 0.041 \\
County FE & Yes & Yes & Yes & Yes \\
Year FE & Yes & Yes & Yes & Yes \\
Region x Year FE & Yes & Yes & Yes & Yes \\
 Robust SE & Cluster & Cluster & Cluster & Cluster \\ \hline
\multicolumn{5}{c}{ Robust standard errors in parentheses} \\
\multicolumn{5}{c}{ *** p$<$0.01, ** p$<$0.05, * p$<$0.1} \\
\end{tabular}

	\end{adjustbox}
\end{table}

\begin{table}[h]
	\caption {CRA Farm Loans and Climate Vulnerability, Middle Income Areas} \label{t5realincomeg3} 
	\begin{adjustbox}{width=\columnwidth,center}
		\begin{tabular}{lcccc} \hline
 & (1) & (2) & (3) & (4) \\
VARIABLES & Num. to large farms & Amount to large farms & Num. to small-mid farms & Amount to small-mid farms \\ \hline
 &  &  &  &  \\
Temp. anomaly & -0.36*** & -27.07*** & 2.10*** & 34.26*** \\
 & (0.07) & (6.14) & (0.21) & (12.79) \\
Temp. anomaly (square) & 0.16*** & 11.39*** & -0.87*** & -9.49* \\
 & (0.03) & (2.24) & (0.09) & (5.19) \\
 &  &  &  &  \\
Observations & 65,603 & 65,603 & 65,603 & 65,603 \\
R-squared & 0.172 & 0.078 & 0.130 & 0.049 \\
County FE & Yes & Yes & Yes & Yes \\
Year FE & Yes & Yes & Yes & Yes \\
Region x Year FE & Yes & Yes & Yes & Yes \\
 Robust SE & Cluster & Cluster & Cluster & Cluster \\ \hline
\multicolumn{5}{c}{ Robust standard errors in parentheses} \\
\multicolumn{5}{c}{ *** p$<$0.01, ** p$<$0.05, * p$<$0.1} \\
\end{tabular}

	\end{adjustbox}
\end{table}

\begin{table}[!]
	\caption {CRA Farm Loans and Climate Vulnerability, High Income Areas} \label{t5realincomeg4} 
	\begin{adjustbox}{width=\columnwidth,center}
		\begin{tabular}{lcccc} \hline
 & (1) & (2) & (3) & (4) \\
VARIABLES & Num. to large farms & Amount to large farms & Num. to small-mid farms & Amount to small-mid farms \\ \hline
 &  &  &  &  \\
Temp. anomaly & 0.03 & -1.68 & 0.21 & -1.32 \\
 & (0.07) & (7.59) & (0.14) & (10.53) \\
Temp. anomaly (square) & 0.01 & 4.22* & -0.22*** & 3.90 \\
 & (0.03) & (2.22) & (0.07) & (5.39) \\
 &  &  &  &  \\
Observations & 29,308 & 29,308 & 29,308 & 29,308 \\
R-squared & 0.154 & 0.060 & 0.054 & 0.026 \\
County FE & Yes & Yes & Yes & Yes \\
Year FE & Yes & Yes & Yes & Yes \\
Region x Year FE & Yes & Yes & Yes & Yes \\
 Robust SE & Cluster & Cluster & Cluster & Cluster \\ \hline
\multicolumn{5}{c}{ Robust standard errors in parentheses} \\
\multicolumn{5}{c}{ *** p$<$0.01, ** p$<$0.05, * p$<$0.1} \\
\end{tabular}

	\end{adjustbox}
	\begin{tablenotes}
		\vspace{5pt}
		\scriptsize
		\item \textit{Note}: Regression specification for Tables \ref{t5realincomeg2} to \ref{t5realincomeg4} is Equation (\ref{e2cen}), 
		$
		y_{ict}=\beta_{1} T_{i t} + \beta_{2} T^2_{i t}+ u_{i}+ \xi_{c} + \eta_{t} +\lambda_{rt} +\gamma_{i} trend + \gamma_{i2} trend^2+e_{i ct}
		$, and for the sample of moderate income area, middle income area, and high income area respectively. 
	\end{tablenotes}
\end{table}

\clearpage
\subsection{Regional Heterogeneity}\label{hetero-main}
The analysis thus far examines farms in the United States as a whole, controlling for unobserved county fixed effects. But these farm systems are not monolithic, and it is useful to examine their heterogeneity. In this section, I conduct such analysis using the Farm Resource Regions categorized by USDA Economic Research Service (ERS). Such regions are constructed based on similarities in land resources, clusters of farming characteristics, and dominant and specialty crops (\cite{USDA2000}). The main reason for such analysis using Farm Resource Regions is to account for the possibility that climate change impact on lending to farm areas is not homogeneous. For example, depending on the dominant crops, the degree of vulnerability in terms of production loss differs between regions, thus potentially leading to differential lending outcomes. 

\begin{figure}[h!]
	\centering
	\includegraphics[width=14.5cm]{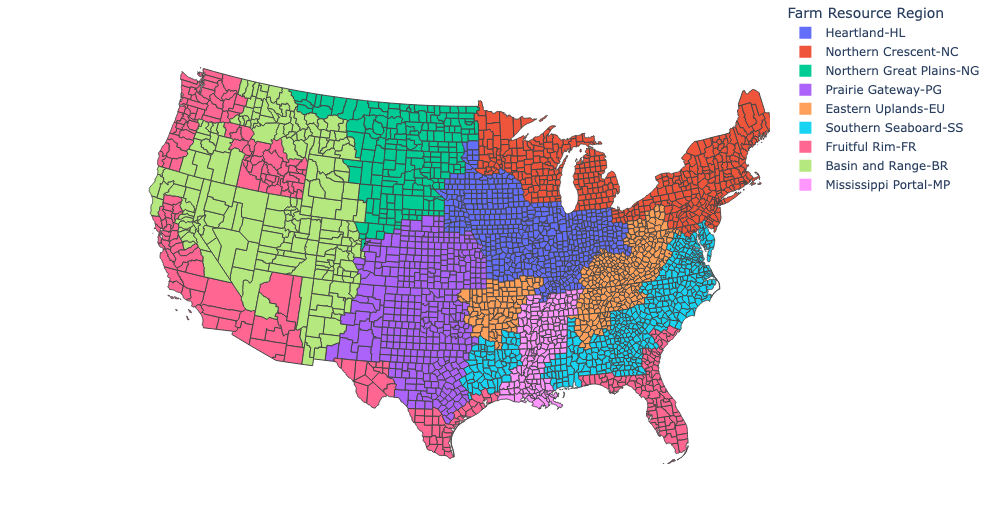}
	\caption{Farm Resource Regions}\label{f7}
	{\scriptsize Source: USDA ERS}
	
\end{figure}

As illustrated by Figure (\ref{f7}), the Farm Resource Regions include:
\begin{inparaenum}[1--]
	\item Heartland (HT),
	\item Northern Crescent (NC),
	\item Northern Great Plains (NG),
	\item Prairie Gateway (PG),
	\item Eastern Uplands (EU),
	\item Southern Seaboard (SS),
	\item Fruitful Rim (FR),
	\item Basin and Range (BR),
	\item Mississippi Portal (MP).
\end{inparaenum}

\begin{figure}[!]
	\centering
	\caption{Marginal Effect of Temperature on CRA Lending, Regional Heterogeneity}\label{fmargin2} 
	\begin{subfigure}{7cm}
		\centering\includegraphics[width=7cm]{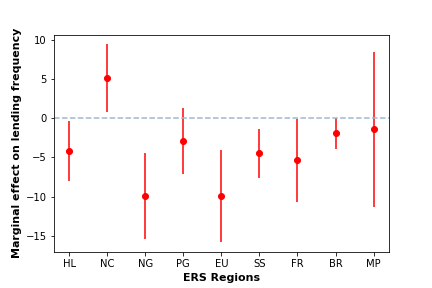}
		\caption{\tiny{Frequencies to Small-Medium Farms}}
	\end{subfigure}%
	\begin{subfigure}{7cm}
		\centering\includegraphics[width=7cm]{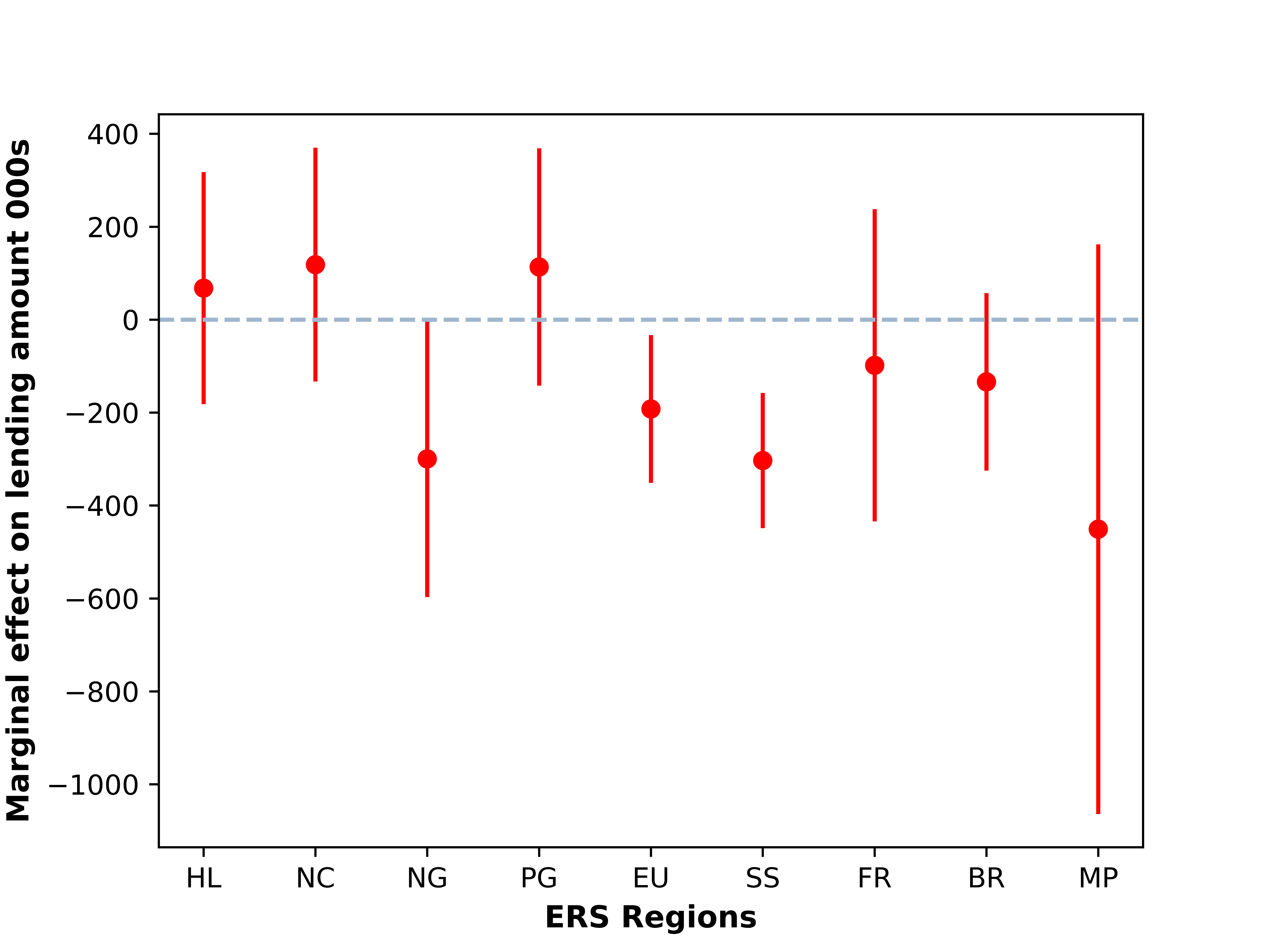}
		\caption{\tiny{Amount to Small-Medium Farms}}
	\end{subfigure}
	\begin{subfigure}{7cm}
		\centering\includegraphics[width=7cm]{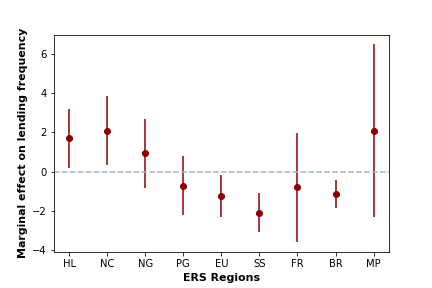}
		\caption{\tiny{Frequencies to Large Farms}}
	\end{subfigure}
	\begin{subfigure}{7cm}
		\centering\includegraphics[width=7cm]{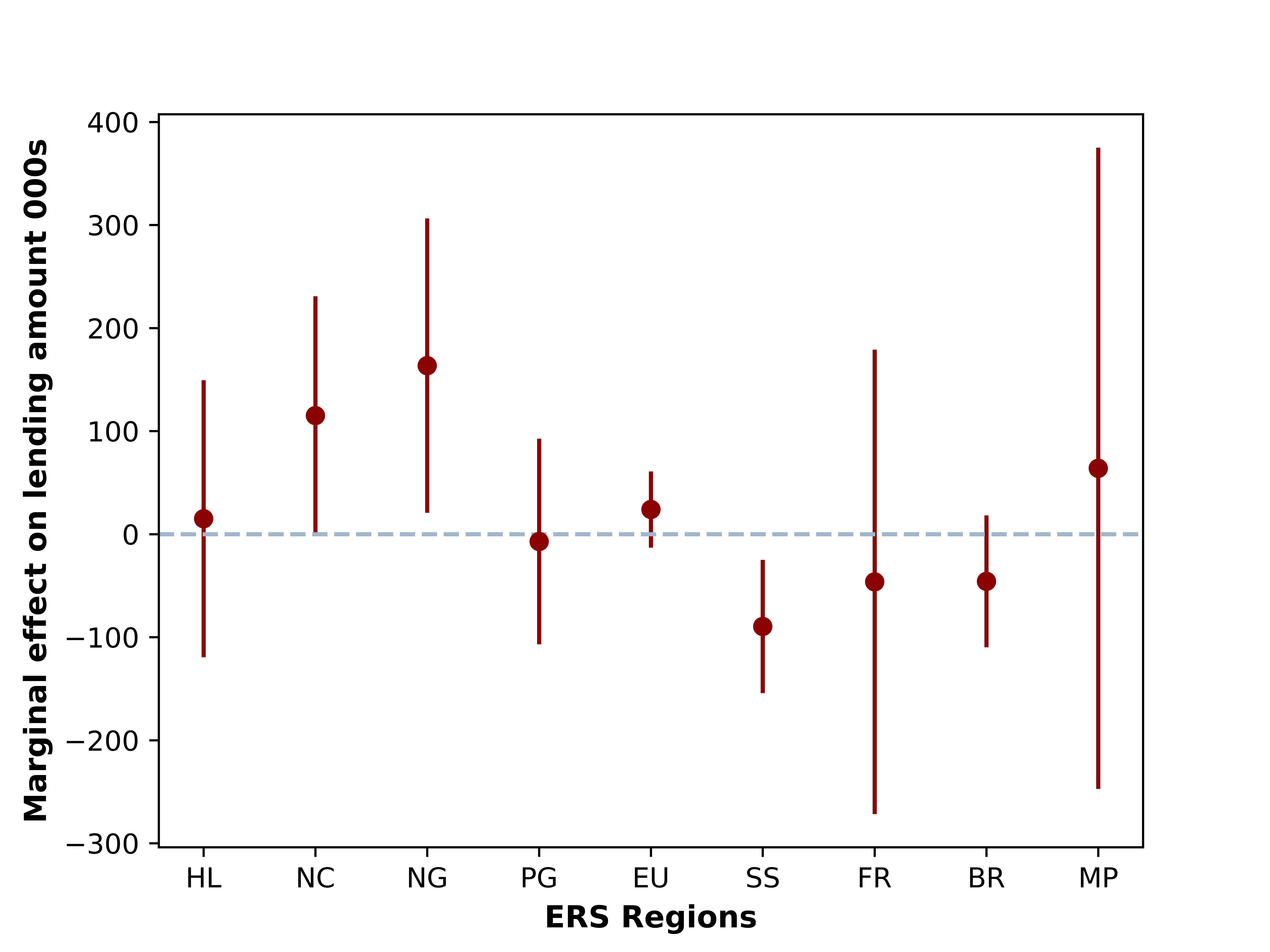}
		\caption{\tiny{Amount to Large Farms}}
	\end{subfigure}\par
	{\tiny	\textit{Note} Climate scenario:  medium term (2041-2060) and `net zero by 2100' (SSP2-4.5), equivalent to 2.2 ${}^{\circ}C$/ 3.96${}^{\circ}F$. Marginal effects with  95\% confidence intervals, calculated based on Equation (\ref{margin1}); The projected temperature estimates are from IPCC. The projections are specific to continuous United States, and refer to permanent increase of annual maximum temperature. The projections are based on CMIP6 Model, taking into account emission uncertainty, and use 1986- 2005 as baseline. \par}
\end{figure}

Following Equation (\ref{margin1}), I estimate the marginal effects for different ERS regions in the  medium term (2041-2060) and under the climate scenario of `net zero by 2100' (SSP2-4.5), equivalent to increase of high temperature by 2.2 Celsius or 3.96 Fahrenheit. The results are in shown Figure (\ref{fmargin2}).  Compare with Figure  (\ref{fmargin1}), the results here are overall consistent: small-medium farms suffer negative impact, while the effects large farms are small and even positive in some cases. However, there are also considerable regional heterogeneity. For example, for all farms in Northern Crescent (NC), increase in temperature anomaly have only positive impact. In contrast, the impact is negative across all categories for the regions of Basin and Range (BR) and Southern Seaboard (SS). 

By matching each county with its region category, I also run the regression specified in Equation (\ref{e5}) for each of the farm regions. The region ``Fruitful Rim" is quite broad, encompassing not only western United States, but also Texas and Florida. However, they are in different climate regions. Thus the analysis decomposes this region into:
\begin{inparaenum}[1)]
	\item parts of Texas and Florida;
	\item the rest of the fruitful rim.\footnote{The regression tables for individual regions are all reported in available upon request. }
\end{inparaenum}


Here I focus on discussing the cumulative effects, calculated using Equation (\ref{e6}). Table (\ref{t11}) presents the calculations based on the scenario in which there is a permanent increase of 3 Fahrenheit degrees of temperature anomaly. The results for small-medium farms are shown in Columns (3) and (4). In terms of the number of loans, there are reductions in almost all the regions except the Northern Crescent. As suggested by the previous results in Table ({\ref{t7}}), the national average of reduced loan number is 14. It seems that the two regions that experience the most drastic decrease are: Northern Great Plains; the Texan and Floridan parts of the Fruitful Rim. The results for the amounts of loans to small-medium farms are more mixed. The effect of temperature anomaly is small or positive for four of the regions. For regions that experience reduction in amount of loans, the effect is particularly pronounced for the Fruitful Rim overall. In sharp contrast, as shown in Columns (1) and (2), for large farms, half of all the regions experience an increase in amount of loans, and 40\% of the regions experience increase in the number of loans. 

Table (\ref{t11}) illustrates the regional heterogeneity of bank lending, and is broadly consistent with the national results in Table (\ref{t7}): \begin{inparaenum}[1)]
	\item The effect of climate change on lending to small-medium farms is generally negative; 
	\item In contrast, large farms seem to fare better;
	\item There seems to be shift of lending from small-medium to large farms in regions such as the Northern Great Plains, southeastern parts of the Fruitful Rim, and the Mississippi Portal;
\end{inparaenum}

\begin{table}[h]
	\centering
	\caption{Cumulative Effects of 3 Fahrenheit Degrees of High Temp. Anomaly}\label{t11}
		\centering
		\begin{tabular}{lcc}
			\toprule
			\toprule
			& (1) & (2) \\
			\textbf{Region} & \textbf{Num. to large} & \textbf{Num. to small-mid} 
			\\ \hline
			\textit{Heartland} & 2.4   & -6.45  \\
			\midrule
			\textit{Northern Crescent} & 2.27  & 8.22   \\
			\midrule
			\textit{Northern Great Plains} & 4.24  & -25.17 \\
			\midrule
			\textit{Prairie Gateway} & -4.5   & -10.46 \\
			\midrule
			\textit{Eastern Uplands} & -0.53  & -9.25  \\
			\midrule
			\textit{Southern Seaboard} & -5.49  & -8.3  \\
			\midrule
			\textit{Fruitful Rim (TX and FL)} & -4.82 & -17.48 \\
			\midrule
			\textit{Fruitful Rim (remaining)} & -13.96  & -12.78  \\
			\midrule
			\textit{Basin and Range} & -1.79  & -8.21 \\
			\midrule
			\textit{Mississippi Portal} & 3.91  & -2.54 \\
			\bottomrule
			\bottomrule
		\end{tabular}%
	\begin{tablenotes}
		\vspace{5pt}
		\scriptsize
		\item \textit{Note}: Results calculated using Equation (\ref{e6}), 
		$
		\sum_{l=0}^{2}{\beta}_{l} + \sum_{m=0}^{2}{\theta}_{m}= \beta_{0}  + 2 \theta_{0} T^* +  \beta_{1}  + 2 \theta_{1} T^*  +  \beta_{2}  + 2 \theta_{2} T^* 
		$
	\end{tablenotes}
\end{table}%

\clearpage
\subsubsection{Extension: Bank-County Level--Census Tract Income Areas}
The analyses of bank-county pair can be expanded further by considering which income areas farms are located in. Thus the unit of analysis here becomes bank-county-income group pair. The econometric specification becomes
\begin{equation}\label{e2bankincome} 
	y_{ibct}=\beta_{1} T_{i t} + \beta_{2} T^2_{i t}+ u_{i}+ \psi_{b} +  \xi_{c} + \eta_{t} +\lambda_{rt} +\gamma_{i} trend + \gamma_{i2} trend^2+e_{ibct}
\end{equation}

The estimation now includes more granular data of how much a bank $b$ lends in income area $c$ of county $i$, and income group fixed effect  $\xi_{c}$ is included. 

The results for all income groups,\footnote{The tables of the results are upon request} estimated by bank sizes\footnote{Note the current estimation includes bank income group fixed effect, not bank-level fixed effect. The income group is classified based on the yearly decile distribution of banks' assets (of the prior year)}, are mostly identical to the results in Tables (\ref{t5realincomeg0bankvs}-\ref{t5realincomeg0bankl}). In short, large banks in general are less willing to lend, small-mid banks lend less to small farms and more to large farms, and very small banks lend more to small farms.

As suggested by similar analysis in Section \ref{countyresults}, it is necessary to understand the heterogeneity of impact by income groups: in general, farms located middle income areas of counties bear the brunt the effect of climate risks. The results in Section \ref{countyresults} are at county level. 
This pattern continues to hold here at bank-county-income group level,
 as seen in additional tables in Appendix \ref{banktables}. 

In summary, the results in this section confirm that at bank level, farms' vulnerability to climate risks still matter for lending. More specifically, as climate risks increase, very small banks maintain or lend more to small farms. Small-mid size banks lend more to large farms, and lend less to small farms. Large banks uniformly lend less to big and small farms in terms of frequencies. The results are largely consistent with the aggregate county-level data, but with richer details by bank characteristics.\footnote{The correspondence between the county-level and bank-county level results is not immediately straightforward. A paper by \cite{Morgan2022} suggests banks themselves are fairly resilient to natural disasters: in general, their balance sheets are not hit hard in a significant way. In short, analysis of climate effect at bank level does not always point to significant results.}

\clearpage 
\subsection{Tables of Bank Level Estimates}\label{banktables}
\begin{table}[ht]
	\caption {Number of CRA Farm Loans and Climate Vulnerability, Bank-County Level by Loan Size} \label{loansize1} 
	\begin{adjustbox}{width=\columnwidth,center}
		\begin{tabular}{lcccc} \hline
 & (1) & (2) & (3) & (4) \\
VARIABLES & Total Num. of Loans & Number of loan (less 100k) & Number of loan (100k to 250k) & Number of loan (250k to 500k) \\ \hline
 &  &  &  &  \\
High temperature anomaly & 0.12 & 0.25*** & -0.07*** & -0.06*** \\
 & (0.08) & (0.07) & (0.01) & (0.01) \\
High temperature anomaly (square) & -0.17*** & -0.22*** & 0.02*** & 0.03*** \\
 & (0.04) & (0.04) & (0.01) & (0.00) \\
Mid-size bank = 1 & 2.62*** & 1.87*** & 0.48*** & 0.26*** \\
 & (0.61) & (0.55) & (0.08) & (0.05) \\
Large-size bank = 1 & 3.72*** & 1.53** & 1.34*** & 0.85*** \\
 & (0.74) & (0.67) & (0.11) & (0.07) \\
Small bank $\times$ Temperature & -0.12 & -0.22 & 0.07** & 0.02* \\
 & (0.20) & (0.18) & (0.03) & (0.01) \\
Small bank $\times$ Temperature (square) & 0.30*** & 0.32*** & -0.01 & -0.01 \\
 & (0.11) & (0.11) & (0.02) & (0.01) \\
Large bank $\times$ Temperature & -0.00 & -0.10 & 0.06*** & 0.04*** \\
 & (0.08) & (0.07) & (0.01) & (0.01) \\
Large bank $\times$ Temperature (square) & 0.12*** & 0.16*** & -0.02*** & -0.02*** \\
 & (0.04) & (0.04) & (0.01) & (0.00) \\
 &  &  &  &  \\
Observations & 480,898 & 480,898 & 480,898 & 480,898 \\
R-squared & 0.018 & 0.023 & 0.011 & 0.016 \\
County FE & Yes & Yes & Yes & Yes \\
Year FE & Yes & Yes & Yes & Yes \\
Bank FE & Yes & Yes & Yes & Yes \\
 Robust SE & Cluster & Cluster & Cluster & Cluster \\ \hline
\multicolumn{5}{c}{ Robust standard errors in parentheses} \\
\multicolumn{5}{c}{ *** p$<$0.01, ** p$<$0.05, * p$<$0.1} \\
\end{tabular}

	\end{adjustbox}
\end{table}

\begin{table}[ht]
	\caption {Amount of CRA Farm Loans and Climate Vulnerability, Bank-County Level by Loan Size} \label{loansize2} 
	\begin{adjustbox}{width=\columnwidth,center}
		\begin{tabular}{lcccc} \hline
 & (1) & (2) & (3) & (4) \\
VARIABLES & Total Amt. of Loans & Amount of loan (less 100k) & Amount of loan (100k to 250k) & Amount of loan (250k to 500k) \\ \hline
 &  &  &  &  \\
High temperature anomaly & -40.80*** & -3.99*** & -14.95*** & -21.86*** \\
 & (5.50) & (1.49) & (2.05) & (3.11) \\
High temperature anomaly (square) & 15.95*** & -0.11 & 5.96*** & 10.10*** \\
 & (2.54) & (0.65) & (0.96) & (1.51) \\
Small-size bank = 1 & -240.40*** & -69.47*** & -81.66*** & -89.27*** \\
 & (36.54) & (10.90) & (13.04) & (18.35) \\
Large-size bank = 1 & 477.37*** & 79.01*** & 168.99*** & 229.38*** \\
 & (31.97) & (8.08) & (12.16) & (15.79) \\
Small bank $\times$ Temperature & 22.71** & 4.53 & 11.96*** & 6.22 \\
 & (10.74) & (3.55) & (4.08) & (4.87) \\
Small bank $\times$ Temperature (square) & -5.05 & 2.06 & -3.38 & -3.72 \\
 & (5.89) & (1.92) & (2.12) & (2.84) \\
Large bank $\times$ Temperature & 29.03*** & 4.64*** & 10.65*** & 13.73*** \\
 & (4.99) & (1.42) & (1.90) & (2.79) \\
Large bank $\times$ Temperature (square) & -12.86*** & -0.47 & -4.63*** & -7.76*** \\
 & (2.54) & (0.66) & (0.97) & (1.50) \\
 &  &  &  &  \\
Observations & 480,898 & 480,898 & 480,898 & 480,898 \\
R-squared & 0.025 & 0.009 & 0.020 & 0.028 \\
County FE & Yes & Yes & Yes & Yes \\
Year FE & Yes & Yes & Yes & Yes \\
Bank FE & Yes & Yes & Yes & Yes \\
 Robust SE & Cluster & Cluster & Cluster & Cluster \\ \hline
\multicolumn{5}{c}{ Robust standard errors in parentheses} \\
\multicolumn{5}{c}{ *** p$<$0.01, ** p$<$0.05, * p$<$0.1} \\
\end{tabular}

	\end{adjustbox}
\end{table}

\begin{table}[ht]
	\caption {CRA Farm Loans and Climate Vulnerability, Bank-County Level} \label{tbankcountyinteract} 
	\begin{adjustbox}{width=\columnwidth,center}
		\begin{tabular}{lcccc} \hline
 & (1) & (2) & (3) & (4) \\
VARIABLES & Num. of loans to large farms & Amount of loans to large farms & Num. of loans to small-medium farms & Amount of loans to small-medium farms \\ \hline
 &  &  &  &  \\
High temperature anomaly & -0.10*** & -13.64*** & 0.22*** & -27.16*** \\
 & (0.03) & (2.59) & (0.08) & (4.03) \\
High temperature anomaly (square) & 0.01 & 7.28*** & -0.18*** & 8.67*** \\
 & (0.01) & (1.30) & (0.04) & (1.74) \\
Mid-size bank (= 1 if yes) & -0.15 & 38.43*** & 2.77*** & 201.97*** \\
 & (0.16) & (13.99) & (0.59) & (28.58) \\
Large-size bank (= 1 if yes) & 0.49** & 186.18*** & 3.23*** & 531.59*** \\
 & (0.20) & (19.92) & (0.71) & (38.12) \\
Small bank $\times$ Temperature & 0.17*** & 10.02** & -0.29 & 12.69 \\
 & (0.05) & (4.00) & (0.19) & (8.37) \\
Small bank $\times$ Temperature (square) & -0.02 & -3.85* & 0.33*** & -1.20 \\
 & (0.02) & (2.08) & (0.11) & (4.66) \\
Large bank $\times$ Temperature & 0.12*** & 10.71*** & -0.12* & 18.31*** \\
 & (0.03) & (2.35) & (0.07) & (3.71) \\
Large bank $\times$ Temperature (square) & -0.04*** & -6.55*** & 0.16*** & -6.31*** \\
 & (0.01) & (1.30) & (0.04) & (1.74) \\
 &  &  &  &  \\
Observations & 480,898 & 480,898 & 480,898 & 480,898 \\
R-squared & 0.005 & 0.021 & 0.025 & 0.016 \\
County FE & Yes & Yes & Yes & Yes \\
Year FE & Yes & Yes & Yes & Yes \\
Bank FE & Yes & Yes & Yes & Yes \\
 Robust SE & Cluster & Cluster & Cluster & Cluster \\ \hline
\multicolumn{5}{c}{ Robust standard errors in parentheses} \\
\multicolumn{5}{c}{ *** p$<$0.01, ** p$<$0.05, * p$<$0.1} \\
\end{tabular}

	\end{adjustbox}
\end{table}

\begin{table}[ht]
	\caption {CRA Farm Loans and Climate Vulnerability, Very Small Banks} \label{t5realincomeg0bankvs} 
	\begin{adjustbox}{width=\columnwidth,center}
		\begin{tabular}{lcccc} \hline
 & (1) & (2) & (3) & (4) \\
VARIABLES & Num. of loans to large farms & Amount of loans to large farms & Num. of loans to small-medium farms & Amount of loans to small-medium farms \\ \hline
&  &  &  &  \\
High temperature anomaly & -0.09 & -4.51 & 0.15 & 9.68 \\
 & (0.13) & (4.38) & (0.34) & (13.20) \\
High temperature anomaly (square) & 0.03 & 1.70 & 0.13 & 7.06 \\
 & (0.04) & (3.34) & (0.14) & (5.34) \\
 &  &  &  &  \\
Observations & 14,055 & 14,055 & 14,055 & 14,055 \\
R-squared & 0.023 & 0.037 & 0.024 & 0.057 \\
County FE & Yes & Yes & Yes & Yes \\
Year FE & Yes & Yes & Yes & Yes \\
Bank FE & Yes & Yes & Yes & Yes \\
Region x Year FE & Yes & Yes & Yes & Yes \\
 Robust SE & Cluster & Cluster & Cluster & Cluster \\ \hline
\multicolumn{5}{c}{ Robust standard errors in parentheses} \\
\multicolumn{5}{c}{ *** p$<$0.01, ** p$<$0.05, * p$<$0.1} \\
\end{tabular}

	\end{adjustbox}
\end{table}

\begin{table}[ht]
	\caption {CRA Farm Loans and Climate Vulnerability, Small-Mid Banks} \label{t5realincomeg0bankm} 
	\begin{adjustbox}{width=\columnwidth,center}
		\begin{tabular}{lcccc} \hline
 & (1) & (2) & (3) & (4) \\
VARIABLES & Num. of loans to large farms & Amount of loans to large farms & Num. of loans to small-medium farms & Amount of loans to small-medium farms \\ \hline
 &  &  &  &  \\
High temperature anomaly & 0.02 & 3.46 & -0.14 & -8.92 \\
 & (0.04) & (2.77) & (0.11) & (5.51) \\
High temperature anomaly (square) & -0.00 & 0.32 & -0.14*** & -0.66 \\
 & (0.01) & (1.11) & (0.05) & (2.05) \\
 &  &  &  &  \\
Observations & 66,647 & 66,647 & 66,647 & 66,647 \\
R-squared & 0.014 & 0.074 & 0.036 & 0.039 \\
County FE & Yes & Yes & Yes & Yes \\
Year FE & Yes & Yes & Yes & Yes \\
Bank FE & Yes & Yes & Yes & Yes \\
Region x Year FE & Yes & Yes & Yes & Yes \\
 Robust SE & Cluster & Cluster & Cluster & Cluster \\ \hline
\multicolumn{5}{c}{ Robust standard errors in parentheses} \\
\multicolumn{5}{c}{ *** p$<$0.01, ** p$<$0.05, * p$<$0.1} \\
\end{tabular}

	\end{adjustbox}
\end{table}

\begin{table}[ht]
	\caption {CRA Farm Loans and Climate Vulnerability, Large Banks} \label{t5realincomeg0bankl} 
	\begin{adjustbox}{width=\columnwidth,center}
		\begin{tabular}{lcccc} \hline
 & (1) & (2) & (3) & (4) \\
VARIABLES & Num. of loans to large farms & Amount of loans to large farms & Num. of loans to small-medium farms & Amount of loans to small-medium farms \\ \hline
 &  &  &  &  \\
High temperature anomaly & 0.04*** & -0.57 & 0.08*** & -4.64*** \\
 & (0.01) & (0.76) & (0.02) & (1.24) \\
High temperature anomaly (square) & -0.03*** & 0.14 & -0.03*** & 0.45 \\
 & (0.00) & (0.25) & (0.01) & (0.40) \\
 &  &  &  &  \\
Observations & 400,196 & 400,196 & 400,196 & 400,196 \\
R-squared & 0.008 & 0.011 & 0.033 & 0.006 \\
County FE & Yes & Yes & Yes & Yes \\
Year FE & Yes & Yes & Yes & Yes \\
Bank FE & Yes & Yes & Yes & Yes \\
Region x Year FE & Yes & Yes & Yes & Yes \\
 Robust SE & Cluster & Cluster & Cluster & Cluster \\ \hline
\multicolumn{5}{c}{ Robust standard errors in parentheses} \\
\multicolumn{5}{c}{ *** p$<$0.01, ** p$<$0.05, * p$<$0.1} \\
\end{tabular}

	\end{adjustbox}
\end{table}

\end{appendix}

	\clearpage
\setcitestyle{numbers}

\bibliographystyle{apalike} 
\bibliography{Bibliography_bankcopy} 

\end{document}